\documentclass[12pt]{article}

\usepackage[margin=1.0in]{geometry}

\usepackage[T1]{fontenc}
\usepackage{baskervald}

\usepackage[font=small, labelfont=bf,skip=1pt]{caption}
\usepackage{graphicx}
\usepackage{hyperref}
\usepackage{amsmath}
\usepackage{amssymb}
\usepackage{longtable}
\usepackage{verbatim}
\usepackage{cite}

\usepackage[usenames,dvipsnames]{xcolor}

\newcommand{\be}{\begin{equation}}
\newcommand{\ee}{\end{equation}}
\newcommand{\ba}{\begin{align}}
\newcommand{\ea}{\end{align}}
\usepackage{titlesec}

\newcommand{\tB}{{\widetilde B}}
\newcommand{\tW}{{\widetilde W}}
\newcommand{\tH}{{\widetilde H}}

\DeclareFontFamily{U}{mathx}{\hyphenchar\font45}
\DeclareFontShape{U}{mathx}{m}{n}{<-> mathx10}{}
\DeclareSymbolFont{mathx}{U}{mathx}{m}{n}
\DeclareMathAccent{\widebar}{0}{mathx}{"73}

\newcommand{\iu}{{\mathrm i}}

\usepackage{tikz}
\usetikzlibrary{trees}
\usetikzlibrary{decorations.pathmorphing}
\usetikzlibrary{decorations.markings}
\tikzset{
    photon/.style={decorate, decoration={snake}, draw=black},
    wino/.style={draw=redwine},    
    electron/.style={draw=black, postaction={decorate},
        decoration={markings,mark=at position .55 with {\arrow[draw=black]{>}}}},
    scalar/.style={draw=black, dashed,postaction={decorate},
        decoration={markings,mark=at position .55 with {\arrow[draw=black]{>}}}},
    gluon/.style={decorate, draw=black,
        decoration={coil,amplitude=4pt, segment length=5pt}}
}

\title{\vspace{-1cm} \bf \Large Electric Dipole Moments in Natural Supersymmetry \vspace{0.25cm}}

\author{Yuichiro Nakai$^a$ and Matthew Reece$^{a,b}$\\
{\small \color{gray} \texttt{ynakai, mreece~(@physics.harvard.edu)} \vspace{0.1cm}}\\
{\small $^a$ Department of Physics, Harvard University, Cambridge, MA, 02138}\\
{\small $^b$ School of Natural Sciences, Institute for Advanced Study, Princeton, NJ 08540}}

\begin{document}
\maketitle

\begin{abstract}

\vspace{1mm}

We discuss electric dipole moments (EDMs) in the framework of CP-violating natural supersymmetry (SUSY). Recent experimental results have significantly tightened constraints on the EDMs of electrons and of mercury, and substantial further progress is expected in the near future. We assess how these results constrain the parameter space of natural SUSY. In addition to our discussion of SUSY, we provide a set of general formulas for two-loop fermion EDMs, which can be applied to a wide range of models of new physics. In the SUSY context, the two-loop effects of stops and charginos respectively constrain the phases of $A_t \mu$ and $M_2 \mu$ to be small in the natural part of parameter space. If the Higgs mass is lifted to 125 GeV by a new tree-level superpotential interaction and soft term with CP-violating  phases, significant EDMs can arise from the two-loop effects of $W$ bosons and tops. We compare the bounds arising from EDMs to those from other probes of new physics including colliders, $b \to s \gamma$, and dark matter searches. Importantly, improvements in reach not only constrain higher masses, but require the phases to be significantly smaller in the natural parameter space at low mass. The required smallness of phases sharpens the CP problem of natural SUSY model building.

\end{abstract}

%
%

\section{Introduction}

Supersymmetry (SUSY) is a promising solution to the hierarchy problem
\cite{Martin:1997ns}.
However, many new parameters induce significant flavor mixing or CP-violating processes
which are severely constrained by experiments.
Hence, for generic SUSY breaking, superparticles are required to be very heavy, reintroducing the (little) hierarchy problem.
One way to address this problem is to consider a carefully chosen mechanism of SUSY breaking
such as gauge mediation (for reviews, see refs.~\cite{Giudice:1998bp,Kitano:2010fa}).
On the other hand, since only the first two generations of squarks and sleptons are relevant for flavor constraints,
another possible path is to assume only stops (and the left-handed sbottom) and higgsinos
are light. Importantly, it is these particles that are essential for the naturalness of electroweak symmetry breaking (EWSB).
This spectrum is also favored from the point of view that direct bounds on stop and higgsino masses do not cause severe fine-tuning
in spite of great advances in SUSY searches at the Large Hadron Collider (LHC).
This framework is called natural SUSY
\cite{Dimopoulos:1995mi,Pomarol:1995xc,Cohen:1996vb}, and is expected to ameliorate both direct collider constraints as well as indirect constraints on CP violation. LHC searches have begun to encroach on the territory favored by natural SUSY, so that even models of quite low-scale mediation of supersymmetry breaking are likely to be tuned by a factor of 10 or 100, though there is some room to evade the most severe bounds (for a wide range of perspectives on the current status of direct superpartner searches, see \cite{Arvanitaki:2013yja,Evans:2013jna,Baer:2015tva,Fan:2015mxp,Kim:2016rsd,Buckley:2016kvr}).

Meanwhile, the last few years have seen significant progress in precision low-energy tests of CP violation. In 2013, the ACME collaboration improved the bound on the electron electric dipole moment (EDM) by an order of magnitude using thorium monoxide \cite{Baron:2013eja}. Within the last year, the bound on the mercury EDM has also been improved by a factor of 4 by Graner et al.~\cite{Graner:2016ses}. These are highly constraining probes of new CP-violating physics, the former being an effective probe of electroweak CP violation and the latter of CP violation in physics coupling to QCD. Because EDMs are dimension six operators, roughly speaking the scale of new physics probed by an experiment scales as the square root of the improvement in the EDM bound, so these results have pushed the mass reach of EDMs up by a factor of 2 or 3. Alternatively, for new physics of fixed mass, the constraint on new CP-violating phases scales directly with the improved bound. Because naturalness favors new physics at low mass, the required smallness of CP-violating phases is an important constraint on model building. Our goal in this paper is to quantify what the current and near-future EDM results tell us about the extent to which CP-violation is allowed in natural SUSY.

The effects of supersymmetric particles on EDMs have been extensively studied over many years. For instance, the two-loop Barr-Zee-type diagrams \cite{Barr:1990vd} generate electron and quark EDMs and chromo-EDMs (CEDMs) \cite{Kadoyoshi:1996bc,Chang:1998uc,Chang:1999zw,Pilaftsis:2002fe,Giudice:2005rz,Li:2008kz}. Possible SUSY contributions to the experimentally measured EDMs have been extensively studied especially in the minimal supersymmetric standard model (MSSM) \cite{Lebedev:2002ne,Demir:2003js,Ellis:2008zy}. However, it is timely to revisit the impact of supersymmetry on EDMs in light of the scenarios that are currently most favored by the LHC. One of the most important revisions to our understanding of supersymmetry is the knowledge that the Higgs boson mass is an unexpectedly heavy 125 GeV. In the context of the MSSM, achieving this mass without completely giving up on naturalness requires large values of the left-right stop mixing parameter $A_t$ (see \cite{Draper:2011aa} and references therein). This tends to enhance the expected EDMs, for generic CP-violating phases. Beyond the MSSM, other new physics operating at tree level could explain why the Higgs is so heavy. Such physics generally introduces new possible sources of CP violation, which have received little attention so far (though see \cite{Blum:2010by,Altmannshofer:2011rm}). Thus, both due to improvements in constraints on EDMs themselves and our altered perspective on the most plausible forms that supersymmetry can take, it is timely to revisit the assessment of what EDM experiments can tell us about SUSY.

The next few sections of this paper introduce some background material and develop general results that will be useful later. They may be skipped by readers who are interested in our conclusions but do not need all of the details of the derivations. In \S\ref{twoloopfermions}, we present a set of very general results for the two loop fermion EDMs (and, with straightforward changes, CEDMs) induced by an inner loop of charged particles connected to an external fermion by one scalar and one vector ($\gamma h$, $\gamma A$, $Zh$, $W^\mp H^\pm$, etc.). Although we are aware of many specific results in the literature that rely on such calculations, we are not aware of a previous reference that presents formulas valid for completely generic couplings of the new particles, so we hope that these results may be useful for readers studying a range of models not limited to SUSY.\footnote{
Several analytic two-loop results in the CP-conserving case have been presented in refs.~\cite{Gribouk:2005ee,Cherchiglia:2016eui}.}
 In \S\ref{naturalSUSY}, we review the basics of natural SUSY including BMSSM effects to lift the Higgs mass \cite{Dine:2007xi}, and establish our notation. In \S\ref{EDMcalculation}, we provide a summary of the experimental status of paramagnetic, diamagnetic, and neutron EDM measurements. In each case, we review which operators contribute: these include not just fundamental fermion EDMs but also CP-violating four-fermion operators. We summarize the computation of all of the ingredients, referring to appendices for some of the details.

With this preparation out of the way, the core results of the paper (where many readers may wish to start their reading) begin in \S\ref{numerical} with a look at how two-loop effects of stops constrain the phase $\arg(A_t \mu)$.\footnote{Cancellation among
several contributions to the EDMs would relax constraints on the parameter space of the MSSM
\cite{Bian:2014zka,Bian:2016zba} although the cancellation is likely to require another tuning of the parameters.}
We consider the MSSM, in which given other SUSY-breaking parameters, the size of $A_t$ is fixed (and large) to obtain the correct Higgs mass of 125 GeV; and extensions of the MSSM in which the Higgs mass is lifted to 125 GeV by unrelated physics, which we assume does not supply a dominant contribution to the EDMs. We find that the present ACME experiment and the mercury EDM measurement give comparable constraints on the stop sector (which is stronger depends on highly uncertain nuclear physics), though an updated ACME result within the next year could surpass them. Already the constraints probe parameter space well beyond the reach of the LHC for order-one phases. In much of the parameter space, EDMs are a stronger indirect probe of the stop sector than $b \to s\gamma$ unless the CP-violating phase is $10^{-3}$ or smaller. In \S\ref{sec:chargino} we consider constraints on the phase $\arg(M_2 \mu)$ from two-loop effects of charginos, which are highly constrained throughout the natural parameter space by ACME. In \S\ref{sec:BMSSM} we consider constraints on phases of a new superpotential operator and soft term introduced to lift the Higgs mass. The analysis of the EDMs in this scenario was initiated in refs.~\cite{Blum:2010by,Altmannshofer:2011rm}. We improve the analysis by including important contributions which have not been discussed in these references and are in fact dominant. We claim that severe constraints from the current EDM measurements lead to fine tuning of the electroweak breaking
when generic CP-violating phases are assumed.

Importantly, as the EDM measurements continue to improve in the future---hopefully with not just a bound but a discovery!---they will not simply continue to probe higher and more unnatural masses, but also probe deeper into the regime of small CP-violating phases at fixed mass. Since the mass reach of the EDMs for order-one phases already significantly surpasses the LHC reach, this is quite important. At one time, the existence of models like gauge mediation could have been taken as an indication that the CP problem is readily solved. However, given that the $\mu$-parameter sits uncomfortably within any model of supersymmetry breaking, one may have had qualms. The situation is now worse: in the MSSM away from the split SUSY limit, gauge mediation must be supplemented with a means of generating a sufficiently large $A_t$. Beyond the MSSM, new interactions must be added to explain the Higgs mass. The more complex the model, the more opportunities there are for nonzero phases to enter. Even if phases enter the low-energy effective theory indirectly, as experiments become more sensitive, subleading effects become visible. As a result, we think that it will become increasingly interesting to carefully study not just the low-energy theory of natural SUSY but detailed models of SUSY breaking to understand which scenarios evade the SUSY CP problem and whether they will continue to escape the downward march of experimental bounds. 

%
%

\section{Two-loop EDMs of elementary fermions}
\label{twoloopfermions}

Many two-loop (``Barr-Zee'') contributions to fermion EDMs have been computed in the literature, but are often presented in special cases \cite{Barr:1990vd,Leigh:1990kf,Pilaftsis:2002fe,Giudice:2005rz,Li:2008kz}. The general structure is an inner loop---possibly a sum of multiple diagrams---generating either an effective coupling of a photon to a scalar and vector or to two vectors. In certain limits, the EDM can be understood through a two-step operator analysis. If we compute the inner loop and take all legs on-shell, these correspond to effective Standard Model operators of higgs--vector--vector type:
\begin{align}
{\rm CP~even}:~{\cal O}_{BB} = g'^2 h^\dagger h B_{\mu \nu}B^{\mu \nu}, {\cal O}_{WW} = g^2 h^\dagger h W^i_{\mu \nu}W^{i\mu\nu}, {\cal O}_{WB} = gg' h^\dagger \sigma^i h W^i_{\mu \nu}B^{\mu \nu}, \nonumber \\
{\rm CP~odd}:~{\cal O}_{B\tB} = g'^2 h^\dagger h B_{\mu \nu}\tB^{\mu \nu}, {\cal O}_{W\tW} = g^2 h^\dagger h W^i_{\mu \nu}\tW^{i\mu\nu}, {\cal O}_{W\tB} = gg' h^\dagger \sigma^i h W^i_{\mu \nu}\tB^{\mu \nu}, \label{SMoperatorhvv}
\end{align}
or of ``W boson dipole moment'' type \cite{Marciano:1986eh,Atwood:1990cm,Kadoyoshi:1996bc}:
\begin{align}
{\rm CP~even}:~{\cal O}_{HB} = \iu g'(D_\mu h)^\dagger (D_\nu h) B^{\mu \nu}, {\cal O}_{HW} = \iu g(D_\mu h)^\dagger \sigma^i (D_\nu h) W^{i\mu \nu}, \nonumber \\
{\rm CP~odd}:~{\cal O}_{H\tB} = \iu g' (D_\mu h)^\dagger (D_\nu h) \tB^{\mu \nu}, {\cal O}_{H\tW} = \iu g(D_\mu h)^\dagger \sigma^i (D_\nu h) \tW^{i\mu \nu}.
\end{align}
Notice that ${\cal O}_{H\tB}$ and ${\cal O}_{H\tW}$ ($W$ boson EDMs) can be rewritten in terms of the first set of CP-odd operators ${\cal O}_{B\tB}, {\cal O}_{W\tW},$ and ${\cal O}_{W\tB}$ using equations of motion (see the appendix of \cite{Fan:2013qn}). The CP-even versions of these operators, ${\cal O}_{HB}$ and ${\cal O}_{HW}$, are traded for the first set of CP-even operators together with four-fermion operators in the Warsaw basis \cite{Grzadkowski:2010es} or for the first set of CP-even operators plus the operators ${\cal O}_W = \frac{\iu g}{2} (h^\dagger \sigma^i \overset{\leftrightarrow}{D^\mu} h)D^\nu W^i_{\mu \nu}$ and ${\cal O}_B =\frac{\iu g'}{2} (h^\dagger \overset{\leftrightarrow}{D^\mu} h)\partial^\nu B_{\mu \nu}$ in other commonly-used bases like that of \cite{Elias-Miro:2013eta}.\footnote{The discrepancy between CP-even operators, for which the second set does {\em not} reduce to the first set, and CP-odd operators, for which it does, is due to the fact that equations of motion set $D_\mu F^{\mu \nu} = j^\nu$ but $D_\mu {\widetilde F}^{\mu \nu}$ is {\em exactly} zero by the Bianchi identity. As a result, in the CP-even case, operators involving matter currents enter the story in a way that they do not for CP-odd operators.} (See \cite{Wells:2015uba} for a clear recent discussion of electroweak physics and translations between various operator bases.) After computing the inner loop in terms of a convenient basis of operators, we could then use the one-loop anomalous dimension matrix \cite{Alonso:2013hga} to compute how these dimension-six operators feed into the dimension-six electron EDM operators.

\begin{figure}[!h]\begin{center}
\includegraphics[clip,width=0.4\textwidth]{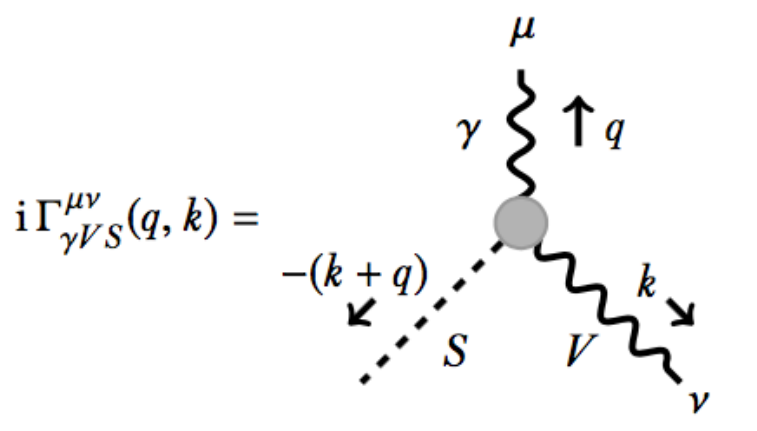}
\end{center}
\caption{Effective vertex of a photon, scalar $S$, and (in general massive) vector $V$ computed from the inner loop.}
\label{fig:innerloopvertex}
\end{figure}

In our analysis, we would like to keep multiple Higgs bosons in the theory---so that $h$ above may represent either $H_u$ or $H_d^\dagger$---and also consider regions of parameter space where particles in the inner loop may be light compared to the Higgs being exchanged in the outer loop. Hence we will work directly with two-loop computations rather than one-loop matching and RG running. In cases with large logarithms, matching and running may give more accurate (resummed) answers, but for now we apply the two-loop formulas uniformly across parameter space. However, it is still convenient to carry out the two-loop calculation by first computing the inner loop and then the outer loop---but to do so, we will leave the particles in the inner loop off-shell. For the inner loop with an on-shell photon of momentum $q$, one off-shell scalar $S$, and one off-shell vector $V$ of momentum $k$, we define the answer as $\iu \Gamma^{\mu \nu}_{\gamma V S}(q,k)$, as shown in Fig.~\ref{fig:innerloopvertex}.  Chromoelectric dipole moments are computed in a very similar manner, with a gluon replacing the photon, and with $V$ required to be a gluon as well.

Gauge invariance for the on-shell photon highly constrains the form that $\Gamma^{\mu \nu}_{\gamma V S}(q,k)$ takes: essentially, it must match onto $SF_{\mu \nu}V^{\mu \nu}$ or $SF_{\mu \nu} {\widetilde V}^{\mu \nu}$. Anticipating the form that the loop integral will take, we parametrize the result as:
\be
\Gamma^{\mu \nu}_{\gamma S V}(q,k) = \int^1_0 dx\frac{1}{x(1-x)}\left[\frac{c_E}{k^2 - \widetilde{\Delta}} (q^\nu k^\mu - \eta^{\mu \nu} q \cdot k) + \frac{c_O}{k^2 - \widetilde{\Delta}} \epsilon^{\mu \nu \rho \sigma} q_\rho k_\sigma\right]. \label{eq:GammaMuNuGeneral}
\ee
Here $c_E$ and $c_O$ (subscripts for ``even'' and ``odd'') are, in general, functions of $x$ and of masses and couplings; $\widetilde{\Delta}$ in general depends on $x$ and on masses but not on $k$ and $q$. We have kept only the linear dependence on $q$, because this is sufficient for computing EDMs. Otherwise, we would not have been able to eliminate $q$-dependence from the denominators. In the special case that $S$ and $V$ are real fields (for instance, a neutral Higgs boson and the photon or $Z$), we can take $c_E$ and $c_O$ to be real. In the case that $S$ and $V$ are complex, we label the vertex by the {\em outgoing} fields, and $\left(\Gamma^{\mu \nu}_{\gamma S V}\right)^* = \Gamma^{\mu \nu}_{\gamma S^* V^*}$.

In the case that the inner loop has only external vector lines---i.e., the $W$ boson EDM contributions---the calculation has already appeared in great generality in the literature \cite{Atwood:1990cm}. In this case the inner loop gives rise to a structure \cite{Chang:2005ac}
\be
\Gamma^{\mu \nu \rho} = \int_0^1 dx\frac{1}{x(1-x)} \frac{{\tilde c}_O}{k^2 - \widetilde{\Delta}} \epsilon^{\mu \nu \rho \sigma} q_\sigma,
\ee 
and adjoining the two $W$ boson lines to the external fermion gives rise to an EDM. 

Let us comment on other contributions to EDMs of elementary fermions and emphasize the importance of Barr-Zee-type contributions.
A one-loop diagram is possible only if a new particle inside the loop has a lepton or baryon quantum number
as in the case of slepton or squark contributions in the MSSM.
Some two-loop diagrams such as the so-called rainbow diagrams also require new particles with lepton or baryon numbers.
If we only add new electrically charged fermions without these quantum numbers,
two-loop Barr-Zee-type diagrams are suppressed by only one power of small SM Yukawa couplings and are likely to give the leading contributions.
It is also the case when we introduce new electrically charged scalars which do not mix with the Higgs boson $h$.
Furthermore, in the case of new scalars, such as stops, which have lepton or baryon quantum numbers but do not (or hardly) couple to
light fermions, the Barr-Zee contributions are dominant.
The Barr-Zee-type contribution from each of these new particles is gauge invariant by itself.
This can be understood from the fact that the inner loop of these particles can be replaced by the gauge-invariant effective vertices
corresponding to the operators of \eqref{SMoperatorhvv} and
the expressions of EDMs are gauge invariant as long as the effective vertices are gauge invariant
\cite{Abe:2013qla}.
On the other hand, if a new scalar is not only CP-violating but also mixes with the Higgs, there are important two-loop diagrams
other than the Barr-Zee type.
In particular, the Barr-Zee contribution from $W$ boson loops gives nonzero EDMs in this case but the inclusion of non-Barr-Zee diagrams
is essential to obtain a gauge-invariant result.
We encounter this situation in the models discussed in section~\ref{sec:BMSSM}.
A generic expression to include these non-Barr-Zee contributions is presented in \eqref{eq:Wtwoloopresult}.
However, in many cases, the Barr-Zee-type diagrams give the most important contributions and are gauge invariant only by themselves.

\subsection{Sign conventions}

Before presenting concrete results, let us specify some conventions. We work in mostly-minus signature and take the covariant derivative to be
\be
D_\mu \equiv \partial_\mu + \iu g T^a A^a_\mu. \label{eq:covariantderiv}
\ee
In particular, for QED the Feynman rule for a charged fermion coupling to the photon is $- \iu e Q \gamma^\mu$ where $Q = -1$ for an electron. In some cases, the literature chooses the opposite sign for the gauge field coupling in the covariant derivative, equivalent to a field redefinition $A_\mu \mapsto - A_\mu$.

The fermion EDM term in the Lagrangian is 
\be
{\cal L} \supset - \frac{\iu}{2} d_f {\widebar f} \sigma_{\mu \nu} \gamma^5 f F^{\mu \nu},   \label{eq:EDMdef}
\ee
leading to a matrix element 
\be
{\cal M}_\mu = - d_f \sigma_{\mu \nu} \gamma^5 q^\nu.
\ee 
Here $\sigma^{\mu \nu} = \frac{\iu}{2} \left[\gamma^\mu, \gamma^\nu\right]$. Notice that if the form of  (\ref{eq:EDMdef}) is kept fixed while the sign of the covariant derivative in (\ref{eq:covariantderiv}) is changed, the overall sign of the EDM $d_f$ will change. Alternatively, one could send $A_\mu \mapsto -A_\mu$ and change {\em both} signs. The computations of EDMs that we present below agree with those in the literature except, in some cases, up to signs. Choices of sign conventions are not always stated explicitly in the literature. We have tried to consistently use (\ref{eq:covariantderiv}) and (\ref{eq:EDMdef}) in all of our studies.

\subsection{The outer loop}

\subsubsection{Neutral particle exchange}

\begin{figure}[!h]\begin{center}
\includegraphics[clip,width=0.65\textwidth]{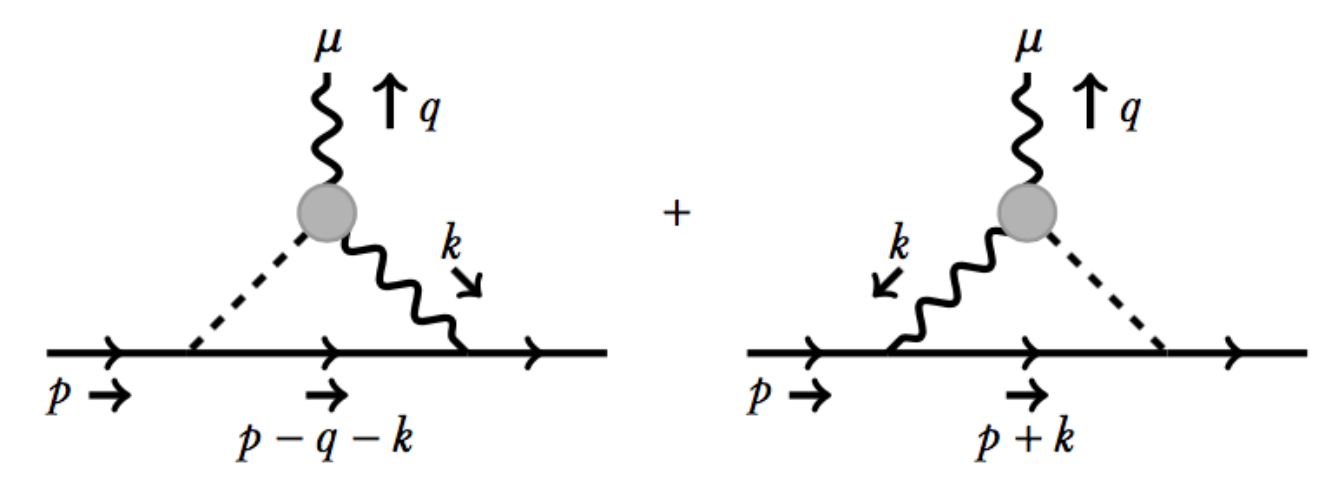}
\end{center}
\caption{Attaching the inner-loop vertex $\iu \Gamma^{\mu \nu}$ to a fermion line to generate an electric dipole moment.}
\label{fig:EDMouterloop}
\end{figure}%

From the general form of $\Gamma^{\mu \nu}_{\gamma S V}(q,k)$, it is straightforward to compute the fermion EDM induced by attaching the $S$ and $V$ lines to the fermion propagator. We treat the fermion $f$ as Dirac. Let us first assume that both $S$ and $V$ are neutral, and coupled to $f$ via the effective Lagrangian
\be
{\cal L} \supset V_\mu {\widebar f} \left(g^V_f \gamma^\mu + g^A_f \gamma^\mu \gamma^5\right) f + S {\widebar f} \left(g^S_f + \iu g^P_f \gamma^5\right) f.
\ee
Thus the vertices have Feynman rules $\iu (g^S_f + \iu g^P_f \gamma^5)$ and $\iu (g^V_f \gamma^\nu + g^A_f \gamma^\nu \gamma^5)$. In this case $g^V_f, g^A_f, g^S_f,$ and $g^P_f$ are all real numbers. Because $S$ and $V$ are neutral, we should add together the two diagrams shown in Figure \ref{fig:EDMouterloop}. A useful identity is:
\be
\int_0^1 dx_1 \int_0^1 dx_2 \int_0^1 dx_3 \frac{\delta(x_1 + x_2 + x_3 - 1)}{x_1 A_1 + x_2 A_2 + x_3 A_3} = \frac{1}{A_3} j\left(\frac{A_1}{A_3}, \frac{A_2}{A_3}\right),
\ee
where
\be
j(r,s) = \frac{1}{r-s} \left(\frac{r\log r}{r-1} - \frac{s\log s}{s-1}\right)
\ee
is the same notation used in, for example, \cite{Giudice:2005rz}.

Computing the inner loop, we find the general result for the $SV$ contribution to the EDM:
\be
d_f^{SV} = -\frac{1}{16 \pi^2 m_S^2} \int_0^1 dx \frac{1}{x(1-x)} j\left(\frac{m_V^2}{m_S^2}, \frac{{\widetilde \Delta}}{m_S^2}\right) g^V_f \left(c_O g^S_f - c_E g^P_f\right).
\ee
In computing the $c_O$ term, it is useful to know that $\epsilon^{\mu \nu \alpha \beta} \gamma_\alpha \gamma_\beta = -\iu \left[\gamma^\mu, \gamma^\nu\right] \gamma^5$. Notice that the axial-vector coupling of $V$ to the fermion drops out of the EDM calculation.

\subsubsection{Charged particle exchange}

\begin{figure}[!h]\begin{center}
\includegraphics[clip,width=0.65\textwidth]{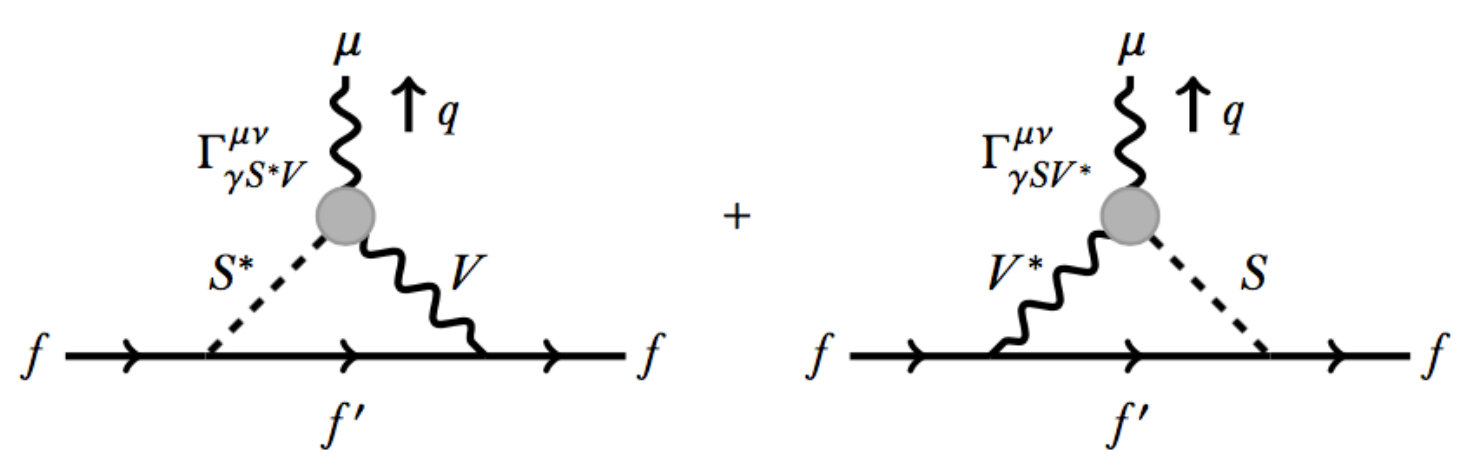}
\end{center}
\caption{Computation of the outer loop in the general case that $S$, $V$ are complex fields. Again we add two diagrams, but now the couplings are complex conjugates of each other.}
\label{fig:EDMouterloopcharged}
\end{figure}%

Next we consider the more general case, illustrated in Figure \ref{fig:EDMouterloopcharged}, in which the fields $S$ and $V$ are complex and their couplings change the incoming fermion $f$ to a different fermion $f'$. This is relevant for considering diagrams where the inner loop produces a vertex like $\gamma W^\pm H^\mp$. In this case our effective Lagrangian is
\be
{\cal L} \supset V_\mu {\widebar f} \left(g^V_{f'f} \gamma^\mu + g^A_{f'f} \gamma^\mu \gamma^5\right) f' + S {\widebar f} \left(g^S_{f'f} + \iu g^P_{f'f} \gamma^5\right) f' + {\rm h.c.},
\ee
and the coupling constants are no longer real: their complex conjugates appear in the hermitian conjugate terms with $f$ and $f'$ interchanged. The novel feature here, relative to the neutral case, is that two diagrams appearing in Fig.~\ref{fig:EDMouterloopcharged} have couplings which are complex conjugates of each other.

In this case, we find
\begin{align}
d_f^{S^*V+SV^*} = -\frac{1}{16 \pi^2 m_S^2} \int_0^1 dx \frac{1}{x(1-x)} j\left(\frac{m_V^2}{m_S^2}, \frac{{\widetilde \Delta}}{m_S^2}\right) & \left[{\rm Re}(c_O^{S^*V} g^{S^*}_{ff'} g^V_{f'f}) + {\rm Im}(c_O^{S^*V} g^{P^*}_{ff'} g^A_{f'f})\right. \nonumber \\
& \left. - {\rm Re}(c_E^{S^*V} g^{P^*}_{ff'} g^V_{f'f}) + {\rm Im}(c_E^{S^*V} g^{S^*}_{ff'} g^A_{f'f})\right].
\end{align}

\subsubsection{Example couplings}

To clarify our conventions, we have listed the couplings $g^{V,A,S,P}$ for gauge and Higgs bosons to electrons, up quarks, and down quarks in Appendix \ref{gVASPclarification}. Among the more important are that $g^V_f = -Q_f e$ for photons, $g^V_f = -\frac{g}{2\cos \theta_W} (T_3 - 2 Q_f \sin^2 \theta_W)$ for $Z$ bosons, and $g^S_f = - m_f/v$ for the Standard Model Higgs (where, throughout the paper, we use the convention $v \approx 246~{\rm GeV}$).

\subsection{The inner loop}

\subsubsection{Scalars in the inner loop}

\begin{figure}[!h]\begin{center}
\includegraphics[clip,width=0.7\textwidth]{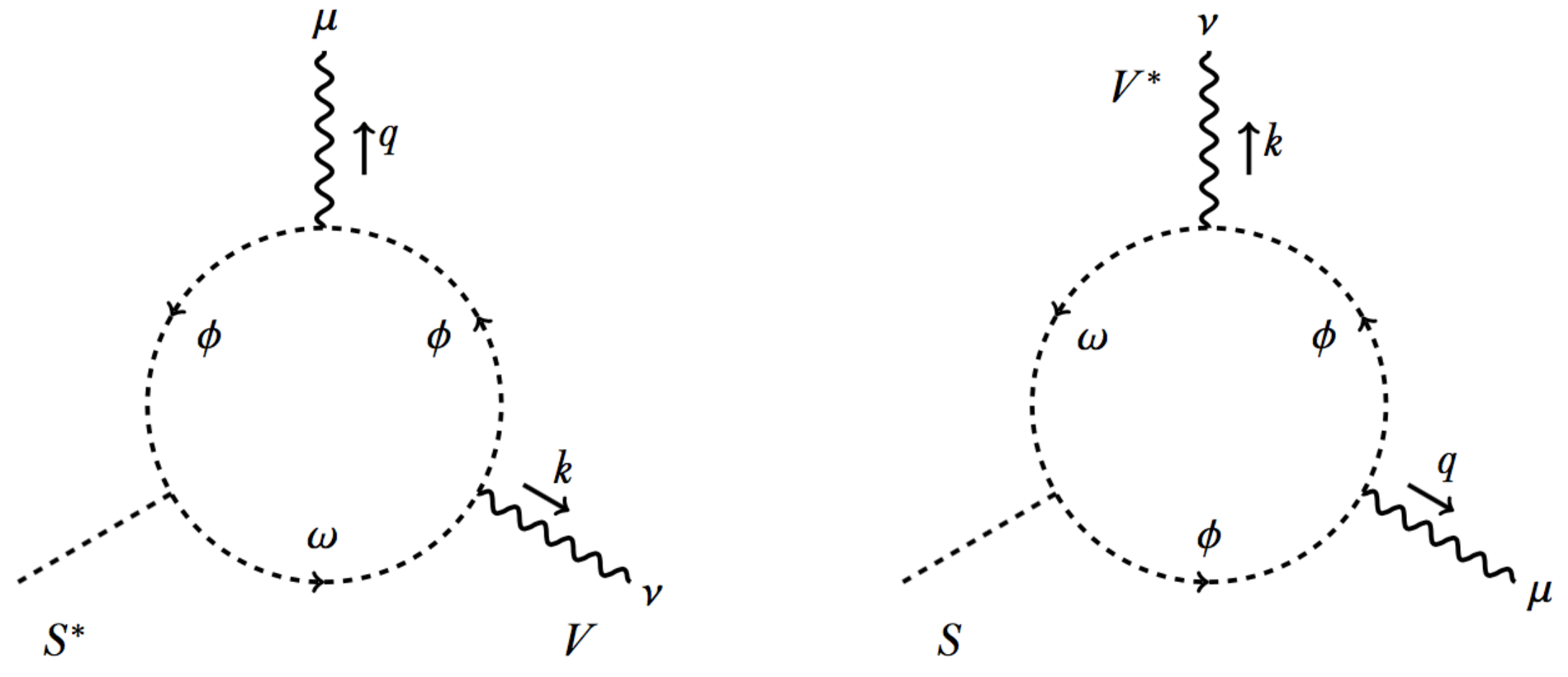}
\end{center}
\caption{Scalars $\phi$ and $\omega$ running in the inner loop to produce an effective $\gamma S^* V$ vertex. If $\phi$ and $\omega$ are indeed distinct, there are two additional diagrams with exchanged labels $\phi \leftrightarrow \omega$. In the case $V = W^\pm$, there are additional diagrams involving a $W^+W^-\gamma$ vertex (see e.g.~\cite{Chang:1999zw}), which are necessary to obtain a gauge-invariant result.}
\label{fig:innerloopscalardiagram}
\end{figure}%

We begin with scalars running in the loop. In this case we always have $c_O = 0$. In general, we can consider a vertex with outgoing photon $A_\mu$, vector $V_\nu$, and scalar $S^*$ generated by two scalars $\phi$ and $\omega$ running in the loop, as depicted in Fig.~\ref{fig:innerloopscalardiagram} at left. (The asterisk here denotes conjugate; both $S^*$ and $V$ are off-shell.) At right is a similar diagram with outgoing $V^*_\nu$ and $S$. (If $V$ and $S$ are neutral, the two diagrams contribute to the same amplitude.) The Feynman vertex for an outgoing vector $V^*_\nu$, incoming $\phi$ with momentum $p_\phi$, and outgoing $\omega$ with momentum $p_\omega$ is taken to be $- \iu g^V_{\phi \omega} (p_\phi + p_\omega)^\nu$. In the general case of distinct scalars $\phi$ and $\omega$ this corresponds to the Lagrangian
\be
{\cal L} \supset \iu g^V_{\phi \omega} V_\mu \left[\omega^\dagger \partial^\mu \phi - (\partial^\mu \omega^\dagger) \phi\right] + \iu g^{V^*}_{\omega \phi} V_\mu^*\left[\phi^\dagger \partial^\mu \omega - (\partial^\mu \phi^\dagger) \omega\right],
\ee
where $g^{V^*}_{\omega \phi} = (g^V_{\phi \omega})^*$. The Feynman vertex for an outgoing scalar $S^*$, incoming $\phi$, and outgoing $\omega$ is denoted $\iu g^S_{\phi \omega}$. In the general case where $\phi$ and $\omega$ are distinct, this corresponds to the couplings 
\be
{\cal L} \supset g^{S}_{\phi \omega} S \omega^\dagger \phi + g^{S^*}_{\omega \phi} S^\dagger \phi^\dagger \omega,
\ee
where $g^{S^*}_{\omega \phi} = (g^S_{\phi \omega})^*$.

Having fixed our conventions, we find that the contribution of the diagram at left to $\Gamma^{\mu \nu}_{\gamma S^* V}$ is given by (\ref{eq:GammaMuNuGeneral}) with the choices
\begin{align}
\left.c_E^{S^*V}\right|_{\phi \omega}&= -\frac{e Q_\phi N_c}{8 \pi^2} \, g^{V^*}_{\omega \phi} \, g^S_{\phi \omega}\, x(1-x)^2, \nonumber \\
\left.{\widetilde \Delta}^{S^* V}\right|_{\phi \omega} &= \frac{x m_\omega^2 + (1-x) m_\phi^2}{x(1-x)}, \label{eq:cEscalarphiomega}
\end{align}
while the diagram at right contributes to the conjugate vertex $\Gamma^{\mu \nu}_{\gamma S V^*}$:
\begin{align}
\left.c_E^{S V^*}\right|_{\phi \omega} &= -\frac{e Q_\phi N_c}{8 \pi^2} \, g^V_{\phi \omega} \,g^{S^*}_{\omega \phi}\, x(1-x)^2, \nonumber \\
\left.{\widetilde \Delta}^{S V^*}\right|_{\phi \omega} &= \frac{x m_\omega^2 + (1-x) m_\phi^2}{x(1-x)}. \label{eq:cEscalarphiomegaconjugate}
\end{align}
In general, when $\phi$ and $\omega$ are distinct scalars and $\omega$ also carries a charge, there will be additional contributions where the roles of the two scalars are interchanged. In the special cases that both $S$ and $V$ are neutral, we may directly add the two contributions (\ref{eq:cEscalarphiomega}) and (\ref{eq:cEscalarphiomegaconjugate}) together. In particular, when $V$ is a photon, we must take $\omega = \phi$, $g^{S^*}_{\omega \phi} = g^S_{\phi \omega} \equiv g_S$, $g^V_{\phi \omega} = g^V_{\omega \phi} = -e Q_\phi$, so that the result collapses to
\be
c_E = N_c \frac{e^2 Q_\phi^2 \, g_S}{4 \pi^2} x (1-x)^2, \quad \widetilde{\Delta} = \frac{m_\phi^2}{x(1-x)}.
\ee
In the case of stop loops, this matches the result of \cite{Chang:1998uc}, although as written it takes a different form: we can exploit the $x \mapsto (1-x)$ symmetry of the $\frac{1}{x(1-x)} j(0,\frac{z}{x(1-x)})$ factor to add odd powers of $(2x-1)^2$ to the numerator of the integrand, replacing $x(1-x)^2$ with $x(1-x)/2$.

\subsubsection{Fermions in the inner loop}

\begin{figure}[!h]\begin{center}
\includegraphics[clip,width=0.65\textwidth]{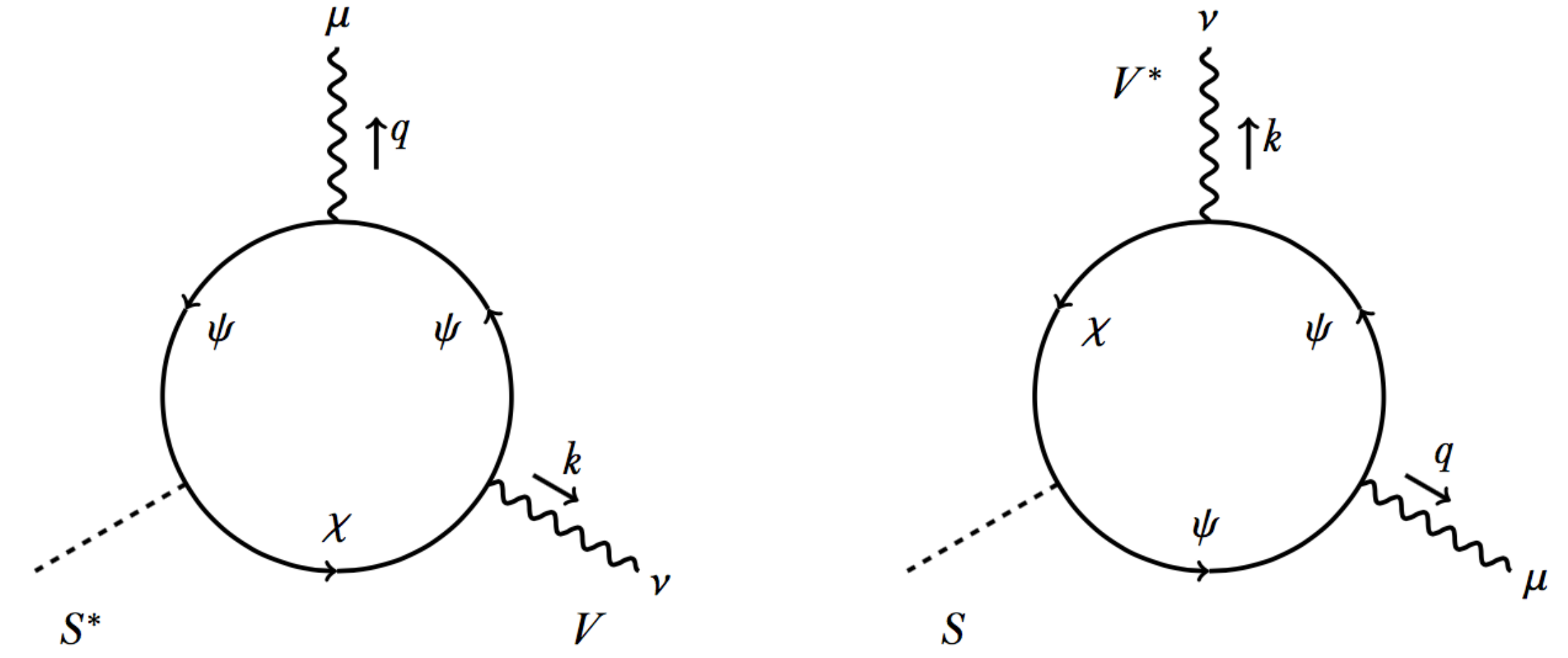}
\end{center}
\caption{Fermions $\psi$ and $\chi$ running in the inner loop to produce an effective $\gamma S^* V$ vertex. The physics is very similar to that of Fig.~\ref{fig:innerloopscalardiagram}.}
\label{fig:innerloopfermiondiagram}
\end{figure}

Now we consider the case of general fermions circulating in the inner loop. Again, in general two distinct fermions can appear, which we label $\psi$ and $\chi$ as shown in Fig.~\ref{fig:innerloopfermiondiagram}. To facilitate comparisons with the literature, against our better judgment we work with four-component Dirac fermion fields. The scalar Feynman rule for a vertex with outgoing scalar $S^*$, incoming fermion $\psi$, and outgoing fermion $\chi$ is taken to be $\iu (g^S_{\psi \chi} + \iu g^P_{\psi \chi} \gamma^5)$, corresponding to the Lagrangian
\be
{\cal L} \supset S {\widebar \chi} (g^S_{\psi \chi} + \iu\, g^P_{\psi \chi} \gamma^5) \psi + S^\dagger {\widebar \psi} (g^{S^*}_{\chi \psi} + \iu\, g^{P^*}_{\chi \psi} \gamma^5) \chi,
\ee
where $g^{S^*}_{\chi \psi} = (g^S_{\psi \chi})^*$ and $g^{P^*}_{\chi \psi} = (g^P_{\psi \chi})^*$. Similarly, the Feynman rule for an outgoing vector $V^*_\nu$, incoming $\psi$, and outgoing $\chi$ is taken to be $\iu (g^V_{\psi \chi} \gamma^\nu + g^A_{\psi \chi} \gamma^\nu \gamma^5)$, corresponding to the Lagrangian
\be
{\cal L} \supset V_\mu {\widebar \chi} (g^V_{\psi \chi} \gamma^\mu + g^A_{\psi \chi} \gamma^\mu \gamma^5) \psi + V^\dagger_\mu {\widebar \psi} (g^{V^*}_{\chi \psi} \gamma^\mu + g^{A^*}_{\chi \psi} \gamma^\mu \gamma^5) \chi,
\ee
where $g^{V^*}_{\chi \psi} = (g^V_{\psi \chi})^*$ and $g^{A^*}_{\chi \psi} = (g^A_{\psi \chi})^*$.

For the general fermion loop at left in Fig.~\ref{fig:innerloopfermiondiagram} we find a contribution to $\Gamma^{\mu \nu}_{\gamma S^* V}$:
\begin{align}
\left. c_E^{S^* V} \right|_{\psi \chi} &= -\frac{e Q_\psi N_c}{4\pi^2} \left[m_\chi x^2(1-x) \left(g^S_{\psi \chi} g^{V^*}_{\chi \psi} + \iu g^P_{\psi \chi} g^{A^*}_{\chi \psi}\right) + (1-x)^3 m_\psi \left(g^S_{\psi \chi} g^{V^*}_{\chi \psi} - \iu g^P_{\psi \chi} g^{A^*}_{\chi \psi}\right)\right], \nonumber \\
\left. c_O^{S^* V} \right|_{\psi \chi} &= -\frac{e Q_\psi N_c}{4\pi^2} \left[m_\chi x(1-x) \left(\iu g^S_{\psi \chi} g^{A^*}_{\chi \psi} - g^P_{\psi \chi} g^{V^*}_{\chi \psi}\right) - (1-x)^2 m_\psi \left(\iu g^S_{\psi \chi} g^{A^*}_{\chi \psi} + g^P_{\psi \chi} g^{V^*}_{\chi \psi}\right)\right], \nonumber \\
\left.{\widetilde \Delta}^{S^* V}\right|_{\psi \chi} &= \frac{x m_\chi^2 + (1-x) m_\psi^2}{x(1-x)}.\label{eq:cEfermionpsichi}
\end{align}
The diagram at right contributes to the conjugate vertex $\Gamma^{\mu \nu}_{\gamma S V^*}$:
\begin{align}
\left. c_E^{S V^*} \right|_{\chi \psi} &= -\frac{e Q_\psi N_c}{4\pi^2} \left[m_\chi x^2(1-x) \left(g^{S^*}_{\chi \psi} g^{V}_{\psi \chi} - \iu g^{P^*}_{\chi \psi} g^{A}_{\psi \chi}\right) + (1-x)^3 m_\psi \left(g^{S^*}_{\chi \psi} g^{V}_{\psi \chi} + \iu g^{P^*}_{\chi \psi} g^{A}_{\psi \chi}\right)\right], \nonumber \\
\left. c_O^{S V^*} \right|_{\chi \psi} &= -\frac{e Q_\psi N_c}{4\pi^2} \left[m_\chi x(1-x) \left(-\iu g^{S^*}_{\chi \psi} g^{A}_{\psi \chi} - g^{P^*}_{\chi \psi} g^{V}_{\psi \chi}\right) - (1-x)^2 m_\psi \left(-\iu g^{S^*}_{\chi \psi} g^{A}_{\psi \chi} + g^{P^*}_{\chi \psi} g^{V}_{\psi \chi}\right)\right], \nonumber \\
\left.{\widetilde \Delta}^{S V^*}\right|_{\chi \psi} &= \frac{x m_\chi^2 + (1-x) m_\psi^2}{x(1-x)}.\label{eq: cEfermionpsichiconjugate}
\end{align}
As in the scalar case, we should also keep in mind that if $\chi$ is charged (and is not the same field as $\psi$) then there are two additional diagrams to consider.

In the special case that $V$ is a photon, we must take $\chi = \psi$. When we include the two diagrams with $\psi$ lines propagating in both directions (or equivalently, cross the external photon lines), we find a total contribution:
\begin{align}
c_E &= N_c \frac{e^2 Q^2 g_S}{2\pi^2} (1-x)(2x^2 - 2x+1) m_\psi \mapsto N_c \frac{e^2 Q^2 g_S}{4\pi^2} \left[(1-x)^2 + x^2\right] m_\psi, \nonumber \\
c_O &= -N_c \frac{e^2 Q^2 g_P}{2\pi^2} (1-x) m_\psi \mapsto -N_c \frac{e^2 Q^2 g_P}{4\pi^2} m_\psi, \quad \widetilde{\Delta} = \frac{m_\psi^2}{x(1-x)},
\end{align}
where the $\mapsto$ indicates not equality but equivalence when integrated; due to the $x \leftrightarrow 1-x$ symmetry of the remaining factors we have added $(x-1/2)$ to $(1-x)$ to produce $1/2$.

%
%

\section{Natural SUSY framework}\label{naturalSUSY}

\subsection{Naturalness and tuning}\label{naturalSUSYtuning}

The framework of natural or effective SUSY is based on assuming that particles that play a key role in electroweak naturalness are relatively light. At tree level, these are the higgsinos and heavy Higgs bosons; at one loop, the stops, winos, and binos; and at two loops, the gluinos. Most other superpartners have small couplings to the Higgs and can be quite heavy.

Naturalness puts stringent constraints on chargino parameters. If we wish to have a supersymmetric explanation for the hierarchy problem without significant fine-tuning, both $|\mu|$ and $|M_2|$ are bounded above. Here we will provide a crude, but useful, characterization of this tuning (for recent detailed comments on tuning and the role of corrections, see \cite{Buckley:2016tbs}). The higgsino mass is controlled by the $\mu$-parameter, which is directly relevant for minimization of the Higgs potential, and the LEP experiments constrain this as $\mu \gtrsim 100 \, \rm GeV$ \cite{Heister:2002mn,Abdallah:2003xe,Abbiendi:2002vz,Acciarri:2000wy}. The parameter $\mu$ appears in the tree-level Higgs potential, implying significant fine tuning when $\mu \gg m_Z$ \cite{Barbieri:1987fn}. In particular, the degree of fine-tuning is approximately \cite{Barbieri:1987fn}
\be
\Delta_{\tH} = \frac{2 \left|\mu\right|^2}{m_h^2} \approx 10\left(\frac{|\mu|}{280~{\rm GeV}}\right)^2.
\ee
On the other hand, the wino mass affects the Higgs potential only at one loop; it is often ignored because it plays little role in collider production, but nonetheless is present in a fully natural spectrum (see e.g.~\cite{Cohen:1996vb}), with:
\be
\Delta_{\tW} = \frac{3 g_2^2}{4 \pi^2 m_h^2} \left|M_2\right|^2 \log\frac{M_{\rm med}}{|M_2|} \approx 10 \left(\frac{|M_2|}{1.0~{\rm TeV}}\right)^2,
\ee
where $m_h \approx 125~{\rm GeV}$ is the light physical Higgs boson mass and $M_{\rm med}$ is the scale at which SUSY breaking is mediated to Standard Model superpartners. In our numerical estimate we have assumed a low mediation scale $M_{\rm med} \approx 100~{\rm TeV}$ (in other words, we expect the tuning to be larger in many models).

The heavy Higgs fields of the MSSM play an underappreciated role in natural SUSY. Their masses can be naturally large, but only when $\tan \beta$ is also large \cite{Perelstein:2007nx,Gherghetta:2014xea,Katz:2014mba}. The tuning associated with taking large $m_A$ depends somewhat on the details of how we lift the Higgs mass to 125 GeV, but for our purposes we will simply estimate
\be \label{eq:Afinetuning}
\Delta_A \approx \frac{2 m_A^2}{m_h^2 \tan^2 \beta} \approx 10 \left(\frac{m_A}{1.4~{\rm TeV}}\right)^2 \left(\frac{5}{\tan \beta}\right)^2.
\ee
This is an additional tuning, independent of the tuning associated with $\mu$ (but perhaps correlated with it, in particular models). In \cite{Katz:2014mba} it was argued that $b \to s \gamma$ constraints shut off the prospect of simultaneously sending $m_A, \tan \beta \to \infty$ consistent with naturalness. As we will see below, if CP phases are generic, similar statements can be made about EDM constraints.

Light stops are also crucial because top/stop loops generate the up-type Higgs soft mass squared $m_{H_u}^2$.
Then, the degree of fine-tuning can be estimated by the sizes of stop soft masses $m_{Q_3}$, $m_{u_3}$
\cite{Kitano:2006gv,Perelstein:2007nx},
\begin{equation} \label{eq:stoptuning}
\begin{split}
\\[-2.5ex]
\Delta_{\tilde{t}} \equiv \left|  \frac{2 \delta m_{H_u}^2}{m_h^2} \right|
\approx \left| \frac{3 y_t^2}{4 \pi^2} \frac{m_{Q_3}^2 + m_{u_3}^2 + |A_t|^2}{m_h^2}
\log \left( \frac{M_{\rm med}}{m_{\rm stop} } \right) \right| \, , \\[1ex]
\end{split}
\end{equation}
where $A_t$ is the $A$-parameter corresponding to the top Yukawa coupling $y_t$ and
$m_{\rm stop} \equiv \left( m_{Q_3} m_{u_3} \right)^{1/2}$ is the geometric mean of the two stop soft masses. 

The gluino also has an important effect on naturalness because its running effect on stop masses is significant.
The lower bound on the gluino mass from direct searches depends on its decay chains, but
even in scenarios relaxing the bound such as R-parity violation, the gluino mass is still required to be heavier than $1.2 \, \rm TeV$
\cite{Evans:2013jna}. Numerically, the naturalness bounds on winos and gluinos are similar for low mediation scales, so the additional attention usually paid to gluinos is mostly due to its large production cross section at hadron colliders. The gluino has little effect on our discussion of EDMs. Conversely, winos are much more difficult to search for at colliders, but will play a key part in EDM constraints. 

The first and second generations of squarks and sleptons, which can give rise to one-loop EDMs of light quarks and electrons, can be quite heavy in natural SUSY. Hence, we will not discuss one-loop EDMs of light fermions in this paper. They can play important roles, even in PeV-scale split SUSY, but the details will hinge on questions of flavor physics involving more assumptions or model-building; see for instance \cite{McKeen:2013dma,Altmannshofer:2013lfa}. In the natural SUSY context it is possible that their effects are quite small. When the cutoff scale of the theory is very high, such as the grand unification scale, these heavy scalars are required to be not heavier than $\mathcal{O}(10) \, \rm TeV$ because $m_{Q_3}^2$ is driven to be tachyonic by the two-loop renormalization group (RG) effect \cite{ArkaniHamed:1997ab}. In this case, we still need some alignment of their soft masses to satisfy flavor constraints. This scenario can be realized, for example, by considering theories with multiple copies of the Standard Model gauge groups and SUSY breaking provided by gauge mediation \cite{Craig:2011yk}. On the other hand, when the cutoff scale of the effective theory with only light multiplets---stops, the left-handed sbottom, higgsinos and gauginos---is of order $10 \, \rm TeV$, the first and second generations of squarks and sleptons can be heavy enough to satisfy flavor constraints. Warped/composite natural SUSY \cite{Sundrum:2009gv,Gherghetta:2011wc,Heidenreich:2014jpa}, where the top and left-handed bottom quarks, the Higgs fields, the gauge fields and their superpartners are composite or partially composite (In the 5D language, they are localized at the IR brane or propagate in the bulk of the extra dimension), corresponds to this case. This class of models does not need to assume any flavor alignment or CP conservation and gives a good example of the scenario discussed in the present work. However, we do not assume any specific model below, and the effects that we discuss may coexist with one-loop effects of squarks and sleptons in some models.

\subsection{Higgs sector physics}\label{naturalSUSYhiggs}

One of the tightest constraints on supersymmetry is the experimental measurement of the 125 GeV Higgs boson mass. In the MSSM, at tree level, the Higgs mass is bounded above by the $Z$ boson mass. To explain why the Higgs has been observed to be much heavier than the $Z$ boson, we must either assume large loop corrections to the Higgs mass or new tree-level physics beyond the MSSM. These different scenarios have implications for the parameter space of SUSY which in turn affect the ways in which EDMs may arise. We will consider a few different scenarios below.

The first scenario is that the Higgs boson mass is lifted by loops of stops with large left-right mixing parameter $A_t$. This is the most natural region of MSSM parameter space in light of the data, although it still requires a high degree of fine tuning (see, for instance, \cite{Hall:2011aa}). As we will see below, two-loop EDMs induced by stops are proportional to ${\rm Im}(A_t \mu)$, so the large $A_t$ scenario can lead to detectably large EDMs. Although we will not discuss it in this paper, it is also possible to lift the Higgs mass through loops of new vectorlike particles {\em beyond} the MSSM \cite{Moroi:1991mg,Babu:2008ge,Martin:2009bg,Graham:2009gy,Nakai:2015swg,Basirnia:2016szw}. Many of the conclusions that we will draw about EDMs induced by stops in the MSSM with large $A$-terms in \S\ref{MSSMlargeAt} below would also apply to new particles beyond the MSSM in these theories.

The second scenario is that the physics that lifts the Higgs mass does not contribute directly to EDMs. An example is new tree-level $D$-term contributions, which can come from extra gauge interactions of the Higgs fields \cite{Randall:2002talk,Batra:2003nj,Maloney:2004rc}. As discussed in ref.~\cite{Dine:2007xi}, after integrating out the massive gauge fields, the effect of the extension is encapsulated in dimension-six K\"ahler potential operators. We do not expect no new CP-violating phases beyond the MSSM in this case (in particular, if the new gauge field is abelian there are not even new charginos that could mix with the usual ones) and ignore the correction to the Higgs potential below. Nonetheless, we can still draw conclusions about the expected size of EDMs based on naturalness arguments in such a scenario. For example, an $A$-term will be generated radiatively from the gluino mass, so unless some tuning cancels it, this will provide a minimum size to the EDM even when the $A$-term is not as large as in the scenario with highly-mixed stops that raise the Higgs mass. 

The third scenario that we will consider is that the Higgs mass is lifted by new tree-level $F$-term contributions. We treat these as dimension-five operators beyond the MSSM, a scenario that has been referred to as the BMSSM \cite{Dine:2007xi}. These effective operators could serve as an approximate stand-in for scenarios like \cite{Espinosa:1991gr,Barbieri:2006bg,Hall:2011aa} with new degrees of freedom in the Higgs sector. We will discuss this scenario in detail in \S\ref{sec:BMSSM} below, but introduce the basic idea here. The BMSSM involves two new operators of effective dimension five. The superpotential of the Higgs sector includes one such  operator,
\begin{equation}
\begin{split}
\\[-2.5ex]
W_{\rm Higgs} = \mu H_u \cdot H_d + \frac{\lambda}{M} (H_u  \cdot H_d  )^2 \, , \label{BMSSMsuperpotential} \\[1ex]
\end{split}
\end{equation}
where $H_u \cdot  H_d = H_u^+ H_d^- - H_u^0 H_d^0$ and $M$ is some cutoff scale of the Higgs sector.
In addition, $\lambda$ is a dimensionless coupling constant which is complex in general.
The other leading higher dimensional operator comes from the corresponding soft SUSY breaking term,
\begin{equation}
\begin{split}
\\[-2.5ex]
\mathcal{L}_{\rm soft} \supset \frac{\lambda m_{\rm SUSY}}{M} (H_u \cdot  H_d  )^2 \, , \label{BMSSMsoft} \\[1ex]
\end{split}
\end{equation}
where $m_{\rm SUSY}$ is a SUSY breaking mass parameter whose absolute value is $\mathcal{O} (100) \, \rm GeV$.
This is also complex in general.
With the new operators of \eqref{BMSSMsuperpotential} and \eqref{BMSSMsoft}, the scalar potential of the Higgs sector is given by
\begin{equation}
\begin{split}
\\[-2.5ex]
V_{\rm Higgs} = V_{\rm MSSM} + \left\{ 2 \epsilon_1 \left(|H_u|^2 + |H_d|^2\right) H_u  \cdot H_d
+ \epsilon_2 \left(H_u  \cdot H_d\right)^2 + \rm  h.c.   \right\} \, ,   \\[1ex]
\end{split}
\end{equation}
where $V_{\rm MSSM}$ is the MSSM Higgs sector scalar potential and 
\begin{equation}
\begin{split}
\\[-2.5ex]
\epsilon_1 \equiv \frac{\lambda \mu^\ast}{M} = |\epsilon_1| e^{\iu \phi_1}
\, , \qquad \epsilon_2 \equiv - \frac{\lambda m_{\rm SUSY}}{M}  = |\epsilon_2| e^{\iu \phi_2} \, .   \\[1ex]
\end{split}
\end{equation}
Note that the new operators provide two additional CP violating phases.
We assume that the cutoff scale of the Higgs sector is larger than several TeV.
Then, (the absolute value of) the two parameters $\epsilon_{1,2}$ are small.
If we turn off $\epsilon_{1,2}$, this scenario reduces to the MSSM.
Hence, this is the most general case which we consider in this paper. A detailed discussion of how the BMSSM scenario modifies the mass spectrum of higgses, charginos, neutralinos, and stops is presented in \S\ref{sec:BMSSMreview}.

%
%

\section{Calculations of the EDMs and experimental constraints}\label{EDMcalculation}

There are three main classes of experiments in EDM searches:
the EDMs of paramagnetic atoms/molecules, diamagnetic atoms and hadrons
(for reviews of the EDMs, see e.g. refs.~\cite{Pospelov:2005pr,Engel:2013lsa}).
They are distinguished in terms of the underlying physics leading to EDMs. In every case, the experimental measurement does not only directly probe EDMs of elementary fermions (as discussed in \S\ref{twoloopfermions}), but a more general combination of CP violating operators, generally including four-fermion operators.
The world record in paramagnetic systems has been given by the measurement of the EDM of the paramagnetic ThO molecule
by the ACME collaboration
\cite{Baron:2013eja}.
For diamagnetic atoms and hadrons, the best limits have been provided by the mercury and neutron EDMs respectively.
Among these three classes of experiments, the EDMs of paramagnetic atoms and molecules
do not suffer from hadronic uncertainty in their calculations. Non-observation of these EDMs gives unambiguous and stringent bounds on possibilities for physics beyond the Standard Model.
We first consider this class of measurements and then discuss the mercury and neutron EDMs.

\subsection{Paramagnetic EDMs: operators probed and experimental status}

The EDMs of paramagnetic systems  are dominated by the electron EDM $d_e$ and the electron-nucleon interaction. The fermion EDM was defined in equation (\ref{eq:EDMdef}). The CP-odd electron-nucleon interaction is
\begin{equation}
\begin{split}
\\[-2.5ex]
\mathcal{L}_{eN} \supset  - \iu\,  C_S \, \bar{e} \gamma_5 e \bar{N} N \, . \label{CS} \\[1ex]
\end{split}
\end{equation}
Here, we have suppressed dependence on the isospin.
The {\em effective} electron EDM as measured with the paramagnetic ThO molecule, which is normalized to reduce to the electron EDM in the case with $C_S=0$, is then given by
\cite{Chupp:2014gka}
\begin{equation}
\begin{split}
\\[-2.5ex]
d_{\rm ThO} \approx d_e + k C_S \, ,  \label{dThO} \\[1ex]
\end{split}
\end{equation}
where $k = 1.6 \times 10^{-15} \, {\rm GeV}^2 \, e \, \rm cm$.
The experimental limit is imposed on the absolute value of $d_{\rm ThO}$.
In paramagnetic systems, the effect of the electron EDM is enhanced and strongly constrained from the measurement.

The current limit of the effective electron EDM measured in paramagnetic systems has been provided by the ACME experiment \cite{Baron:2013eja}:
\be
|d_{\rm ThO}| < d_e^{\rm max} \equiv 8.7 \times 10^{-29} \, e \, \rm cm.
\ee
Projected future limits from improvements of ACME are \cite{Doyletalk}
\begin{align}
{\rm ACME~II:} &\quad |d_{\rm ThO}| < 0.5 \times 10^{-29} \, e \, \rm cm, \\
{\rm ACME~III:} &\quad  |d_{\rm ThO}| < 0.3 \times 10^{-30} \, e \, \rm cm.
\end{align}
Some of the important improvements that will play a role in ACME II have already been demonstrated \cite{Panda:2016ifg}, and a new result is expected within the next year.

\subsection{Paramagnetic EDMs: computation}

\subsubsection{The electron EDM} \label{electronEDM}

\begin{figure}[!t]
\vspace{-1cm}
 \begin{minipage}{0.33\hsize}
  \begin{center}
   \includegraphics[clip, width=6.5cm]{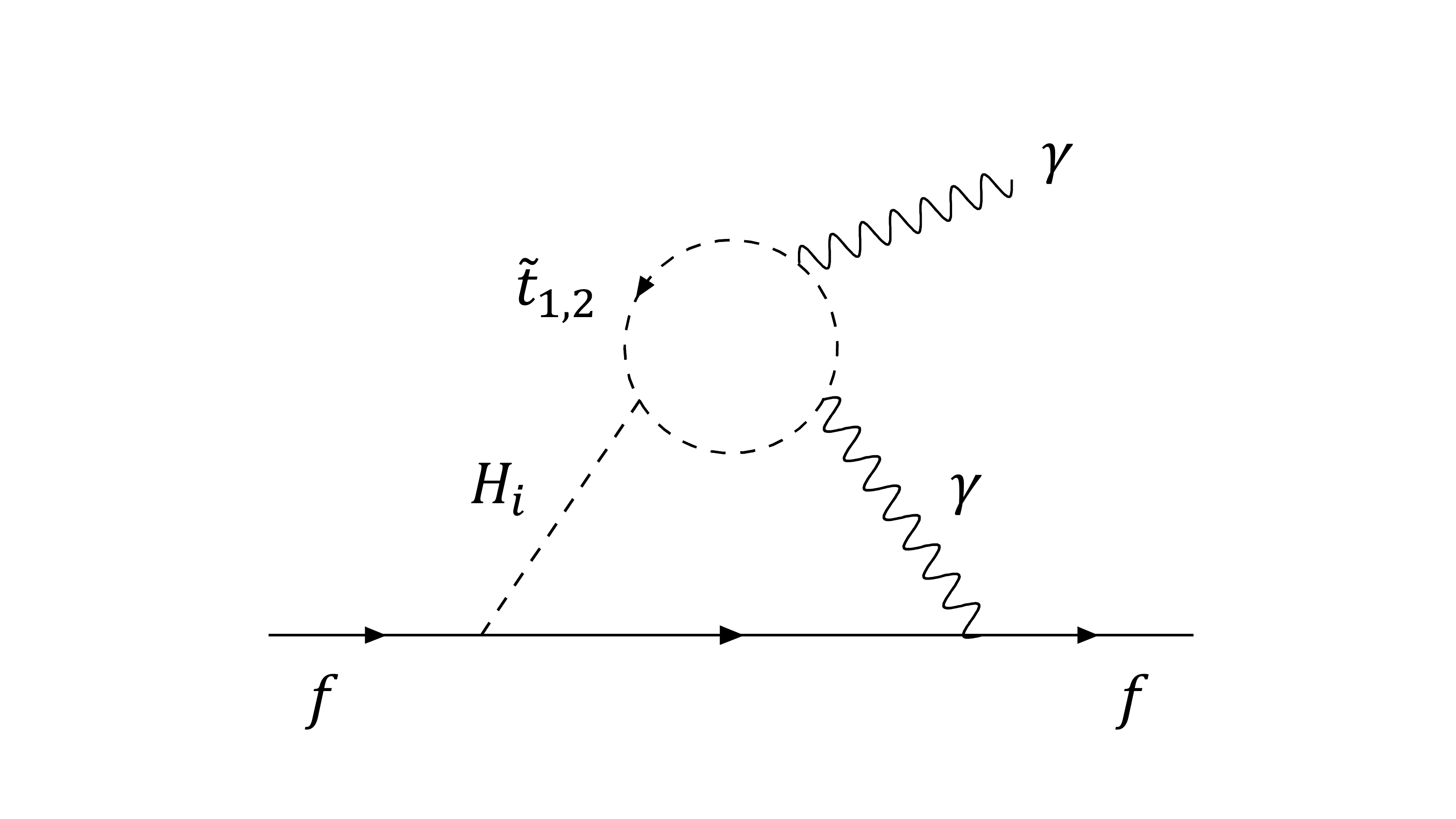}
  \end{center}
 \end{minipage}
 \begin{minipage}{0.33\hsize}
  \begin{center}
   \includegraphics[clip, width=6.5cm]{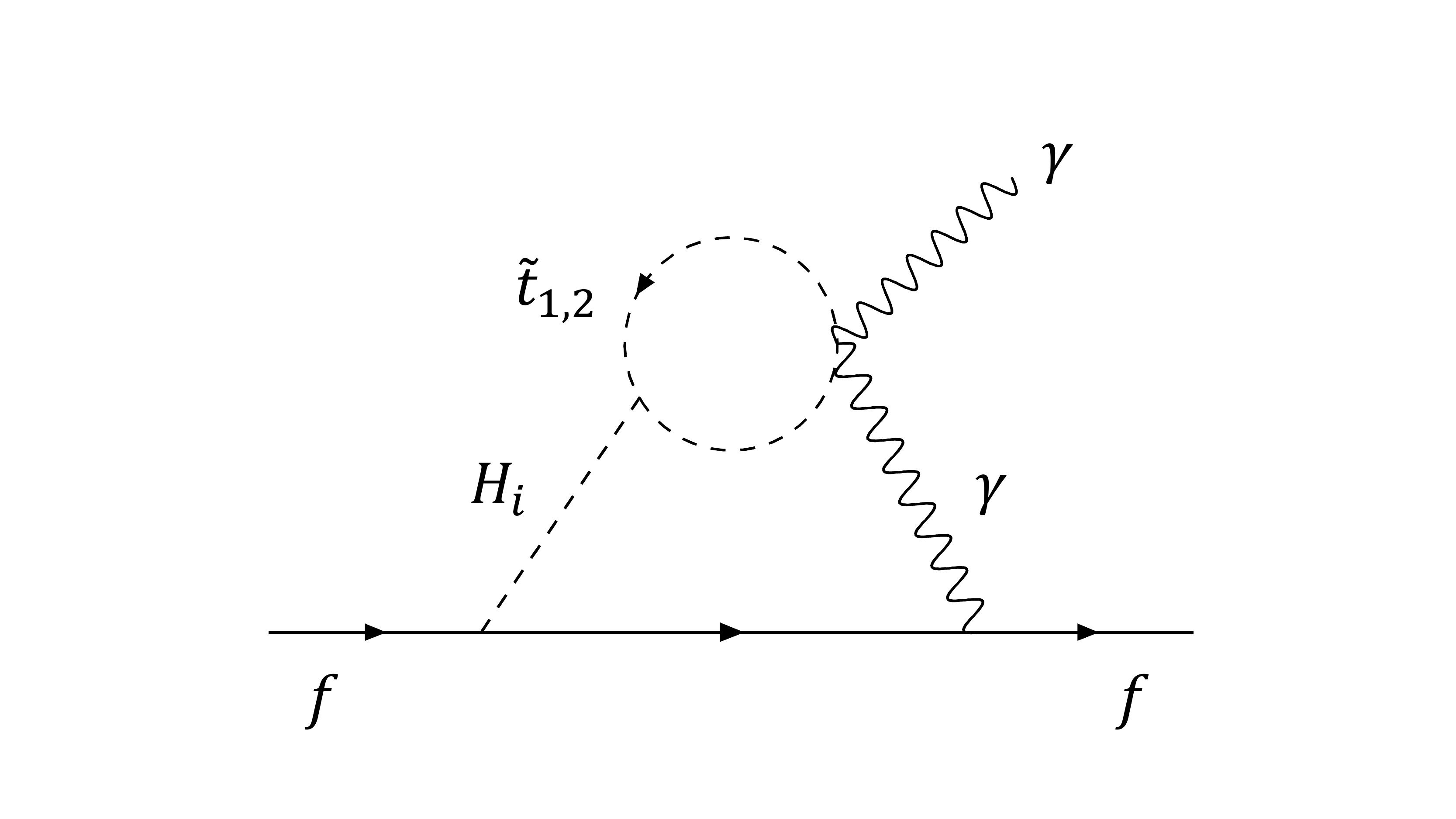}
  \end{center}
 \end{minipage}
 \begin{minipage}{0.2\hsize}
  \begin{center}
  \includegraphics[clip, width=6.5cm]{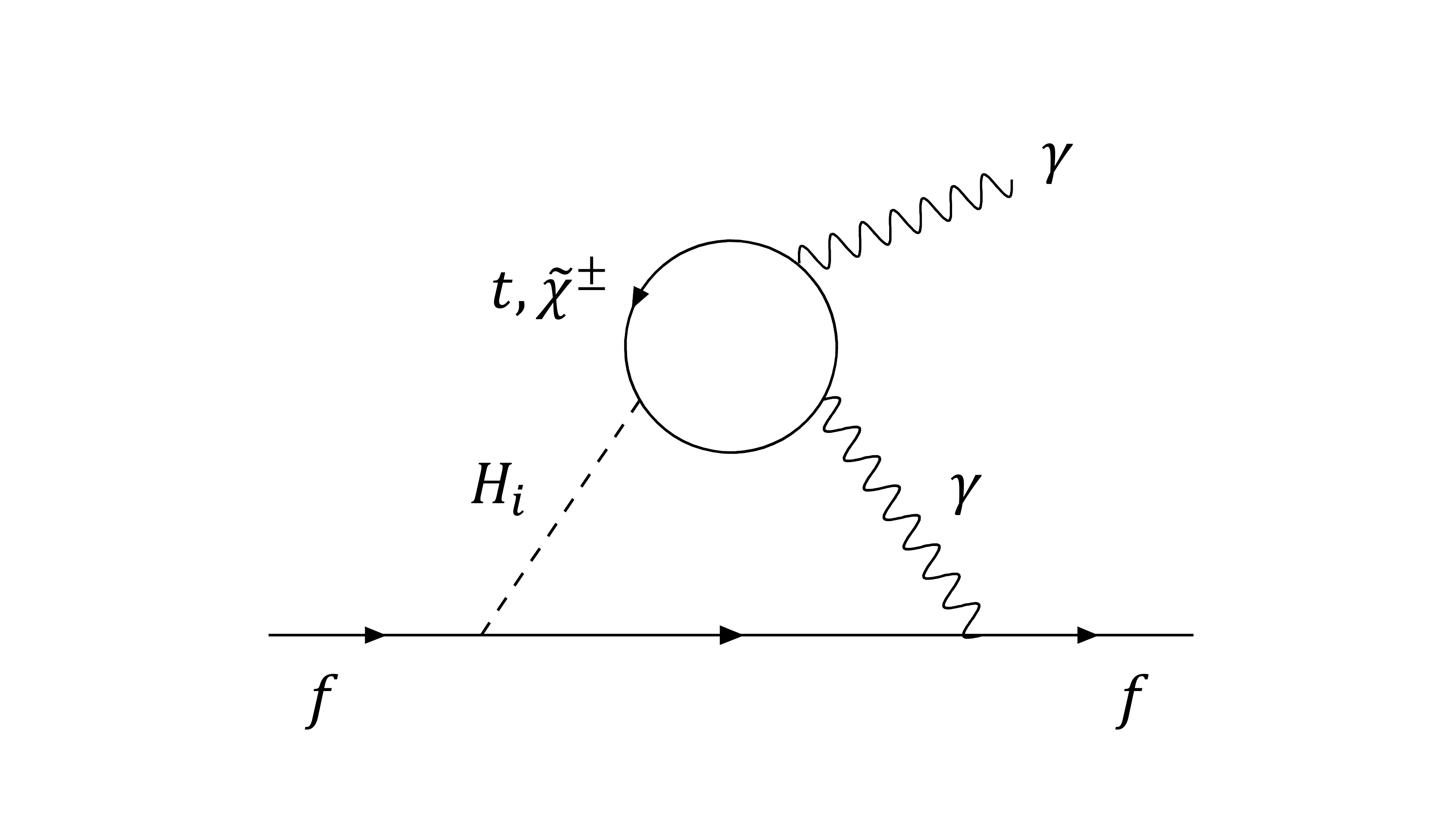}
  \end{center}
 \end{minipage}

 \begin{minipage}{0.33\hsize}
  \begin{center}
   \includegraphics[clip, width=6.5cm]{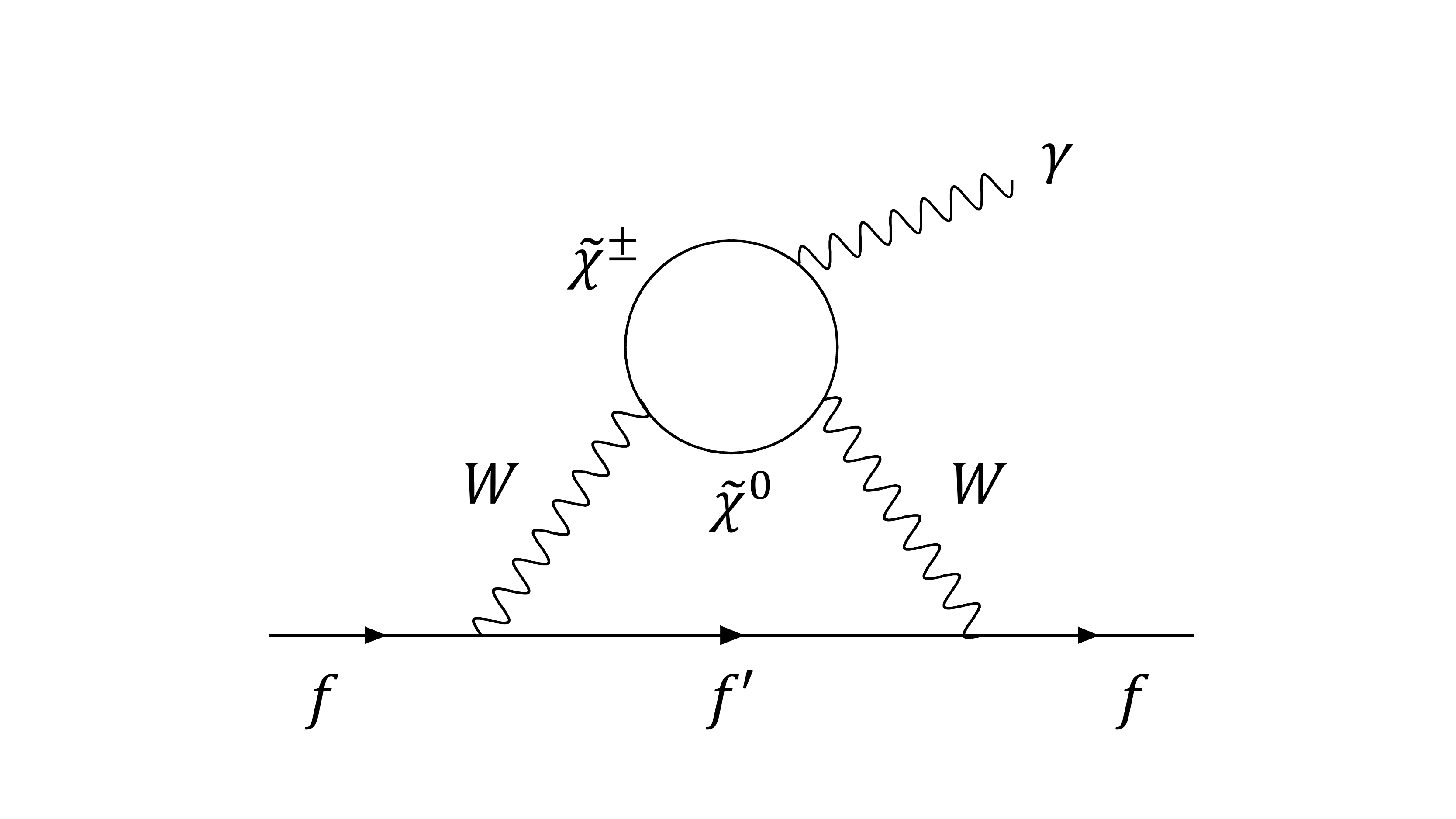}
  \end{center}
 \end{minipage}
 \begin{minipage}{0.33\hsize}
  \begin{center}
   \includegraphics[clip, width=6.5cm]{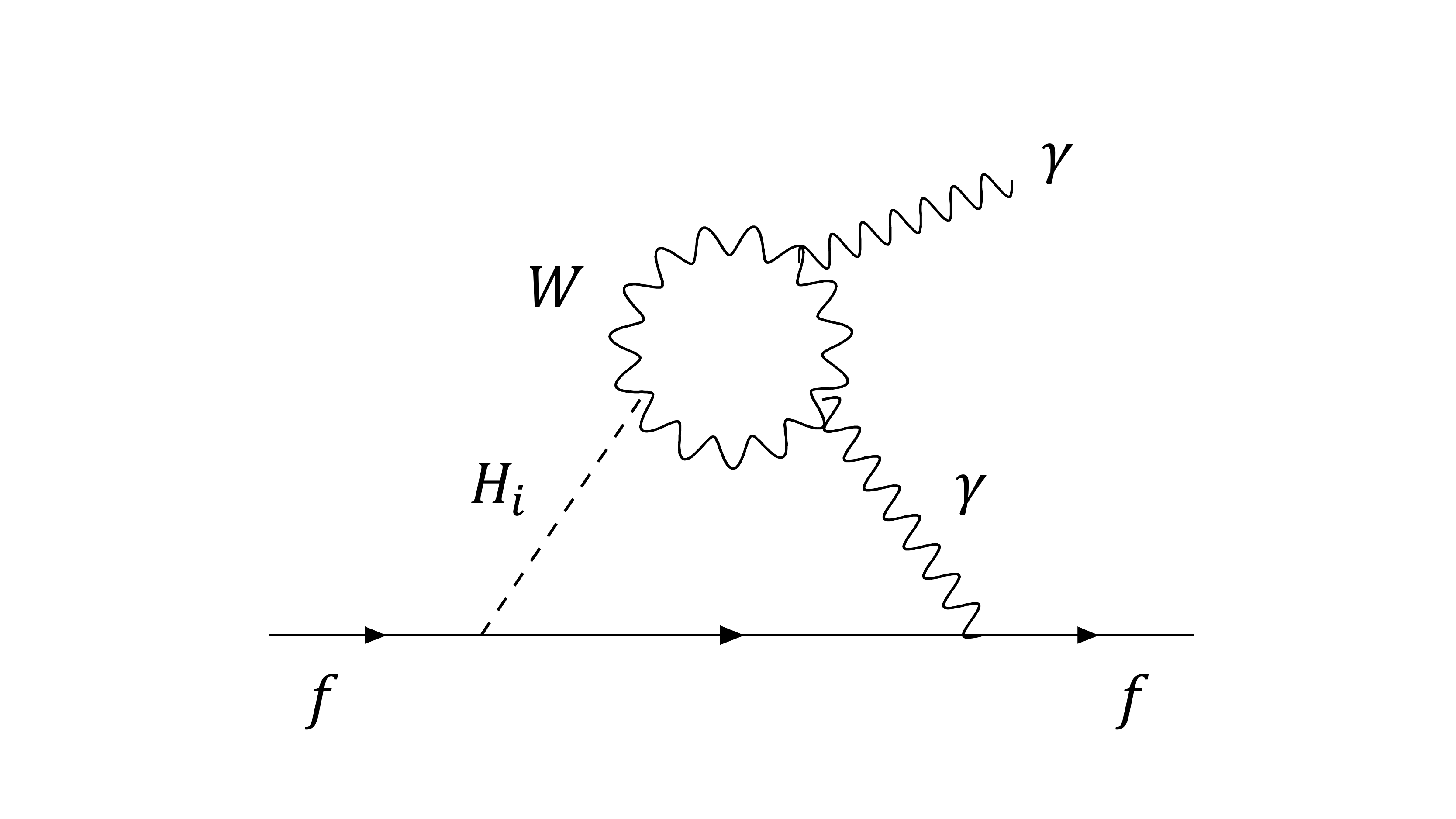}
  \end{center}
 \end{minipage}
 \begin{minipage}{0.2\hsize}
  \begin{center}
  \includegraphics[clip, width=6.5cm]{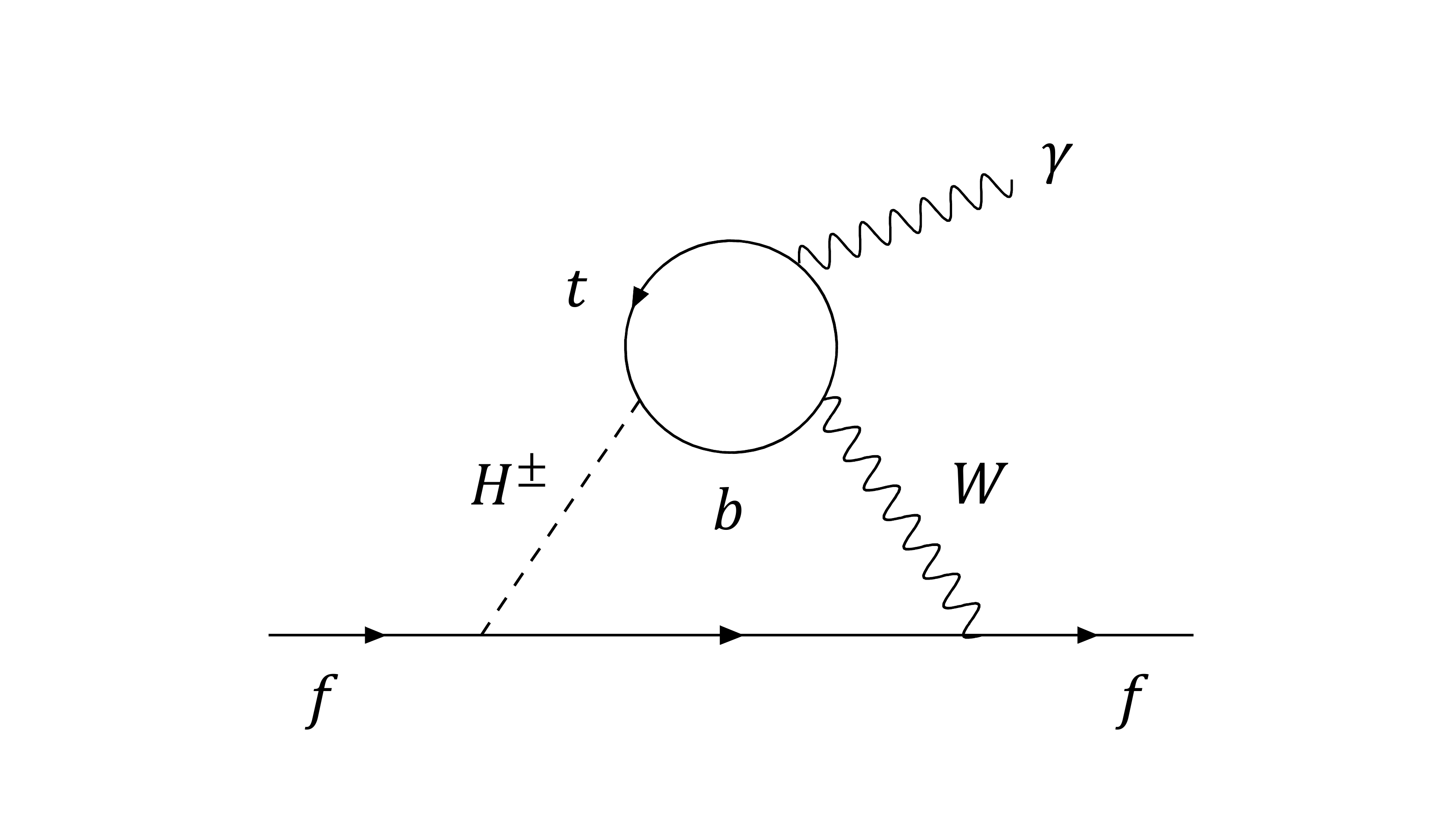}
  \end{center}
 \end{minipage}
  \caption{The two-loop diagrams dominantly contributing to the EDM of a fermion $f$
(there are also mirror graphs).
The two top left diagrams include stop loops. The top right diagram denotes contributions of top quark and chargino loops.
The bottom left diagram shows the $W$ EDM contribution.
The bottom middle diagram includes $W$ boson loops.
The bottom right diagram shows the contribution of top/bottom quark loops through charged Higgs bosons.}
  \label{fig:BarrZee}
\end{figure}

Let us consider contributions to the electron EDM. We will not consider the one-loop diagrams with a selectron, as mentioned above, because sleptons play little role in natural SUSY and incorporating them would require a careful treatment of flavor physics (but see \cite{McKeen:2013dma,Altmannshofer:2013lfa}). There are many two-loop contributions to the electron EDM which can be computed using the results we presented in \S\ref{twoloopfermions} and arise from diagrams shown in Figure~\ref{fig:BarrZee} (there are also mirror graphs).\footnote{
Ref.~\cite{Yamanaka:2012ia}
has considered the rainbow diagrams. However, these diagrams give sub-dominant contributions in our setups.}
These include effects of stops, tops, charginos, and $W$ bosons running in the inner loop, connecting to the outer loop by gauge fields and Higgs bosons. Contributions with a $Z$ boson \cite{Li:2008kz} are small for the electron EDM, due to the small vector coupling of the electron to $Z$s; they are more important for the EDMs of quarks.

The diagrams with stops, tops, charginos, and charged Higgs bosons in the inner loop are all special cases of the results presented in \S\ref{twoloopfermions}; specific choices of couplings to plug in to the formulas are presented in more detail in Appendix \ref{sec:coupling} and some selected results for EDM contributions are presented in Appendix \ref{sec:edmformulas}. The diagram with $W$ bosons in the inner loop cannot be derived from the general results of \S\ref{twoloopfermions}, due to additional complications associated with gauge invariance and the $W$ couplings to the external fermions. In the MSSM, this contribution appears at a higher loop order,
but in the BMSSM with tree-level CP violation in the Higgs sector there is a nonzero two-loop contribution in the present setup and we cannot neglect it.
The calculation of this contribution was initiated by Barr and Zee in their original paper
\cite{Barr:1990vd}.
Ref.~\cite{Abe:2013qla} has considered contributions from Nambu-Goldstone modes
and non-Barr-Zee-type diagrams and obtained a gauge invariant result for the EDM.
With our notation, the expression is 
\begin{equation} \label{eq:Wtwoloopresult}
\begin{split}
\\[-2.5ex]
\frac{d_f}{e} \biggr|_{W} = & \,\, -Q_f \frac{4 \alpha}{(4 \pi)^3} \frac{1}{v} \sum_{i=1}^3 
\left\{ \left( 3 + \frac{m_{H_i}^2}{2m_W^2} \right)  f \left( \frac{m_W^2}{m_{H_i}^2} \right)
+ \left( 5 - \frac{m_{H_i}^2}{2m_W^2}  \right) g \left( \frac{m_W^2}{m_{H_i}^2} \right)   \right\} \\[1.5ex]
&\qquad \times {\rm Im} \left\{ \left( g^S_{H_i \bar{f} f } + i g^P_{H_i \bar{f} f } \right)
(- \sin (\alpha - \beta) \, O_{1i} + \cos (\alpha - \beta) \, O_{2i}) \right\} \, , \\[1ex]
\end{split}
\end{equation}
where the loop functions $f$ and $g$ are
\begin{align}
f(z) &= \frac{z}{2} \int_0^1 dx\, \frac{1-2x(1-x)}{x(1-x)-z} \log\frac{x(1-x)}{z}\, , \nonumber \\
g(z) &= \frac{z}{2} \int_0^1 dx\, \frac{1}{x(1-x)-z} \log\frac{x(1-x)}{z} \, . \label{fzgz}
\end{align}

\subsubsection{The CP-odd electron-nucleon interaction}

\begin{figure}[!t]
\vspace{-1cm}
 \begin{minipage}{0.5\hsize}
  \begin{center}
   \includegraphics[clip, width=7.5cm]{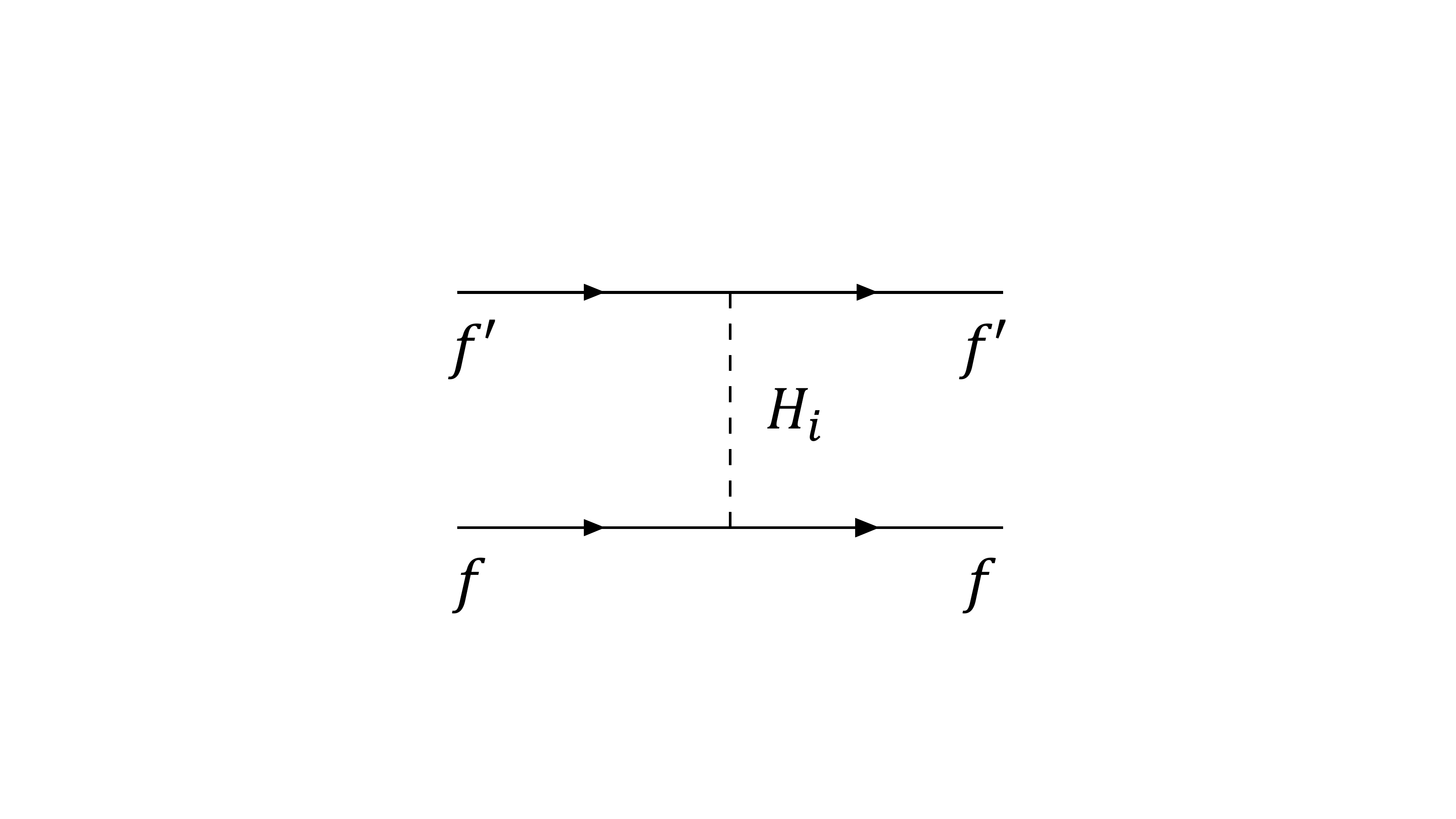}
  \end{center}
 \end{minipage}
\hspace{-1cm}
 \begin{minipage}{0.5\hsize}
  \begin{center}
   \includegraphics[clip, width=7.5cm]{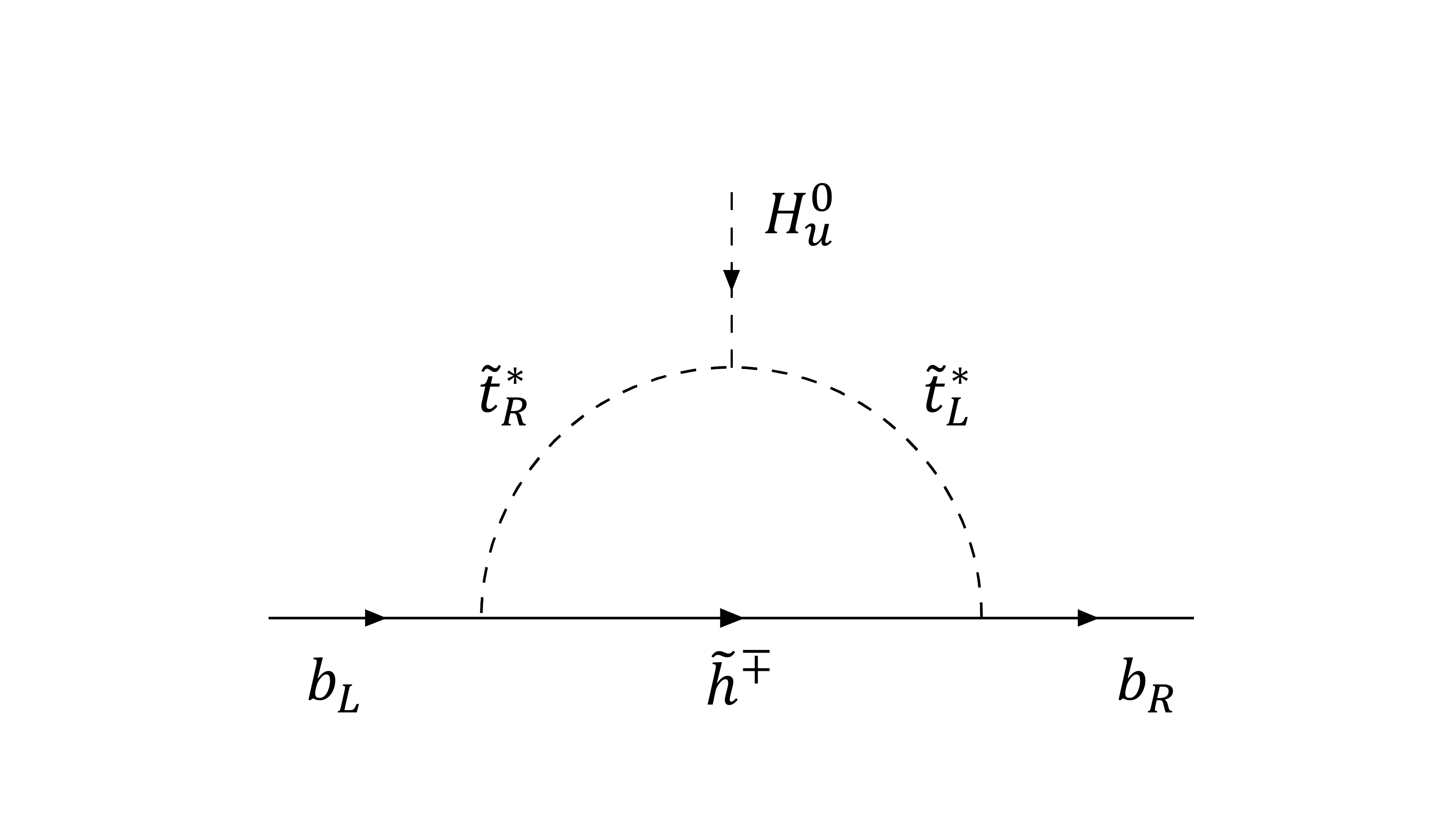}
  \end{center}
 \end{minipage}
  \caption{The CP-odd four fermion operators induced by the tree-level Higgs boson exchange (left).
Without new CP phases beyond the MSSM, the four fermion operators
come from the wrong-Higgs Yukawa coupling of bottom quarks induced by a stop/higgsino loop (right).}
  \label{fig:CPodd}
\end{figure}

The electron-nucleon interaction also contributes to the EDMs of paramagnetic atoms/molecules and also 
the EDMs of diamagnetic atoms and hadrons.
To derive the coefficient $C_S$, we first consider CP-odd four fermion operators,
\begin{equation}
\begin{split}
\\[-2.5ex]
\mathcal{L}_{\rm Four-Fermi} = \sum_{f,f'}
C_{ff'} \, \left( \bar{f} f \right) \left(\bar{f}' \, \iu\gamma_5 f' \right) \, , \label{fourfermi} \\[1ex]
\end{split}
\end{equation}
which generate the electron-nucleon interaction.
From the physical Higgs boson couplings with a Standard Model fermion $f$ \eqref{fermionint}, these operators are induced by
the tree-level Higgs boson exchange as shown in the left diagram of Figure~\ref{fig:CPodd}.
The coefficients $C_{ff'}$ are given by 
\begin{equation}
\begin{split}
\\[-2.5ex]
C_{ff'} = \sum_{i = 1}^3 \frac{g^S_{H_i \bar{f}f} g^P_{H_i \bar{f}' f'}}{m_{H_i}^2} \, . \label{cffprime} \\[1ex]
\end{split}
\end{equation}
From these interactions, we find the coefficient of the electron-nucleon interaction,
\begin{equation}
\begin{split} 
\\[-2.5ex]
C_S \approx C_{de} \frac{29 \, \rm MeV}{m_d} + C_{se} \frac{49 \, \rm MeV}{m_s} + C_{be} \frac{74 \, \rm MeV}{m_b}  \, ,
\label{Csexp} \\[1ex]
\end{split}
\end{equation}
where we have used $m_s \langle N |\bar{s} s | N \rangle \simeq 49 \, \rm MeV$ and
$m_b \langle N |\bar{b} b | N \rangle \simeq 74 \, \rm MeV$
\cite{Junnarkar:2013ac}.

Without new CP phases beyond the MSSM, the CP-odd four fermion operators
come from the wrong-Higgs Yukawa couplings
\cite{Lebedev:2002ne}.
As shown in the right diagram of Figure~\ref{fig:CPodd},
finite one-loop corrections of gauginos or higgsinos induce the wrong-Higgs Yukawa coupling of bottom quarks,
\begin{equation}
\begin{split}
\\[-2.5ex]
\mathcal{L}_{\rm bottom} = - \, y_b H_d^0 \, b^c b_L - y'_b H_u^{0 \, \dagger} \, b^c b_L + {\rm h.c. } \\[1ex]
\end{split}
\end{equation}
We present the detailed formulas in Appendix \ref{app:wronghiggsformula} based on \cite{Pilaftsis:2002fe}. The contribution from the CP-odd four fermion operators is important for the EDMs of paramagnetic atoms/molecules for a large $\tan \beta$ in the MSSM.

\subsection{The neutron EDM: operators probed and experimental status}

The most famous contribution to the neutron EDM comes from the theta term in the Standard Model.
This contribution can be removed by the usual Peccei-Quinn (PQ) symmetry
\cite{Peccei:1977hh} via a dynamical axion, which we assume here.
In this case, the neutron EDM is dominated by the EDMs and the CEDMs,
\begin{equation}
\begin{split}
\\[-2.5ex]
\mathcal{L}_{\rm CEDM} = - \iu g_s \frac{\tilde{d}_f}{2} \bar{f} \sigma_{\mu\nu} \gamma_5 f G^{\mu\nu} \, , \\[1ex]
\end{split}
\end{equation}
of up and down quarks.
Here, $g_s$ is the $SU(3)_C$ gauge coupling and $G_{\mu\nu} \equiv G_{\mu\nu}^a T^a$ $(a = 1, \cdots , 8)$ is the gluon field strength.
In addition, the dimension-six Weinberg operator
\cite{Weinberg:1989dx},
\begin{equation}
\begin{split}
\\[-2.5ex]
\mathcal{L}_w = \frac{1}{3} w f^{abc} G^a _{\mu\nu} \tilde{G}^{\nu\rho b} G_\rho^{\,\, \mu c} \, , \\[1ex]
\end{split}
\end{equation}
where $f^{abc}$ is the structure constant and $\tilde{G}^{\mu\nu} = \frac{1}{2} \epsilon^{\mu\nu\lambda\sigma} G_{\lambda\sigma}$,
and the four-fermion operators \eqref{fourfermi} may give important contributions.
Although these contributions suffer from large QCD uncertainties, consistent calculations using QCD sum rule techniques have been developed \cite{Pospelov:2000bw,Demir:2002gg,Demir:2003js}.
The result of the neutron EDM is summarized as
\cite{Ellis:2008zy}
\begin{eqnarray}
&&d_n = \Delta d_n (d_q , \tilde{d}_q) + \Delta d_n (w) + \Delta d_n (C_{ff'}) \, , \label{neutronEDMtotal} \\[2ex]
&&\Delta d_n (d_q , \tilde{d}_q) = (1.4 \pm 0.6) \left(d_d - 0.25 \, d_u \right)
+ (1.1 \pm 0.5) \, e \, (\tilde{d}_d + 0.5 \, \tilde{d}_u )   \, , \label{EDMpart} \\[2ex]
&&\Delta d_n (w) = \pm e \left(20 \pm 10 \right) \, {\rm MeV} \times w \, , \label{Weinbergpart} \\[2ex]
&&\Delta d_n (C_{ff'}) = \pm e \, 2.6 \times 10^{-3} \, {\rm GeV}^2 \left( C_{bd} + 0.75 \, C_{db} \right)/m_b \, , \label{FourFermipart}
\end{eqnarray}
where $d_q , \tilde{d}_q$ $(q = u,d)$ are evaluated at the electroweak scale
and these contributions have been reliably calculated.
The lattice calculation has also presented the result of the EDM part
\cite{Bhattacharya:2015esa}
which is consistent with the QCD sum rule calculation.
The calculations of the Weinberg operator \eqref{Weinbergpart}
and the four-fermion operators \eqref{FourFermipart} still have large uncertainty
and even the signs are not determined.
The Weinberg operator is evaluated at $1 \, \rm GeV$.

The current limit on the neutron EDM is \cite{Baker:2006ts}:
\be
|d_n| <  d_n^{\rm max} \equiv 2.9 \times 10^{-26}  \, e \, \rm cm.
\ee

\subsection{The neutron EDM: computation}

\begin{figure}[!t]
\vspace{-1cm}
\begin{center}
   \includegraphics[clip, width=7.5cm]{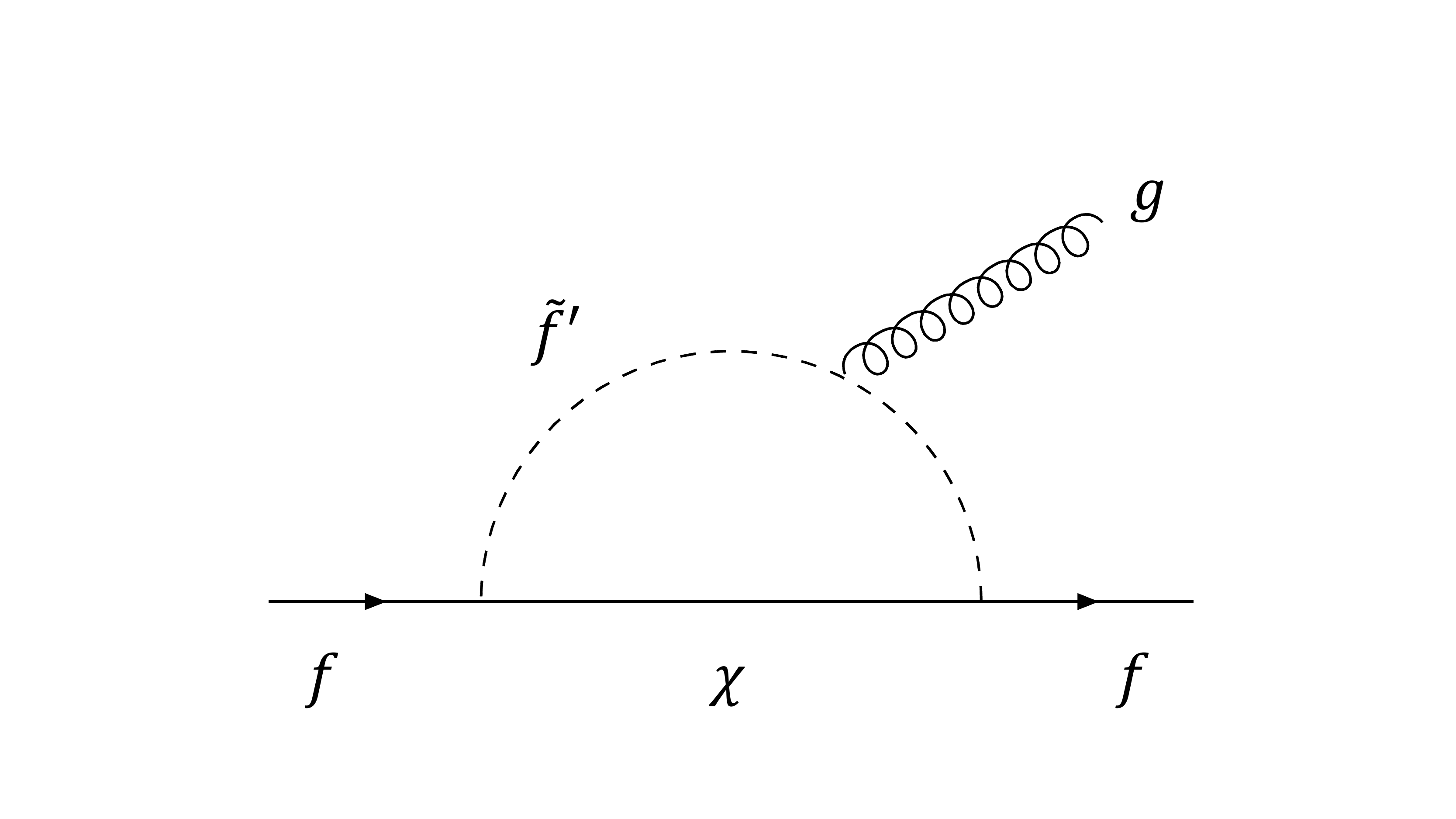}
\end{center}
\vspace{-0.7cm}
  \caption{The one-loop contribution to the CEDM of a fermion $f$.}
  \label{fig:OneLoopCEDM}
\end{figure}

The quark EDMs are generated by two-loop diagrams shown in Fig.~\ref{fig:BarrZee}, as in the case of the electron EDM. For the CEDMs, the diagrams are similar to the ones shown in Figure~\ref{fig:BarrZee} with photons replaced by gluons.
For instance, the stop contribution is expressed as 
\begin{equation}
\begin{split}
\\[-2.5ex]
\tilde{d}_q = \frac{\alpha_s}{64\pi^3} \sum_{i=1}^3 \frac{g^P_{H_i \bar{q} q}}{m_{H_i}^2} \sum_{a=1}^2
\Gamma_{H_i \tilde{t}_a^\ast \tilde{t}_a} F(m_{{\tilde t}_a}^2/m_{H_i}^2) \, . \\[1ex]
\end{split}
\end{equation}
The couplings $\Gamma_{H_i \tilde{t}_a^\ast \tilde{t}_a}$ are defined in Appendix \ref{StopHiggs}.
The Weinberg operator is generated by the two-loop gluino or higgsino exchange diagram
\cite{Dai:1990xh}.
However, the RG evolution of this operator suppresses its contribution
\cite{Braaten:1990gq,Arnowitt:1990eh}.
Another contribution comes from the CEDM of the bottom quark, which generates the Weinberg operator through a threshold correction at the scale $m_b$ \cite{Demir:2003js},
\begin{equation}
\begin{split}
\\[-2.5ex]
\Delta w (1\, {\rm GeV}) = 0.72 \, w (m_b) = - 0.72 \times \frac{g_s^3 \tilde{d}_b (m_b)}{32 \pi^2 m_b}
= - 0.68 \times \frac{g_s^3 \tilde{d}_b (m_Z)}{32 \pi^2 m_b}  \, .  \\[1ex]
\end{split}
\end{equation}
The $b$-quark CEDM is generated by one-loop diagrams with a higgsino exchange
\cite{Ibrahim:1997nc} as shown in Figure~\ref{fig:OneLoopCEDM}. The details of the calculation are presented in Appendix \ref{app:CEDMbottomquark}. The four-fermion operators are generated by the neutral Higgs boson exchange and the coefficients $C_{bd,db}$ are given by \eqref{cffprime}.

\subsection{The mercury EDM: operators probed and experimental status}

The most important contribution to the EDMs of diamagnetic systems such as mercury
is given by the Schiff moment which mainly comes from
CP-odd pion-nucleon interactions, $\mathcal{L}_{\pi NN} \supset \bar{g}^{(0)} \bar{N} \tau^a N \pi^a + \bar{g}^{(1)} \bar{N} N \pi^0$.
The coupling $\bar{g}^{(1)}$ is induced by the CEDMs $\tilde{d}_{u,d}$
\cite{Pospelov:2001ys} and the four quark operators $C_{q_1q_2}$.
The coupling $\bar{g}^{(0)}$ is induced by the CEDMs $\tilde{d}_{u,d}$.
The mercury EDM also has contributions from the electron EDM $d_e$ and the CP-odd electron-nucleon interactions.
The resulting expression of the mercury EDM has been estimated as
\cite{Ellis:2008zy}
\begin{equation}
\begin{split}
\\[-2.5ex]
d_{\rm Hg} = & \,\, 7 \times 10^{-3} \, e \, ( \tilde{d}_u - \tilde{d}_d ) + 10^{-2} \, d_e \\[1ex]
&- 1.4 \times 10^{-5} \, e \, {\rm GeV}^2 
\left( \frac{0.5 \, C_{dd}}{m_d} + 3.3 \kappa \, \frac{C_{sd}}{m_s} + \left(1 -0.25 \kappa \right) \frac{C_{bd}}{m_b} \right) \\[1ex]
& + 3.5 \times 10^{-3} \, e \, {\rm GeV} \,  C_S + 4 \times 10^{-4} \, e \,  {\rm GeV} \, \left( C_P - \frac{\langle \sigma_n \rangle_{\rm Hg} - \langle \sigma_p \rangle_{\rm Hg}}{\langle \sigma \rangle_{\rm Hg}} \, C'_P \right)  \, , 
\label{mercuryEDM}\\[1ex] 
\end{split}
\end{equation}
where $\kappa \equiv \langle N | m_s \bar{s} s | N \rangle / 220 \, {\rm MeV} = 0.22 \pm 0.045 \pm 0.068$
\cite{Junnarkar:2013ac}. We have corrected the relative coefficient of $C'_P$ compared to $C_P$ from that of \cite{Ellis:2008zy} according to \cite{Jung:2013hka}; it is approximately $-1$.
The electron-nucleon interactions $C_P $, $C'_P$ are defined as $\mathcal{L}_{eN} \supset C_P \bar{e} e \bar{N} \iu \gamma_5 N
+ C'_P \bar{e} e \bar{N} \iu \gamma_5 \tau_3 N$.
They are given by
\begin{equation}
\begin{split}
\\[-2.5ex]
&C_P \simeq - \, 0.38 \, {\rm GeV} \sum_{q = c,s,t,b} \frac{C_{eq}}{m_q} \, , \\[1ex]
&C'_P \simeq  - \, 0.81 \, {\rm GeV} \, \frac{C_{ed}}{m_d}
 - \, 0.18 \, {\rm GeV} \sum_{q = c,s,t,b} \frac{C_{eq}}{m_q} \label{Cp}  \, . \\[1ex] 
\end{split}
\end{equation}
Note that the central value is shown in each term of \eqref{mercuryEDM} which has large uncertainty. For example, the coefficient of $d_e$ in the expression is highly uncertain \cite{Jung:2013mg}. Below, we will use the estimated coefficient of ${\tilde d}_d$ in the above formula to make approximate comparisons of the relative reach of the mercury EDM, electron EDM, and $b \to s \gamma$ for new physics. However, in presenting exclusion plots in figures, our numerical results for the mercury EDM use a more conservative constraint on the quark CEDMs $\tilde{d}_u, \tilde{d}_d$ obtained by a likelihood analysis explained in Appendix~\ref{sec:mercuryuncertainty}. We present results for two different nuclear physics calculations of how $d_{\rm Hg}$ depends on pion--nucleon couplings: case (i) based on \cite{deJesus:2005nb} and case (ii) based on \cite{Ban:2010ea}. In both cases we marginalize over other uncertainties, for instance arising from QCD sum rule estimates of how pion--nucleon couplings depend on quark CEDMs. The two scenarios (i) and (ii) produce significantly different results, indicating that the interpretation of the mercury EDM is overwhelmingly dominated by uncertainties in nuclear physics that will have to be addressed by theorists in order to understand what any future experimental observation of a nonzero result is telling us about new physics. The bound that we find from scenario (ii) could be interpreted as conservative.

The most recent experimental limit on the mercury EDM is \cite{Graner:2016ses}
\be
|d_{\rm Hg}| < d_{\rm Hg}^{\rm max} \equiv 7.4 \times 10^{-30} \, e \, \rm cm. \label{mercuryEDMupperbound}
\ee
All of the necessary ingredients to compute the mercury EDM have already been discussed in the previous subsections.

%
%

\section{Stop contributions}\label{numerical}

We now numerically study implications on the parameter space of CP-violating natural SUSY
from the present and projected EDM measurements.
As discussed in \S\ref{naturalSUSYhiggs}, we consider three scenarios to lift up the Higgs boson mass.
The first scenario is the MSSM with near-maximal stop mixing.
The large $A_t$-term with a CP-violating phase 
\be
\phi_t \equiv \arg(A_t \mu b_\mu^*)
\ee
leads to sizable EDMs (because we can always rotate $H_d$ to remove a phase in $b_\mu$, we will sometimes simply denote this phase by $\arg(A_t \mu)$).
The second scenario is to introduce extra gauge interactions of the Higgs fields.
After integrating out the massive gauge fields,
the Higgs mass is increased by dimension six operators in the K\"ahler potential of the Higgs fields.
As in the first scenario, there is no new CP-violating phase beyond the MSSM. The $A$-term can now be relatively small, but it will nonetheless be induced by renormalization group effects, so the phase $\phi_t$ originates from $\arg(M_{\tilde g} \mu)$.
Finally, we consider the scenario where the new Higgs interactions of \eqref{BMSSMsuperpotential} and \eqref{BMSSMsoft}
lift the Higgs boson mass.
In this scenario, we concentrate on two new phases associated with the new Higgs interactions. Because the physics of these phases is rather different from that of $\phi_t$, we postpone the discussion until \S\ref{sec:BMSSM}. Thus, in the remainder of this section, we consider $\phi_t \neq 0$ and set all other CP-violating phases in the MSSM to zero for simplicity.

We find that one of the most important effects of the stops is the generation of quark CEDMs, which are strongly constrained by the experimental tests of the mercury EDM. These contributions may be readily understood through the general 2-loop formulas of \S\ref{twoloopfermions}. (The first calculations may be found in \cite{Chang:1998uc}.) The CEDMs are generated with an inner loop with two external gluons and one pseudoscalar Higgs boson $A^0$. In terms of the stop mass eigenstates, the relevant couplings are the diagonal ones:
\be
{\cal L} \supset g^{A^0}_{{\tilde t}_a {\tilde t}_a} A^0 {\tilde t}_a^\dagger {\tilde t}_a,
\ee
where in the small-mixing limit,
\be
g^{A^0}_{{\tilde t}_a {\tilde t}_a} \approx (-1)^{a+1} y_t^2 v \frac{\left| \mu A_t\right|}{m_{{\tilde t}_1}^2 - m_{{\tilde t}_2}^2} \arg(\mu A_t).
\ee
This coupling may be substituted where the general $g^S_{\phi \omega}$ appears in equation (\ref{eq:cEscalarphiomega}). More general expressions for the couplings of stops to the Higgs, valid also in the presence of additional Higgs-sector phases, are given in the Appendix \ref{StopHiggs}.

From this we can extract, based on our general results, the CEDM of the down quark:
\be
{\tilde d}_d = \frac{\alpha_s}{64 \pi^3} \frac{m_d}{m_A^2} y_t^2 \tan \beta \frac{|\mu A_t|}{m_{{\tilde t}_1}^2-m_{{\tilde t}_2}^2} \sin(\phi_t) \left[F(m_{{\tilde t}_1}^2/m_A^2) - F(m_{{\tilde t}_2}^2/m_A^2)\right],\label{eq:stopCEDM}
\ee
where $F(z) = \int_0^1 dx\, \frac{x(1-x)}{z-x(1-x)} \log \frac{x(1-x)}{z}$.
A similar expression holds for the up quark, but is suppressed in the large $\tan \beta$ limit. These expressions may be easily understood in the limit $m_{\tilde t} \gg m_A$ by first integrating out the stops to produce an effective $A^0 G^a_{\mu \nu} G^{a\mu\nu}$ operator and then computing the mixing of this operator with the quark CEDM, using the limit $F(z) \rightarrow - \frac{1}{6z} \log z$ for $z \gg 1$.
The mercury EDM is approximately $d_{\rm Hg}/e \approx -7 \times 10^{-3} {\tilde d}_d$ (assuming the central value) whereas the electron EDM $d_e/e$ is controlled by a formula similar to the right-hand side of (\ref{eq:stopCEDM}) but with $\alpha_s m_d \to 2 N_c Q_t^2 \alpha m_e$. (The factor of 2 comes from ${\rm tr}(T^a T^b) = \frac{1}{2}\delta^{ab}$ in the CEDM case.) In other words, for a given point in parameter space, we expect the ratio of mercury EDM to electron EDM (as inferred from ThO) to be about
\be
\left|\frac{d_{\rm Hg}}{d_e}\right| \approx \frac{7 \times 10^{-3} \alpha_s m_d}{2 N_c Q_t^2 \alpha m_e} \approx 0.4,
\ee
whereas the measured bound for mercury currently reaches to EDMs that are smaller by a factor of about
\be
\left|\frac{d_{\rm Hg}^{\rm max}}{d_e^{\rm max}}\right| \approx \frac{7.4 \times 10^{-30} e\, {\rm cm}}{8.7 \times 10^{-29} e\, {\rm cm}} \approx 0.09,
\ee
so the mercury measurement is currently more constraining for stops by a factor of slightly more than 4 in the EDM or $\sim 2$ in stop mass.
However, as described before, the mercury EDM suffers from large theoretical uncertainty
and it may not be appropriate to assume the formula using the central value like $d_{\rm Hg}/e \approx -7 \times 10^{-3} {\tilde d}_d$.
More careful treatments will be performed in numerical studies.

Although the CEDM, via the measurement of mercury, is likely to be currently the strongest EDM constraint on stops, in the near future an improved bound on the electron EDM from the ACME collaboration is expected to surpass the current constraint from mercury.

\subsection{The MSSM with the near-maximal stop mixing}\label{MSSMlargeAt}

\begin{figure}[!h]\begin{center}
\includegraphics[clip,width=0.47\textwidth]{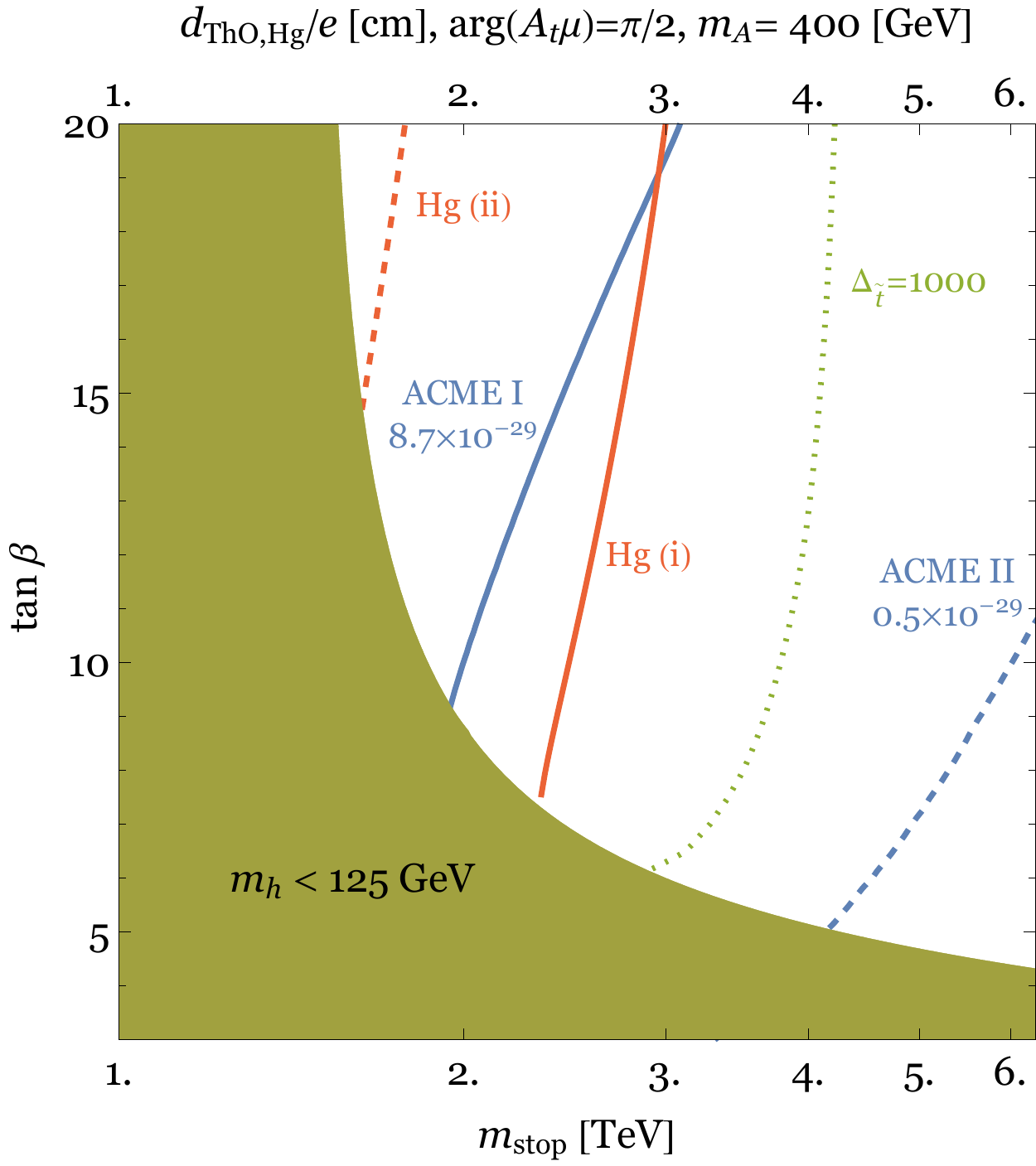}~~\includegraphics[clip,width=0.5\textwidth]{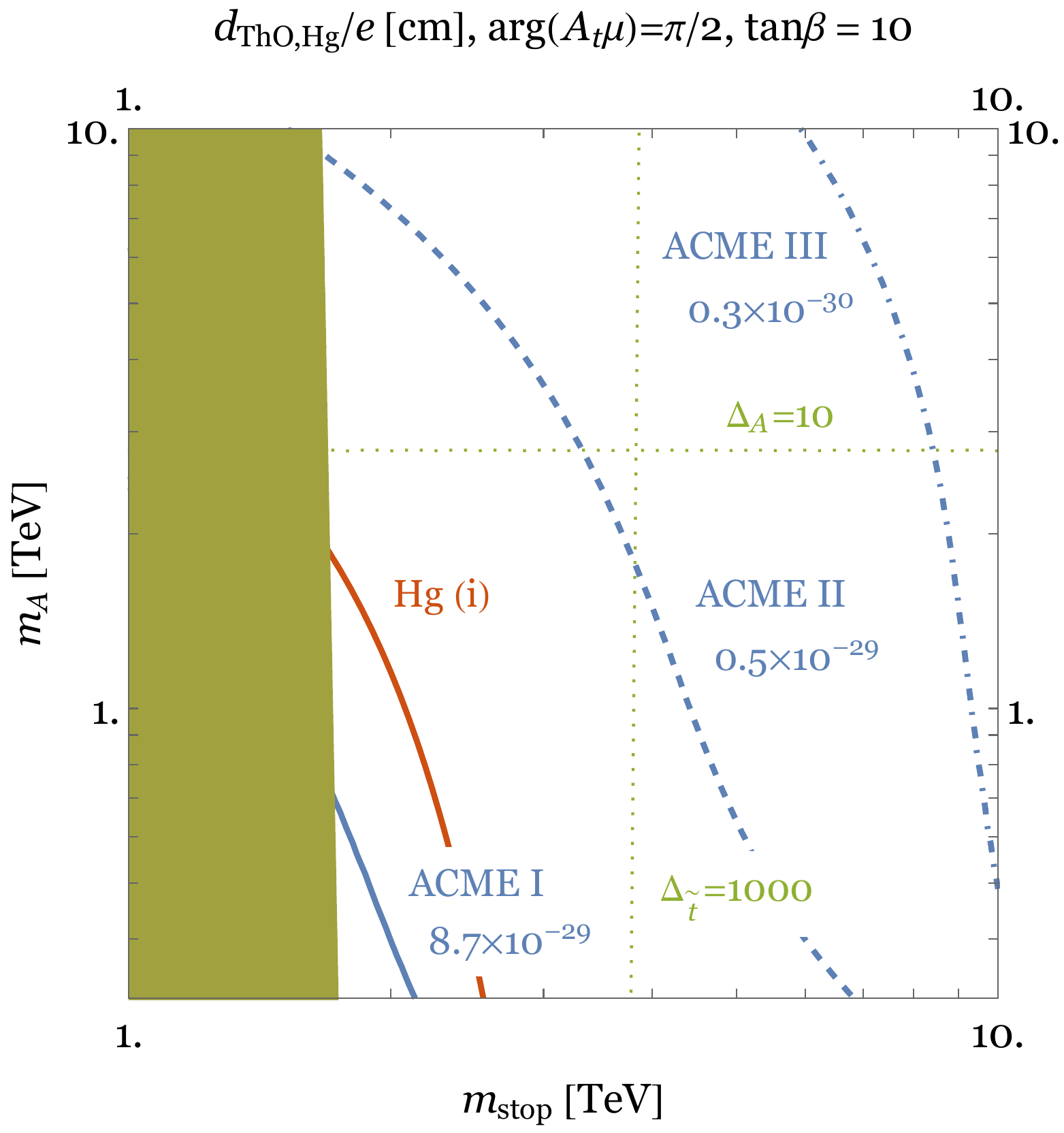}
\end{center}
\caption{EDM constraints on the stop parameter space in the MSSM, where stop loops with large $A$-term lift the Higgs mass to 125 GeV. The horizontal axis shows the common stop soft mass $m_{\rm stop} = {\widetilde m}_{Q_3} = {\widetilde m}_{u_3}$. At left we fix $m_A = 400$ GeV and vary $\tan \beta$ on the vertical axis; at right we fix $\tan \beta = 10$ and vary $m_A$ on the vertical axis. In the brown/green shaded region, no choice of $A_t$ is sufficient to achieve the correct Higgs mass. In the rest of the parameter space, at each point we choose $A_t$ to achieve $m_h = 125$ GeV. Regions of parameter space to the left of the solid blue contours are excluded by measurements of ThO. Red solid and dashed contours denote the mercury EDM constraints for the cases (i) and (ii) discussed in Appendix~\ref{sec:mercuryuncertainty}), respectively.
The blue dashed and dot-dashed contours (``ACME II'' and ``ACME III'') are future projections. The dotted green lines display the stop fine tuning (\ref{eq:stoptuning}) and tree-level Higgs fine tuning (\ref{eq:Afinetuning}). We have fixed $|\mu| = 350$ GeV in these figures.}
\label{fig:mssmEDMs}
\end{figure}%

We first consider the MSSM with large $A_t$ terms in order to explain the $125~{\rm GeV}$ Higgs boson mass with relatively light stops (see, for instance, ref.~\cite{Draper:2011aa} and references therein). As discussed above, the dominant effect will be two-loop EDMs and CEDMs, which for down-type fermions are $\tan \beta$-enhanced. In addition, a stop/higgsino loop induces the wrong-Higgs Yukawa coupling of bottom quarks
which generates a CP-odd four fermion operator as in \eqref{cffprime} and \eqref{MSSMfourfermi}.
From the expression of \eqref{Csexp}, the contribution from the CP-odd electron-nucleon interaction is given by
\begin{equation}
\begin{split} 
\\[-2.5ex]
C_S \approx C_{be} \frac{74 \, \rm MeV}{m_b}  \, . \\[1ex]
\end{split}
\end{equation}
Then, we can discuss current and projected constraints on stop masses from the paramagnetic ThO molecule
in terms of the effective EDM \eqref{dThO}.
For the neutron EDM, there are the stop contributions to the quark EDMs and the CEDMs, together with CP-odd four fermion operators, with total given in \eqref{neutronEDMtotal}.
The mercury EDM constraint can be calculated from the likelihood analysis given in Appendix~\ref{sec:mercuryuncertainty}.

We now numerically analyze current and projected constraints on stop masses from the EDM of the paramagnetic ThO molecule and the neutron and mercury EDMs. For simplicity, we assume $m_{Q_3} = m_{u_3} = m_{\rm stop}$. For each choice of $m_{\rm stop}$, $\tan \beta$, and $m_A$, the absolute value of $A_t$ is fixed to obtain the correct Higgs boson mass. We use the SusyHD code \cite{Vega:2015fna} for this calculation of the Higgs boson mass. There are still moderately large theoretical uncertainties in Higgs mass calculations; see \cite{Draper:2016pys,Athron:2016fuq} for recent discussions. If an EDM is detected in the future, and we wish to interpret it in the MSSM, a careful assessment of such uncertainties would be desirable. For now, we expect that our results give a reasonably accurate view of the constraints on parameter space.

Figure~\ref{fig:mssmEDMs} shows constraints on stop masses, $\tan \beta$, and $m_A$ in the MSSM with near-maximal stop mixing. In the dark shaded region, no choice of $A_t$ suffices to obtain $m_h = 125~{\rm GeV}$. Away from this region, $A_t$ is always selected to obtain the correct Higgs mass. We take $|\mu| = 350 \, \rm GeV$ and assume that the masses of the first and second generations of squarks, the right-handed sbottom and sleptons are $10 \, \rm TeV$ and the three gaugino masses are $2 \, \rm TeV$. When varying $\tan \beta$ we fix $m_A = 400~{\rm GeV}$ and when varying $m_A$ we fix $\tan \beta = 10$.  The plot illustrates that the strongest current constraints come from measurements of ThO and Hg, ruling out stops above 2 to 3 TeV over a wide range of moderately large $\tan \beta$ and pseudoscalar Higgs masses below the TeV scale. One can evade the constraint by lifting the pseudoscalar Higgs mass above about 3 TeV, but at the price of introducing additional tree-level fine-tuning in the Higgs sector. In the MSSM, achieving a large enough Higgs mass always requires worse than percent-level tuning from stop loops, but for large CP phases this is made significantly worse by the EDM constraint. The figures also illustrate that the upcoming update from ACME II is expected to significantly improve over the current constraint from mercury, while ACME III will push the bounds on stop masses out to nearly 10 TeV. The neutron EDM constraint is not shown because it is currently substantially weaker than the ThO and Hg constraints.

Because EDMs are dimension six operators, bounds on masses of new particles roughly scale as the square root of the sensitivity to the EDM. However, for {\em fixed} masses, the EDM measurements directly probe smaller CP violating phases. Stop and pseudoscalar Higgs masses near 1 to 2 TeV are currently constrained only for nearly-maximal CP violating phases, but ACME III constrain such masses to have percent-level phases or smaller. This is potentially a very powerful constraint on models of supersymmetry breaking.

\subsection{Extra gauge interactions}

\begin{figure}[!h]\begin{center}
\includegraphics[clip,width=0.48\textwidth]{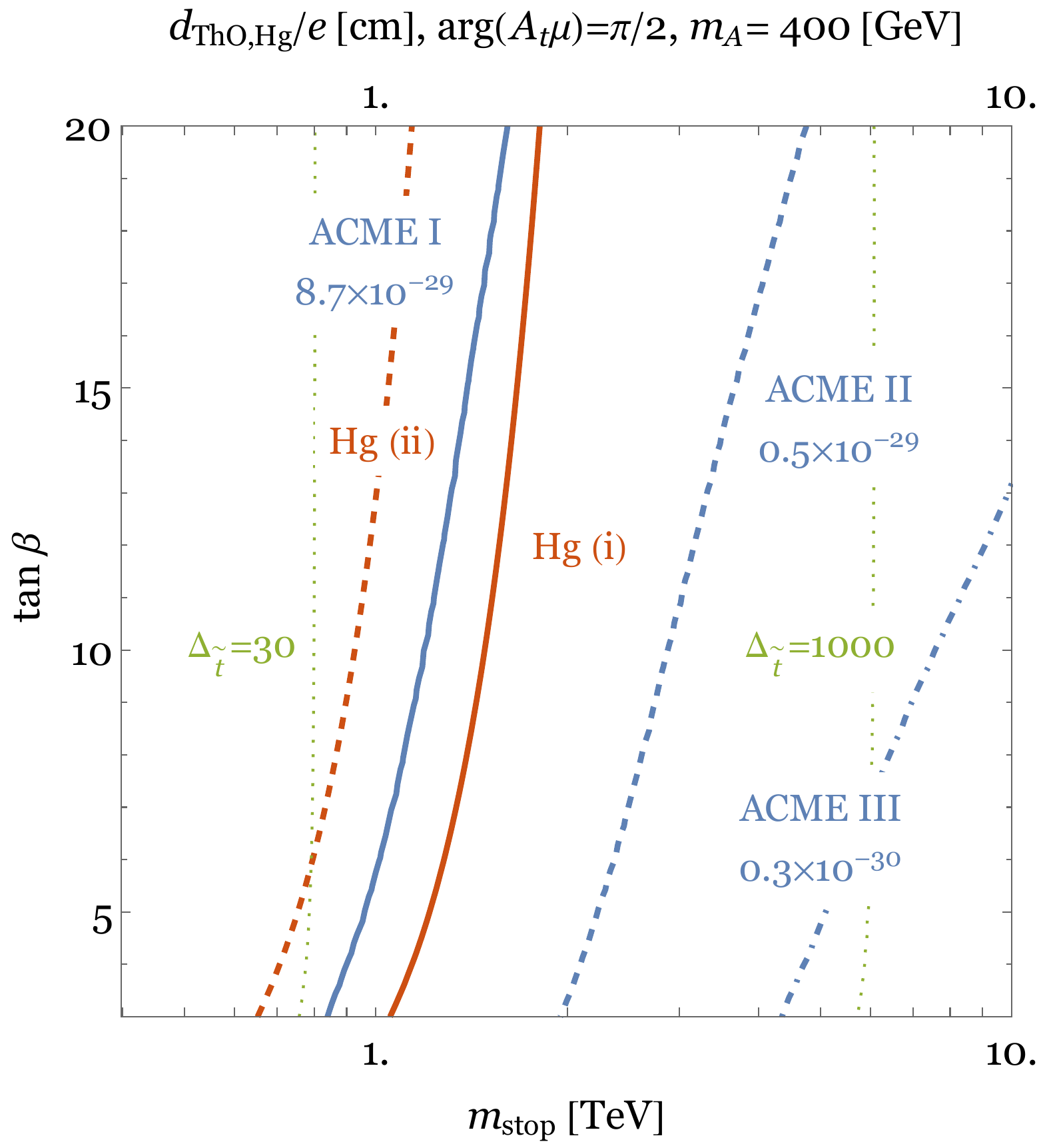}\includegraphics[clip,width=0.5\textwidth]{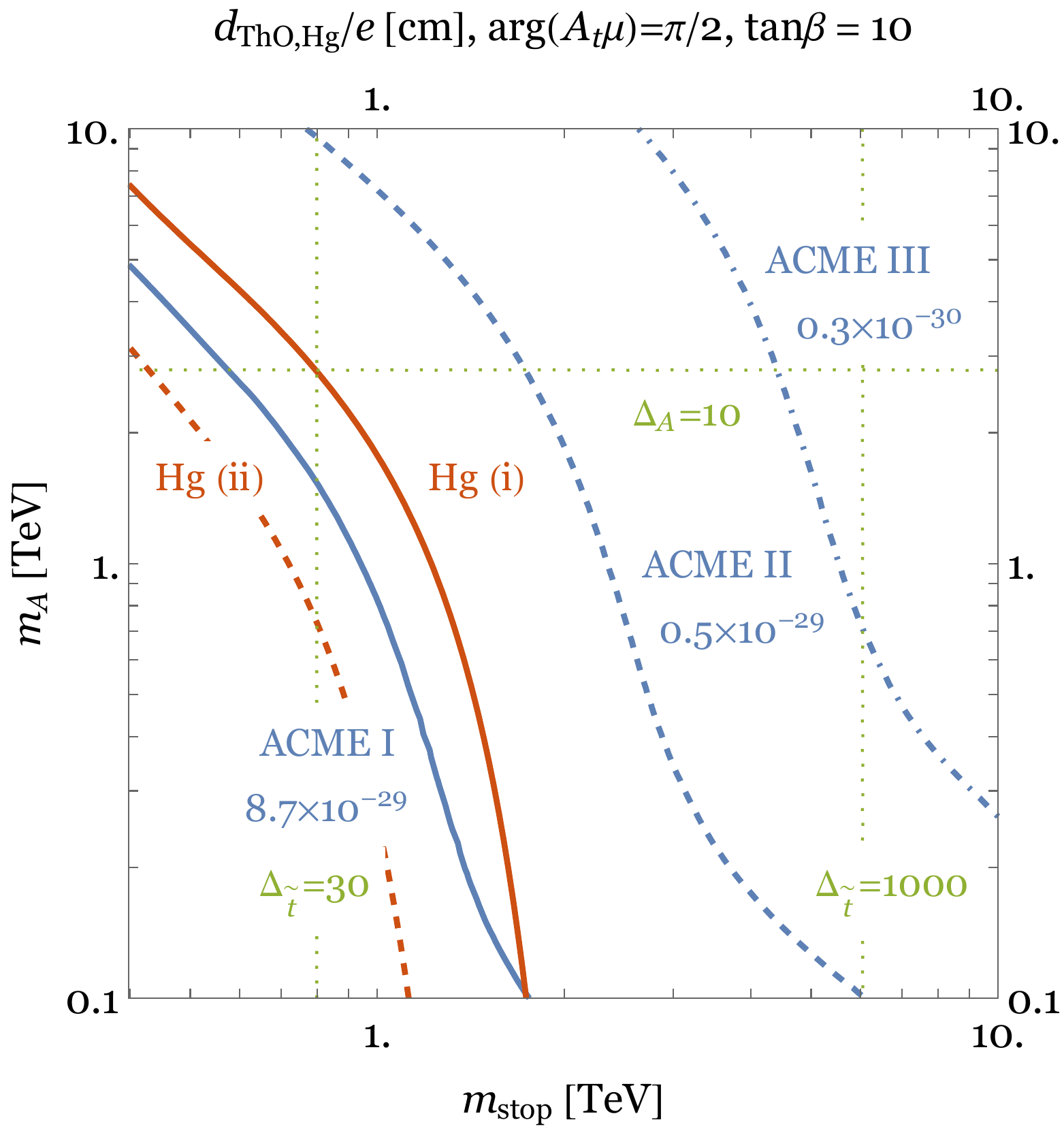}
\end{center}
\caption{EDM constraints on stop parameter space. This is very similar to Fig.~\ref{fig:mssmEDMs}, except that we no longer assume the MSSM, so the $A$-term is not constrained by the Higgs boson mass. Instead, we take into account the expected size of an $A$-term induced by RG running from the gluino mass, as in (\ref{radiativeA}).
The red solid and dashed contours (``Hg (i)'' and ``Hg (ii)'') denote the mercury EDM constraint for the two nuclear physics computations discussed in
Appendix~\ref{sec:mercuryuncertainty}. }
\label{fig:DtermEDMs}
\end{figure}%

Our second scenario raises the Higgs mass through new interactions that do not significantly influence the EDMs, for example by new abelian gauge interactions of the Higgs fields.  After integrating out the massive gauge fields, the Higgs mass is uplifted by dimension six operators in the K\"ahler potential of the Higgs fields.
As in the first scenario, there is no new CP-violating phase beyond the MSSM. Now we no longer have the constraint that $A_t$ must be chosen to achieve a correct Higgs mass. However, the gluino mass $M_{\tilde{g}}$ leads to a sizable $A_t$ by the running correction, 
\begin{equation}
\begin{split}
\\[-2.5ex]
\delta A_t \approx - \frac{2}{3 \pi^2} g_s^2 M_{\tilde{g}} \log \left( \frac{M_{\rm med}}{M_{\tilde{g}}} \right) 
\approx - 790 \, {\rm GeV} \left( \frac{M_{\tilde{g}}}{2 \, \rm TeV} \right) \frac{\log (M_{\rm med}/M_{\tilde{g}})}{\log (50)} \, . 
\label{radiativeA} \\[1ex]
\end{split}
\end{equation}
Here, $g_s$ is the color gauge coupling. The choice of $M_{\tilde g}$ at 2 TeV is motivated by the approximate upper end of current collider searches, and involves mild tuning for electroweak symmetry breaking \cite{Buckley:2016kvr,Buckley:2016tbs}.
That is, we necessarily expect this size of $A_t$ without unnatural tuning. (See \S3.1 of \cite{Katz:2014mba} for a more detailed discussion of the expected size of $A_t$ induced from the gluino mass, including a figure showing results including a more careful solution of the RGEs.)

Figure~\ref{fig:DtermEDMs} shows the constraints on stop masses and $\tan \beta$ from the paramagnetic EDM and the mercury EDM in the scenario with extra gauge interactions of the Higgs fields. The curves are similar to those shown in the MSSM case, but due to the smaller values of the $A$-terms considered, the reach is more modest.
Nonetheless, again the mercury EDM (for the case (i)) already constrains the region where stops and pseudoscalars are below 1 TeV, and future ACME improvements (labeled ``ACME II'' and ``ACME III'') will push toward large mass regions with tuning of a part in a thousand, and to CP violating phases of order $10^{-2}$ in the sub-TeV mass region. In this scenario, because the Higgs mass is assumed to be decoupled from the spectrum of stops, there is at least a chance that a fully natural model could be realized. Fig.~\ref{fig:DtermEDMs} shows that in the region of possibly low fine-tuning, order-one CP violating phases are already excluded.

\subsection{Comparison to the $b \to s\gamma$ constraint on stops and higgsinos}

The EDM induced from stop loops depends on the off-diagonal terms in the stop mass matrix, being proportional to $\arg(A_t \mu b_\mu^*)$. A number of other precision CP-even observables depend on the value of $A_t \mu$, and it is interesting to ask how the new physics reach of EDMs compares to that of such observables. Interestingly, these observables depend crucially on left-right stop mixing, and thus are nonzero even in the ``stop blind spot'' region of parameter space in which the lighter stop mass eigenstate decouples from the Higgs boson. The blind spot is difficult to probe with traditional precision electroweak observables \cite{Craig:2014una,Fan:2014axa}.

The deviation of the $b \to s \gamma$ branching ratio from that predicted by the Standard Model is one interesting new physics observable induced by loops of stops and higgsinos. The leading dependence is roughly \cite{Altmannshofer:2012ks}
\begin{align}
\Delta_{bs\gamma} \equiv \frac{\delta {\rm Br}(B \to X_s \gamma)}{{\rm Br}(B \to X_s \gamma)_{\rm SM}} &\approx 1.28 \tan \beta \frac{A_t \mu m_t^2}{m_{{\widetilde Q}_3}^2 m_{{\widetilde u}_3}^2} \log \frac{m_{{\widetilde Q}_3} m_{{\widetilde u}_3}}{\mu^2}.
\end{align}
The recently updated Standard Model prediction \cite{Misiak:2015xwa} is ${\rm Br}(B \to X_s \gamma)_{\rm SM} \approx (3.36 \pm 0.23) \times 10^{-4}$ while the experimental result is ${\rm Br}(B \to X_s \gamma)_{\rm exp} \approx (3.43 \pm 0.21 \pm 0.07) \times 10^{-4}$ \cite{Lees:2012ufa,Amhis:2014hma}. (The Standard Model result has moved closer to the central value of the experiments.) Ignoring the slight asymmetry between positive and negative corrections, we can take the experimental bound to be $|\Delta_{b s\gamma}| \lesssim \Delta_{bs\gamma}^{\rm max} \equiv 0.2$ at 95\% CL.

Thus, in the ratio of the deviation of the $b \to s\gamma$ rate from the Standard Model prediction and the mercury (or electron) EDM, the magnitude of $A_t \mu$ and $\tan \beta$ drop out. In the limit $\mu \approx m_A$, $m_{{\tilde t}_1} \approx m_{{\tilde t}_2} \gg m_A$, the expressions simplify even more, and we find:
\be
\left|\frac{d_{\rm Hg}}{\Delta_{bs\gamma}}\right| \approx 2.3 \times 10^{-26} \sin(\phi_t)\, e\, {\rm cm} \approx 620 \sin(\phi_t) \left|\frac{d_{\rm Hg}^{\rm max}}{\Delta_{bs\gamma}^{\rm max}}\right|.
\ee
In other words, in the region of parameter space where the stops are significantly heavier than the pseudoscalar Higgs, the CP phase $\phi_t$ must be of order $10^{-3}$ or smaller in order for the $b \to s\gamma$ constraint to be as important as the mercury EDM constraint. In the opposite limit $m_A \gg m_{{\tilde t}_1}, m_{{\tilde t}_2}$, the EDM decouples while the $b \to s\gamma$ constraint remains unchanged. As a result, the $b \to s\gamma$ constraint is stronger in this region of parameter space, though for order-one CP violating phase and stop masses near a TeV it does not dominate until $m_A \gtrsim 40~{\rm TeV}$, well outside the natural range of parameters even for quite large values of $\tan \beta$.

The experimental bound on $b \to s\gamma$ is often viewed as one of our most important indirect constraints on natural realizations of supersymmetry \cite{Ishiwata:2011ab,Blum:2012ii,Katz:2014mba}. Here we have shown that, in any scenario in which the phase of $A_t \mu$ is larger than about $10^{-3}$, the EDM constraint will be even more important. This tightens the argument of \cite{Katz:2014mba} that there can be no natural decoupling of the heavy Higgs bosons at large $\tan \beta$ consistent with precision bounds, unless one can build a model in which the CP phase is naturally very small.

%
%
\section{EDM constraints in the chargino sector}\label{sec:chargino}

In this section we will discuss constraints from electric dipole moments arising from the relative phase between wino and higgsino masses, i.e.~those proportional to ${\rm arg}(\mu M_2 b_\mu^*)$. These arise from charginos running in 2-loop Barr-Zee-type diagrams \cite{Chang:2002ex,Pilaftsis:2002fe}. In the context of split supersymmetry, in which we decouple all new particles except neutralinos, charginos, and gluinos, the dominant contributions arise from two-loop Barr-Zee diagrams involving exchange of the light Higgs boson \cite{ArkaniHamed:2004yi,Giudice:2005rz}. As emphasized in \cite{Fan:2013qn}, similar bounds would play a role in the much more general context of vectorlike fermion dark matter coupled to the Higgs boson. In the larger context of natural supersymmetry, the additional Higgs bosons play a major role and can mediate the dominant chargino effects at large $\tan \beta$ \cite{Li:2008kz}. We will take these two scenarios in turn, first focusing on the split-SUSY-like limit of a single light Higgs boson and then turning on the effects of additional Higgs bosons.

Because charginos carry no color charge or flavor charge, they do not lead to two-loop CEDMs of quarks or to four-fermion operators. Only their two-loop contributions to the EDMs $d_e$ and $d_q$ will play a role. As a result, they influence the mercury EDM only through $d_e$, which is more tightly constrained by the ThO EDM.

\subsection{EDM constraints on charginos alone}

If we have only one light Higgs, as in split supersymmetry, the chargino EDM is induced through two-loop diagrams of $\gamma h$, $Zh$, and $WW$ type, as computed in \cite{Giudice:2005rz}. The $Zh$ diagram is highly subleading for the electron EDM, but relevant for quark EDMs. We include all contributions. As explained in \cite{ArkaniHamed:2004yi}, the leading logarithmic dependence of the calculation can be understood by first integrating out the charginos at one loop to obtain
\be
\frac{e^2}{16\pi^2} (\arg \det {\cal M}_{\tilde C}) F_{\mu \nu} {\widetilde F}^{\mu \nu} = \frac{e^2}{8\pi^2} \frac{{\rm Im}(g^2 M_2 \mu H_u \cdot H_d)}{\left|M_2 \mu - g^2 H_u \cdot H_d\right|^2} F_{\mu \nu} {\widetilde F}^{\mu \nu}
\ee
and then considering the one-loop anomalous dimension mixing this operator with a fermionic EDM. 

\begin{figure}[!h]\begin{center}
\includegraphics[clip,width=1.0\textwidth]{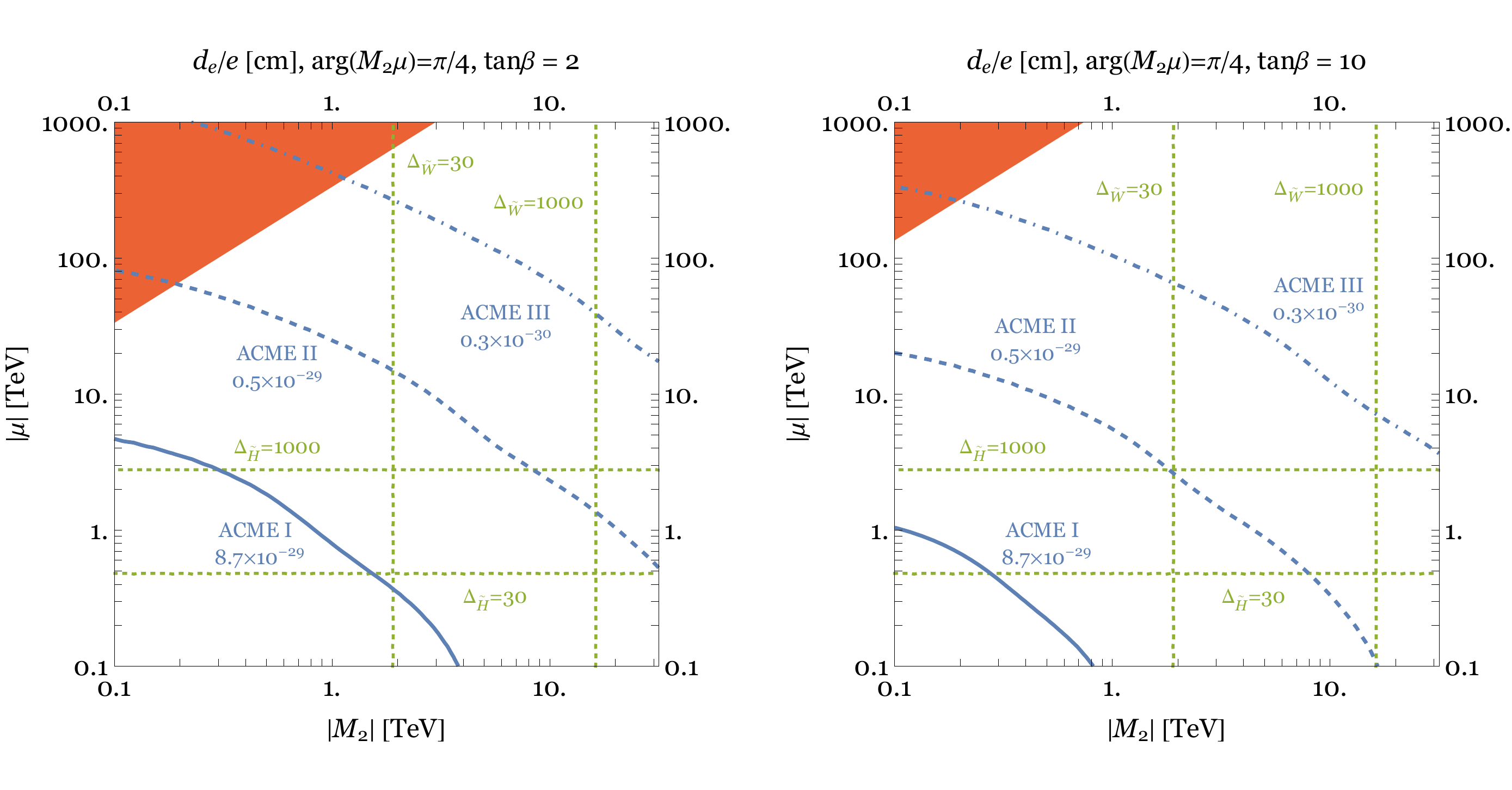}
\end{center}
\caption{Constraints on chargino parameter space from the ACME constraint (and future projections) on the electron EDM $d_e$, in the limit when all particles except charginos and neutralinos are decoupled. Notice that masses are in {\em TeV}. In both figures the CP-violating phase is taken to be $\pi/4$. The left figure shows the case $\tan \beta = 2$ whereas the right figure shows $\tan \beta = 10$. Regions to the lower left of the blue curves are (or will be) probed by the ACME measurement. We have $d_e \propto \sin 2\beta \sim 1/\tan \beta$, so the constraints are weaker at large $\tan \beta$. Dotted green lines show contours of fine tuning arising from the effects of the higgsino and wino masses on the Higgs potential at tree level and one loop, respectively. The dark orange shaded regions involve a tuning purely within the chargino sector (unrelated to electroweak naturalness) and so are disfavored.}
\label{fig:CharginoFirstLook}
\end{figure}

In Figure \ref{fig:CharginoFirstLook} we show contours in the $(M_2, \mu)$ plane corresponding to the current ACME bound, $|d_e| < 8.7 \times 10^{-29} e\, {\rm cm}$, as well as to projected future results (labeled ``ACME II'' and ``ACME III''). Because the EDM contribution decreases at large $\tan \beta$, we present two different results at $\tan \beta = 2$ and $10$ respectively. We can see that in either case, the parameter space with phase ${\rm arg}(\mu M_2 b_\mu^*) = \pi/4$ and chargino masses below 1 TeV is already excluded. The next improvement will probe masses above 10 TeV. This is an extremely powerful constraint on supersymmetric parameter space. Some clues to the interesting range of values for $M_2$ and $\mu$ come from naturalness, gauge coupling unification, and dark matter. The higgsinos play a crucial role in precision unification of gauge couplings, but they can be as heavy as 1000 TeV while maintaining precise unification \cite{ArkaniHamed:2012gw}.

As discussed in \S\ref{naturalSUSYtuning}, naturalness puts stringent constraints on chargino parameters, preferring higgsinos below 300 GeV and winos below 1 TeV if we wish to avoid more than a factor of 10 tuning. To illustrate the degree of fine-tuning, we depict $\Delta_{\tH,\tW} = 30, 1000$ with dotted green lines in Fig.~\ref{fig:CharginoFirstLook}. We see that in the natural region of chargino parameter space, large CP-violating phases are already excluded!

A secondary naturalness consideration is that if the higgsino is much heavier than the wino, there is a one-loop threshold correction to the wino mass \cite{Pierce:1996zz,Giudice:1998xp,Gherghetta:1999sw},
\be
\delta M_2^* = \frac{g^2}{16\pi^2} \mu \sin(2\beta) \left[\frac{\log(m_A^2/|\mu|^2)}{1-|\mu|^2/m_A^2}\right],
\ee
where we have taken $b_\mu$ real so that $\sin 2\beta = 2 b_\mu/m_A^2$. In theories where $\mu$ is on the same order as the scalar soft masses---e.g. when the Giudice-Masiero mechanism makes it of order $m_{3/2}$ and the soft masses are as well---the term in brackets is order-one. Even in the \mbox{(mini-)}split supersymmetry context, where we assume a fine tuning of the Higgs mass, it would take an additional fine tuning---and one without obvious anthropic motivation---to make the wino much lighter than this threshold correction. The shaded dark orange region at the upper left in Fig.~\ref{fig:CharginoFirstLook} is the region excluded by this ``naturalness'' consideration (taking $m_A^2 = 2|\mu|^2$ for concreteness). The region in which $|M_2| \gg |\mu|$ is disfavored by similar logic, but it has generally been of less theoretical interest and so we cut off the plot at values of $M_2$ too low for it to be relevant.

A final consideration is dark matter. In the limit $|M_1|, |\mu| \gg |M_2|$, the thermal relic abundance of wino dark matter will overclose the universe unless $|M_2| \lesssim 3~{\rm TeV}$. In the opposite limit $|M_1|,|M_2| \gg |\mu|$, the thermal relic abundance of higgsino dark matter will overclose the universe unless $|\mu| \lesssim 1~{\rm TeV}$. In more general mixed scenarios, a sizable admixture of at least one of wino or higgsino should be present in the lightest neutralino, because binos alone have no significant annihilation channels. As a result, thermal neutralino dark matter is always expected to have a mass below about 3 TeV. In many nonthermal dark matter scenarios, for instance in cases of moduli or gravitino decays, one produces {\em more} dark matter than in the thermal context. As a result, if the lightest neutralino is stable, this provides a strong preference for the parameter space in the lower-left corner of Fig.~\ref{fig:CharginoFirstLook}. (The literature on neutralino dark matter is vast; some useful entry points for physics mentioned in this paragraph include references \cite{Mizuta:1992qp,Moroi:1999zb,Chattopadhyay:2005mv,Hisano:2006nn,ArkaniHamed:2006mb,Moroi:2013sla}.)

\begin{figure}[!h]\begin{center}
\includegraphics[width=1.0\textwidth]{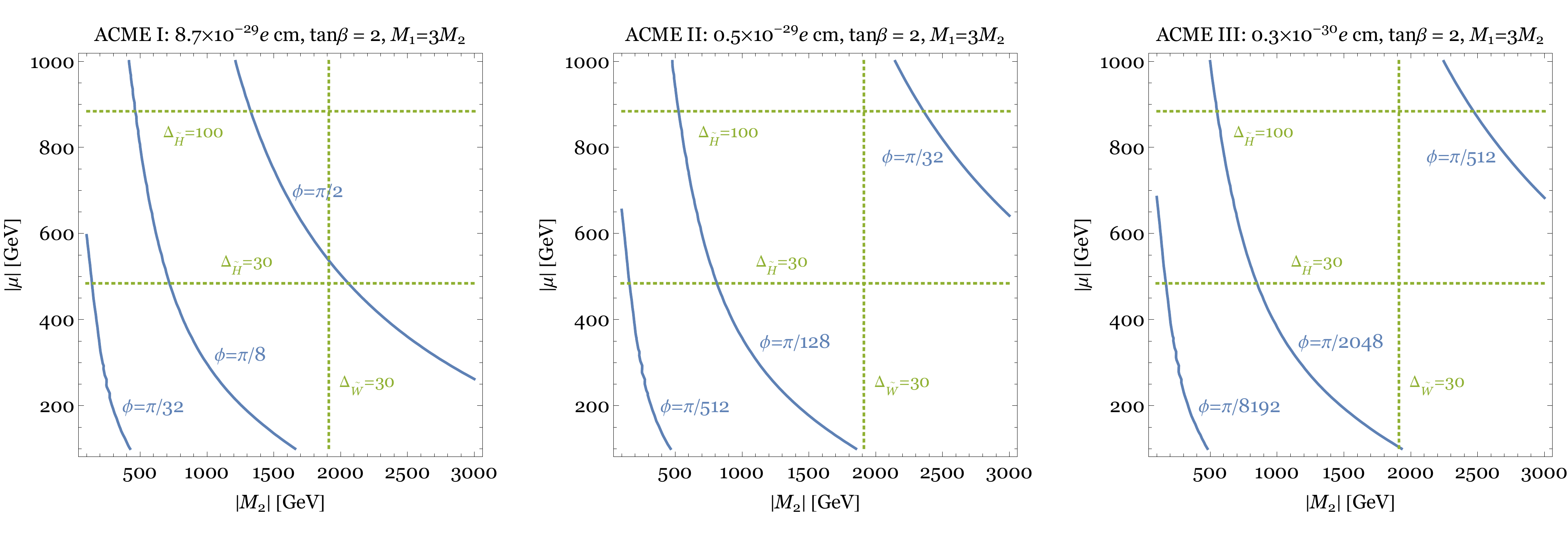}
\end{center}
\caption{A closer look at the low mass region: constraints on the chargino phase $\phi = \arg(M_2 \mu b_\mu^*)$ as a function of the chargino mass parameters, with the current ACME electron constraint (left) and two future projections. Regions below and to the left of the blue curves are excluded. Again, we have superimposed dotted green contours of electroweak fine tuning associated with large higgsino or wino masses.}
\label{fig:acmechargino}
\end{figure}%

Since naturalness and dark matter both prefer that we have charginos in the range below a few TeV, we will now zoom in on the lower-left corner of our plot and examine the future ACME reach in more detail. In particular, because ACME I has already ruled out a large part of the parameter space with large CP-violating phases, it is useful to phrase the constraints in terms of the {\em largest allowed phase} for a given point in parameter space. We show this in Figure \ref{fig:acmechargino} for the case of $\tan \beta = 2$. The light Higgs contribution we plot scales as $\sin(2\beta)$, so the case $\tan \beta = 10$ would be roughly a factor of 4 more pessimistic than $\tan \beta = 2$, essentially shifting each label to the curve to its left. The left-hand panel shows the current constraint: all of the parameter space with electroweak tuning a factor of 30 or less is already at least mildly constrained at small $\tan \beta$. ACME II will bring the constraint on the phase $\arg(M_2 \mu b_\mu^*)$ to percent level in this region of low electroweak tuning, and the next generation will probe phases at the $10^{-4}$ to $10^{-3}$ level. Of course, such a small phase does not necessarily require {\em tuning} in the sense that a large $\mu$ requires tuning for electroweak symmetry breaking. Nonetheless, the requirement of such small phases will be a strong constraint on possible mechanisms of supersymmetry breaking. Alternatively, there is great discovery potential in this low-mass, small-phase region of parameter space. The improvement in mass reach for a fixed phase scales as the square root of the improvement in measurement of $d_e$, but the improvement in phase for fixed mass scales linearly.

\subsection{EDM constraints on charginos in the 2HDM context}

The EDM constraint on pure charginos is very important in the context of split supersymmetry, motivated by gauge coupling unification and dark matter independent of naturalness. But this is not the full story. In versions of mini-split supersymmetry with scalars below the PeV scale, one-loop EDMs arising from phases in sfermion mass matrices are potentially as important as the chargino effects \cite{McKeen:2013dma,Altmannshofer:2013lfa}. On the other hand, in theories of {\em natural} SUSY the chargino contributions can be even larger than what we have computed so far due to the effects of the heavy Higgs fields, which have $\tan \beta$-enhanced couplings to the electron. This point has previously been emphasized in \cite{Li:2008kz}. As we have discussed in \S\ref{naturalSUSYtuning}, the heavy Higgs fields of the MSSM play an underappreciated role in natural SUSY, leading to tree-level fine tuning if $m_A$ is large unless $\tan \beta$ is correspondingly large. However, large $\tan \beta$ can be constrained through predictions of enhanced new physics effects in processes like $b \to s\gamma$. Here we will explore how EDM constraints behave as a function of $m_A$ and $\tan \beta$.

\begin{figure}[!h]\begin{center}
\includegraphics[width=1.0\textwidth]{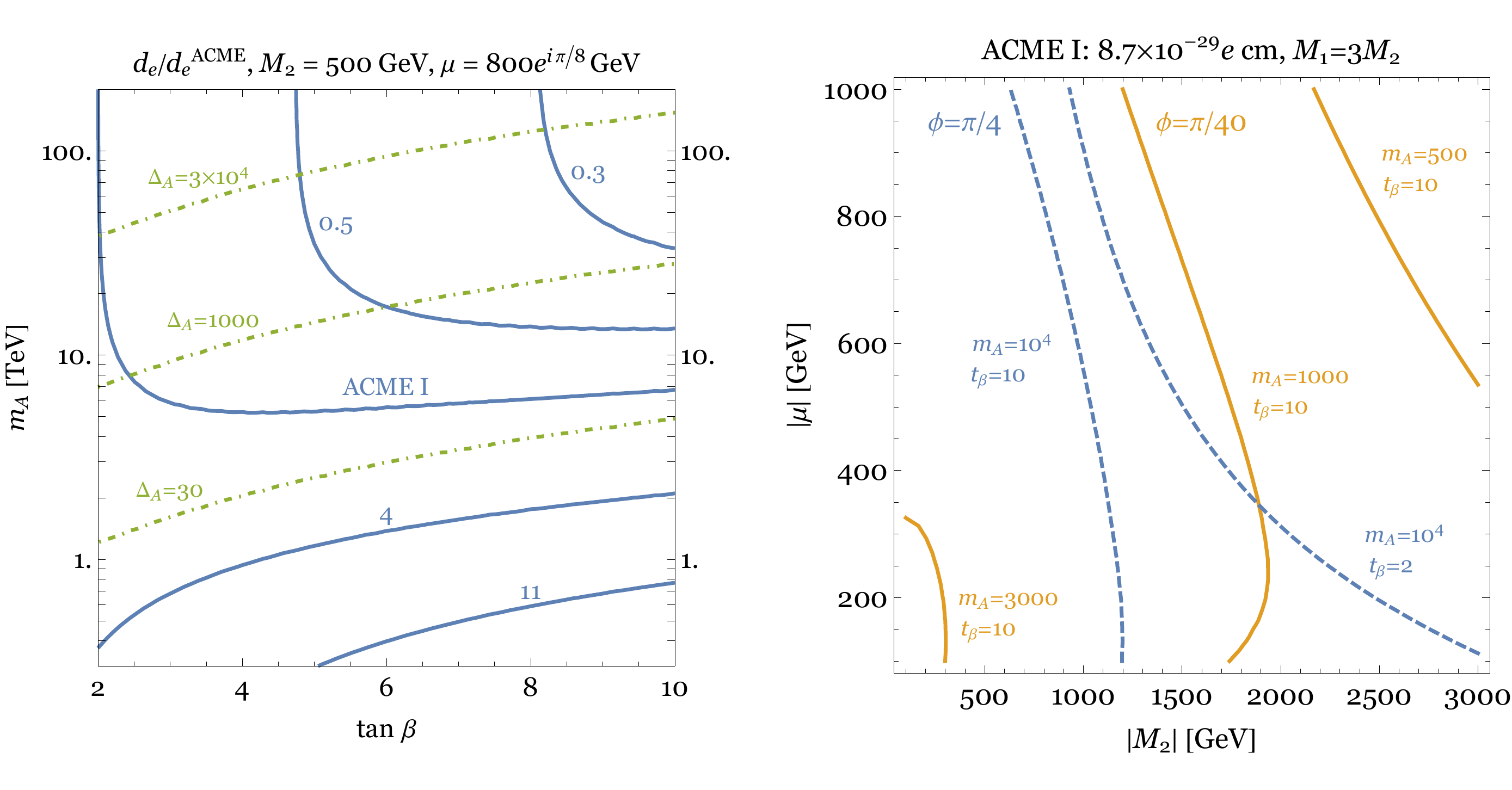}
\end{center}
\caption{Effect of including the full set of Higgs fields in two-loop chargino EDM contributions. {\em Left:} Blue contours show the ratio of the electron EDM to the value $8.7 \times 10^{-29} e\, {\rm cm}$ marginally excluded by ACME. We have fixed $M_2$ and $\mu$ to values for which the electron EDM saturates the current ACME constraint at $\tan \beta = 2$ and $m_A \to \infty$. When $m_A$ is in the natural region, the electron EDM is significantly larger than when the heavy Higgses are decoupled. Green dotted lines indicate tree level tuning associated with the heavy Higgs fields. {\em Right:} Here we show the current exclusion contours in the $(M_2, \mu)$ plane for several different choices of phase ($\pi/4$ for blue dashed curves, $\pi/40$ for orange solid curves), $m_A$ (in GeV), and $\tan \beta$. Regions to the left and below the curves are excluded.}
\label{fig:de2hdmplots}
\end{figure}%

The left-hand panel of Figure \ref{fig:de2hdmplots} shows, for a fixed choice of $M_2$ and $\mu$, how the electron EDM varies with $m_A$ and $\tan \beta$. The parameters are chosen so that this point is marginally excluded by the current ACME measurement in the $m_A \to \infty$ decoupling limit when $\tan \beta = 2$. We see that even at $\tan \beta = 10$, when the EDM in the $m_A \to \infty$ limit is safe by about a factor of 4, this choice of $M_2$ and $\mu$ is excluded when the heavy Higgses are lighter than about 8 TeV. Furthermore, notice that taking the chargino-only limit that we plotted in Fig.~\ref{fig:CharginoFirstLook} is associated with significant tuning cost; keeping $m_A$ light enough to pay only a factor of 30 tree level tuning, for example, increases the EDM by at least a factor of 2 (and much more when $\tan \beta$ is large). 

The right-hand panel of Figure \ref{fig:de2hdmplots} shows {\em current} exclusion contours in the $(M_2, \mu)$ plane as the phase $\arg(M_2 \mu b_\mu^*)$, $m_A$, and $\tan\beta$ are varied. The case $m_A = 10~{\rm TeV}, \tan \beta = 2$ is comparable to the result shown above in Fig.~\ref{fig:acmechargino}; for larger $\tan \beta$ the effects of the extra Higgses are very important, and for smaller values of $m_A$ at or below the TeV scale the current measurement probes phases an order of magnitude smaller than are accessible in a theory with only a single light Higgs boson.

\subsection{Comparison to collider and dark matter search constraints}

\begin{figure}[!h]\begin{center}
\includegraphics[clip,width=0.45\textwidth]{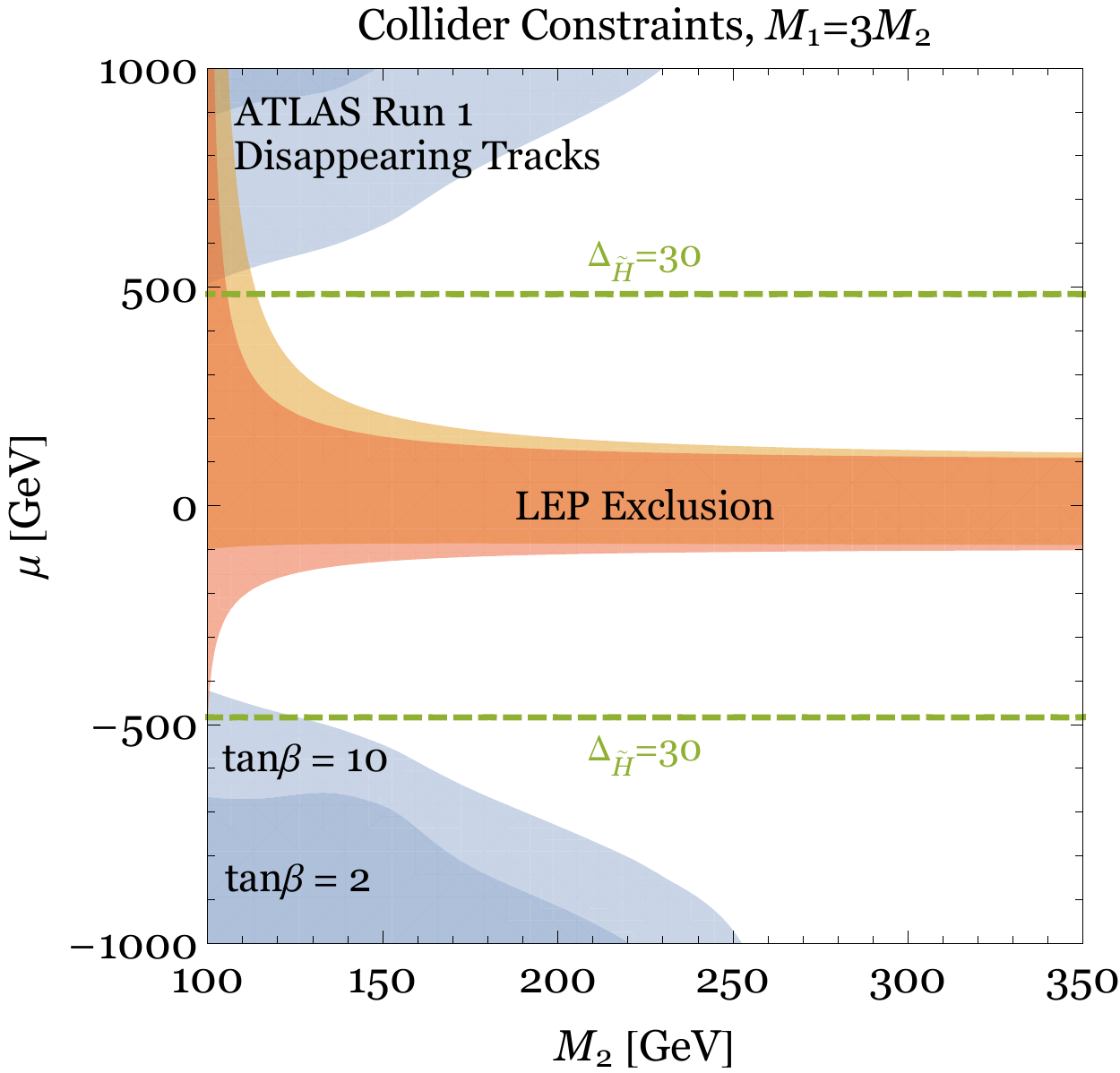}
\end{center}
\caption{Constraints on chargino parameter space from collider searches. We take $M_1 = 3 M_2$, as in anomaly mediation, so that the LSP is either mostly wino-like or mostly higgsino-like. Highly wino-like LSPs are constrained by the ATLAS and CMS searches for disappearing tracks (left-hand corners of the plot), while points with charginos lighter than about 100 GeV are constrained by LEP (band in the center of the plot). The dashed green line is the higgsino mass at which the tree level tuning for electroweak symmetry breaking is a factor of 30.}
\label{fig:charginocollider}
\end{figure}

Collider searches have so far covered only a small slice of chargino parameter space. LEP ruled out theories with light charginos; roughly speaking, we can view the LEP bound as stating that the lightest chargino should have mass above 100 GeV (though the precise bound depends on further details of the spectrum and decay chain). One of the more successful LHC searches is for the ``disappearing track'' signature of a wino LSP \cite{Aad:2013yna,CMS:2014gxa}, which constrains $|M_2| \gtrsim 280~{\rm GeV}$ in the limit $|M_1|, |\mu| \to \infty$. The key point is that the charged wino is very nearly degenerate with the neutral wino and hence has a somewhat long lifetime, leaving a track that vanishes when it decays and the charge is transferred to a soft pion or electron \cite{Chen:1995yu,Chen:1999yf,Feng:1999fu}. To reinterpret the LHC Run 1 bound as a constraint on chargino parameter space, we use the lifetime computation and two-loop radiative wino mass splitting from \cite{Ibe:2012sx}. However, note that their expression for the approximate tree-level splitting between wino mass eigenstates substantially underestimates the mixing induced when the higgsinos and binos are both below the TeV scale; we have instead computed the tree level splitting by directly diagonalizing the mass matrices. In much of the parameter space the tree level mixing-induced wino mass splitting is large enough that the decay is relatively prompt and the particle escapes the LHC disappearing track constraints. Interestingly, we find that the precise shape of the excluded region is highly sensitive to the relative phase of $M_1$ and $\mu$, although we only show results for real $M_1$ and $\mu$ because the EDM constraint is far stronger than the LHC bound in the presence of generic CP-violating phases. We show the constraints from LEP's chargino exclusion and from the ATLAS Run 1 disappearing track search (which gives a very slightly stronger bound than the corresponding CMS search) in Fig.~\ref{fig:charginocollider}. Additional searches that constrain electroweakinos, such as searches for multi-lepton final states, may place weak additional bounds but have not yet dramatically exceeded the reach of LEP, except in cases with light sleptons or in certain cascades with a light bino. Fully interpreting the collider constraints goes beyond the scope of this paper (see, for instance, \cite{Martin:2014qra} or the very recent \cite{Han:2016qtc}, which also looks at the disappearing track constraint across parameter space).

Mixed gaugino--higgsino dark matter is significantly constrained by direct detection results, of which the latest and most powerful are from LUX \cite{Akerib:2016vxi}, PandaX-II \cite{Tan:2016zwf}, and Xenon1T \cite{Aprile:2017iyp}. Mostly-wino and mostly-higgsino dark matter are strongly constrained by searches for gamma rays from their annihilation, either directly (line searches) or via electroweak gauge bosons. Significant constraints come from Fermi-LAT dwarf galaxy searches \cite{Ackermann:2015zua,Abdallah:2016ygi}. Our choice of $|\mu| = 350~{\rm GeV}$ for plots in \S\ref{numerical} is in part based on having the possibility of higgsino dark matter heavy enough to have escaped Fermi-LAT bounds. A full update of the current constraints on electroweakino dark matter, or review of the literature, is beyond the scope of this paper, but we note that nearly pure higgsino-like dark matter with mass near a TeV is significantly outside the reach of any current or near-future dark matter detection or collider experiment. EDMs, especially the future updates from ACME, provide a uniquely powerful window on this region of parameter space.

Recent work drawing similar conclusions about the interplay between dark matter direct and indirect detection, EDM measurements, and collider searches may be found in \cite{Fan:2013qn,Nagata:2014wma}. Vectorlike electroweak particles, with chargino-like quantum numbers but more general couplings, may play a role in theories more general than SUSY. Adding such particles to a supersymmetric theory could help explain the Higgs mass and provide new dark matter candidates \cite{Basirnia:2016szw}. Because such a scenario involves larger Yukawa couplings than the higgsinos have, the EDM contributions would be even larger than the chargino contributions in the MSSM.

%
%

\section{Higgs interactions beyond the MSSM} \label{sec:BMSSM}

In this section, we discuss the case that the Higgs mass is lifted by new tree-level interactions beyond the MSSM that can involve CP-violating phases. First, we derive the mass spectrum of this extension of the MSSM. Then, we present the bounds on EDMs in this scenario arising from the new phases.

\subsection{Spectrum and couplings in the BMSSM}\label{sec:BMSSMreview}

Recall from \S\ref{naturalSUSYhiggs} that the BMSSM involves two new operators,
\begin{align}
W & \supset \frac{\lambda}{M} (H_u \cdot H_d)^2, \nonumber \\
{\cal L}_{\rm soft} & \supset \frac{\lambda m_{\rm SUSY}}{M} (H_u \cdot H_d)^2. \nonumber
\end{align}
We now summarize the mass spectrum of the Higgs sector, following the discussion
of refs.~\cite{Dine:2007xi,Blum:2010by,Altmannshofer:2011rm}.
The Higgs scalar fields can be parametrized as
\begin{equation}
\begin{split}
\\[-2.5ex]
H_u = 
\begin{pmatrix}
H_u^+ \\[1ex]
\frac{1}{\sqrt{2}} \left( v_u + h_u + i a_u \right) 
\end{pmatrix}  \, ,
\qquad H_d = 
e^{\iu \theta} \begin{pmatrix}
\frac{1}{\sqrt{2}} \left( v_d + h_d + i a_d \right)  \\[1ex]
H_d^-
\end{pmatrix}  \, .  \\[1ex]
\end{split}
\end{equation}
Here, $v_{u,d}$ are the vacuum expectation values (VEVs) 
satisfying $v^2 = v_u^2 + v_d^2 \simeq (246 \, \rm GeV)^2$ and $\tan \beta = v_u/ v_d$.
We have put a physical phase $\theta$ to $H_d$ by using a $U(1)_Y$ transformation.
For later purposes, we define
\begin{equation}
\begin{split}
\\[-2.5ex]
&\epsilon_{1r} \equiv |\epsilon_1| \cos (\phi_1 + \theta) \, , \qquad
\epsilon_{1i} \equiv |\epsilon_1| \sin (\phi_1 + \theta) \, ,  \\[2ex]
&\epsilon_{2r} \equiv |\epsilon_2| \cos (\phi_2 + 2\theta) \, , \qquad
\epsilon_{2i} \equiv |\epsilon_2| \sin (\phi_2 + 2\theta) \, .  \\[1ex]
\end{split}
\end{equation}
Without CP violation in the Higgs sector, the mass eigenstates are divided into the CP-even and odd parts,
\begin{equation}
\begin{split}
\\[-2.5ex]
\begin{pmatrix}
h \\[1ex]
H
\end{pmatrix} = 
\begin{pmatrix}
c_\alpha & - s_\alpha \\[1ex]
s_\alpha & c_\alpha
\end{pmatrix}
\begin{pmatrix}
h_u \\[1ex]
h_d
\end{pmatrix} \, ,
\qquad \begin{pmatrix}
G^0 \\[1ex]
A
\end{pmatrix} = 
\begin{pmatrix}
s_\beta & - c_\beta \\[1ex]
c_\beta & s_\beta
\end{pmatrix}
\begin{pmatrix}
a_u \\[1ex]
a_d
\end{pmatrix} \, ,  \\[1ex]
\end{split}
\end{equation}
where $h$, $H$ are the lighter and heavier CP-even mass eigenstates,
$A$ is the CP-odd eigenstate and $G^0$ is the would-be Nambu-Goldstone mode.
We write $\sin \beta$ as $s_\beta$ for simplicity of expressions.
The mixing angle $\alpha$ is given by
\begin{equation}
\begin{split}
\\[-2.5ex]
\sin 2 \alpha = - \frac{(m_A^2 + m_Z^2) \sin 2 \beta + 4 \epsilon_{1r} v^2}{m_H^2 - m_h^2 }  \, ,   \qquad
\cos 2 \alpha = - \frac{m_A^2 - m_Z^2 + 2 \epsilon_{2r} v^2}{m_H^2 - m_h^2} \cos 2 \beta \, .  \\[1ex]
\end{split}
\end{equation}
Here, $m_Z$ is the $Z$ boson mass, $m_A^2$ is the eigenvalue corresponding to the state $A$ and
$m_h^2$, $m_H^2$ are the lighter and heavier eigenvalues of the following $2 \times 2$ matrix,
\begin{equation}
\begin{split}
\\[-2.5ex]
\mathcal{M}_S^2 = &\,\,m_A^2
\begin{pmatrix}
c_\beta^2 & - s_\beta c_\beta \\[1ex]
- s_\beta c_\beta & s_\beta^2
\end{pmatrix}
+ m_Z^2
\begin{pmatrix}
s_\beta^2 & - s_\beta c_\beta \\[1ex]
- s_\beta c_\beta & c_\beta^2
\end{pmatrix} \\[1.5ex]
&\quad\qquad- 2 v^2 \epsilon_{1r}
\begin{pmatrix}
s_{2\beta} & 1 \\[1ex]
1 & s_{2\beta}
\end{pmatrix}
+ 2 v^2 \epsilon_{2r}
\begin{pmatrix}
c_\beta^2 & 0 \\[1ex]
0 & s_\beta^2
\end{pmatrix} \, . \\[1ex]
\end{split}
\end{equation}
In the case with a moderate $\tan \beta$ and $m_A^2 \gg m_Z^2$, we have $\alpha \approx \beta - \frac{\pi}{2}$.
We assume this case below.

However, CP violation enables the three physical Higgs bosons $h$, $H$, $A$ to mix with each other
while the would-be Nambu-Goldstone mode $G^0$ remains unchanged.
The mass-squared matrix is then given by
\begin{equation}
\begin{split}
\\[-2.5ex]
\mathcal{M}_H^2 = 
\begin{pmatrix}
m_h^2 & 0 & m_{hA}^2 \\[1ex]
0 & m_H^2 & m_{HA}^2 \\[1ex]
m_{hA}^2 & m_{HA}^2 & m_A^2
\end{pmatrix}  \, , \label{masssquaredmatrix} \\[1ex]
\end{split}
\end{equation}
where the off-diagonal elements are
\begin{equation}
\begin{split}
\\[-2.5ex]
&m_{hA}^2 = 2 v^2 s_{\beta - \alpha} \epsilon_{1i} - v^2 c_{\beta + \alpha} \epsilon_{2i}  \, ,  \\[1.5ex]
&m_{HA}^2 = 2 v^2 c_{\beta - \alpha} \epsilon_{1i} - v^2 s_{\beta + \alpha} \epsilon_{2i} \, , \\[1ex]
\end{split}
\end{equation}
at the order of $\epsilon_{1,2}$.
The mass-squared matrix \eqref{masssquaredmatrix} can be diagonalized by an orthogonal matrix $O$,
\begin{equation}
\begin{split}
\\[-2.5ex]
O^T \mathcal{M}_H^2 O = {\rm diag} \left(m_{H_1}^2 , m_{H_2}^2 , m_{H_3}^2 \right) \, . \\[1ex]
\end{split}
\end{equation}
Here, $m_{H_1}^2 \leq m_{H_2}^2 \leq m_{H_3}^2$ are the three eigenvalues 
corresponding to the eigenstates $H_1, H_2, H_3$.
At the order of $\epsilon_{1,2}$, the charged Higgs boson mass matrix is diagonalized as
\begin{equation}
\begin{split}
\\[-2.5ex]
\begin{pmatrix}
H_u^+ \\[1ex]
{H_d^-}^\ast
\end{pmatrix} = 
\begin{pmatrix}
s_\beta + \iu\, c_\beta \eta & c_\beta + \iu\, s_\beta \eta \\[1ex]
-c_\beta + \iu\, s_\beta \eta & s_\beta - \iu\, c_\beta \eta
\end{pmatrix}
\begin{pmatrix}
G^+ \\[1ex]
H^+
\end{pmatrix} \, , \\[1ex]
\end{split}
\end{equation}
where $H^+$ is the physical mass eigenstate, $G^+$ is the would-be Nambu-Goldstone mode
and $\eta$ is given by
\begin{equation}
\begin{split}
\eta = \frac{v^2 \left(\epsilon_{1i} - s_\beta c_\beta \, \epsilon_{2i} \right)}{m_A^2} \, .   \\[1ex]
\end{split}
\end{equation}
The charged Higgs boson mass is approximately obtained as
\begin{equation}
\begin{split}
\\[-2.5ex]
m_{H^\pm}^2 \simeq m_A^2 + m_W^2 + \epsilon_{2r} v^2 \, ,   \\[1ex]
\end{split}
\end{equation}
where $m_W$ is the $W$ boson mass.
At the order of $\epsilon_{1,2}$, the minimum condition $\frac{\partial V_{\rm Higgs}}{\partial \theta} = 0$ gives
\begin{equation}
\begin{split}
\\[-2.5ex]
\tan \theta = \frac{v^2 \left( \epsilon_{2i} s_{2 \beta} - 2 \epsilon_{1i} \right)}
{s_{2 \beta} \left( m_{H^\pm}^2 - m_W^2 \right) + v^2 \left( \epsilon_{2r} s_{2 \beta} - 2 \epsilon_{1r} \right) } \, .   \\[1ex]
\end{split}
\end{equation}
For a small $\tan \beta$, the phase $\theta$ is $\mathcal{O}(\epsilon_{1,2})$.  

Let us next consider the chargino and neutralino mass spectra.
In the presence of the new operator of \eqref{BMSSMsuperpotential}, 
we have new higgsino interactions
\cite{Dine:2007xi},
\begin{equation}
\begin{split}
\\[-2.5ex]
\mathcal{L}_{\rm Higgs} \supset - \frac{\epsilon_1}{\mu^\ast} \biggl[ \,
&2 ( H_u \cdot H_d ) ( \widetilde{H}_u  \cdot \widetilde{H}_d ) + 2 ( \widetilde{H}_u  \cdot H_d ) ( H_u  \cdot \widetilde{H}_d ) \\[1.5ex]
&+ ( H_u  \cdot  \widetilde{H}_d ) ( H_u   \cdot \widetilde{H}_d ) +  ( \widetilde{H}_u  \cdot H_d ) ( \widetilde{H}_u  \cdot H_d ) \biggr] + \rm h.c.
\label{higgsinoint} \, , \\[1ex]
\end{split}
\end{equation}
where $\widetilde{H}_{u,d}$ denote the fermion components of the Higgs chiral superfields.
These terms provide not only additional contributions to the chargino and neutralino masses
but also new Higgs-chargino/neutralino interactions which contribute to the two-loop EDMs induced by chargino loops.
The chargino mass matrix in the basis of $(\widetilde{W}^+, \widetilde{H}_u^+, \widetilde{W}^-, \widetilde{H}_d^-)$ is given by
\begin{equation}
\begin{split}
\\[-2.5ex]
\mathcal{M}_{\tilde{C}} = \begin{pmatrix}
0 & X^T \\[1ex]
X & 0
\end{pmatrix} \, , \qquad
X = \begin{pmatrix}
M_{\tilde{W}} & \frac{g}{\sqrt{2}} v s_\beta \\[1ex]
\frac{g}{\sqrt{2}} v c_\beta e^{- \iu \theta}  & \mu - \frac{\epsilon_1 v^2}{\mu^\ast} s_\beta c_\beta e^{\iu \theta}
\end{pmatrix} \, . \\[1ex]
\end{split}
\end{equation}
Here, $M_{\tilde{W}}$ is the Wino mass and $g$ is the $SU(2)_L$ gauge coupling.
The matrix $X$ can be diagonalized by a singular value decomposition,
\begin{equation}
\begin{split}
\\[-2.5ex]
{C^R}^\dagger X C^L = {\rm diag} \left( m_{\tilde{\chi}_1} , m_{\tilde{\chi}_2} \right) \, , \\[1ex]
\end{split}
\end{equation}
where $C^{L,R}$ are unitary matrices and $m_{\tilde{\chi}_1} \leq m_{\tilde{\chi}_2}$.
The neutralino mass matrix in the basis of $(\widetilde{B}, \widetilde{W}^0, \widetilde{H}_d^0, \widetilde{H}_u^0)$ is given by
\begin{equation}
\begin{split}
\\[-2.5ex]
\mathcal{M}_{\tilde{N}} = \begin{pmatrix}
M_{\tilde{B}} & 0 & - \frac{g'}{2} v c_\beta e^{- \iu \theta} & \frac{g'}{2} v s_\beta  \\[1ex]
0 & M_{\tilde{W}} & \frac{g}{2} v c_\beta e^{- \iu \theta} & - \frac{g}{2} v s_\beta \\[1ex]
 - \frac{g'}{2} v c_\beta e^{- \iu \theta} & \frac{g}{2} v c_\beta e^{- \iu \theta} & \frac{\epsilon_1}{\mu^\ast} v^2 s_\beta^2 
& -\mu + 2 \frac{\epsilon_1 v^2}{\mu^\ast} s_\beta c_\beta e^{\iu \theta} \\[1ex]
\frac{g'}{2} v s_\beta & - \frac{g}{2} v s_\beta & -\mu + 2 \frac{\epsilon_1 v^2}{\mu^\ast} s_\beta c_\beta e^{\iu \theta} &
\frac{\epsilon_1 v^2}{\mu^\ast} c^2_\beta e^{2 \iu \theta}
\end{pmatrix} \, , \\[1ex]
\end{split}
\end{equation}
where $M_{\tilde{B}}$ is the Bino mass and $g'$ is the $U(1)_Y$ gauge coupling.
This symmetric complex matrix can be diagonalized by a unitary matrix $N$,
\begin{equation}
\begin{split}
\\[-2.5ex]
N^T \mathcal{M}_{\tilde{N}} N = {\rm diag} \left(m_{\tilde{\chi}^0_1} , m_{\tilde{\chi}^0_2} ,
m_{\tilde{\chi}^0_3}, m_{\tilde{\chi}^0_4} \right) \, . \\[1ex]
\end{split}
\end{equation}
Here, $m_{\tilde{\chi}^0_1} \leq m_{\tilde{\chi}^0_2} \leq m_{\tilde{\chi}^0_3} \leq m_{\tilde{\chi}^0_4}$
are four real, positive eigenvalues.

Finally, we present the masses and physical eigenstates of the third generation squarks. 
In the MSSM with the new superpotential interaction \eqref{BMSSMsuperpotential}, the stop mass-squared matrix is given by
\begin{equation}
\begin{split}
\\[-2.5ex]
&\mathcal{M}^2_{\tilde{t}} =  \\[1ex]
&\begin{pmatrix}
m_{Q_3}^2 + m_t^2 + \Delta_{\tilde{u}_L} &&
\!\!\!\!\!\! -m_t \left| \mu \cot \beta \, e^{\iu \theta} - A_t^\ast - \frac{\epsilon_1}{\mu^\ast} v^2 c^2 _\beta \, e^{2 \iu\theta} \right|
e^{-\iu \delta_t }\\[1ex]
-m_t \left| \mu^\ast \cot \beta \, e^{-\iu \theta} - A_t - \frac{\epsilon_1^\ast}{\mu} v^2 c^2_\beta \, e^{-2 \iu \theta} \right|
e^{\iu \delta_t }
 && m_{u_3}^2 + m_t^2 + \Delta_{\tilde{u}_R} 
\end{pmatrix}
\, , \label{stopmatrix} \\[1ex]
\end{split}
\end{equation}
where
$\Delta_{\tilde{u}_L} = m_Z^2 \cos 2\beta \left( \frac{1}{2} - \frac{2}{3} \sin^2 \theta_W \right)$,
$\Delta_{\tilde{u}_R} = m_Z^2 \cos 2\beta \left( \frac{2}{3} \sin^2 \theta_W \right)$ and $m_t$ is the top quark mass.
The phase $\delta_t = \arg(\mu^\ast \cot \beta \, e^{-\iu \theta}
- A_t - \frac{\epsilon_1^\ast}{\mu} v^2 c^2_\beta \, e^{-2\iu \theta} )$ in the mass-squared matrix can be absorbed by redefinition of the right-handed stop, $\tilde{t}_R \rightarrow e^{\iu \delta_t} \, \tilde{t}_R $.
Diagonalizing this matrix, we write smaller and larger eigenvalues as
$m_{\tilde{t}_1}^2$ and $m_{\tilde{t}_2}^2$ respectively.
These correspond to physical stop masses.
The eigenstates are given by
\begin{equation}
\begin{split}
\\[-2.5ex]
\tilde{t}_1 = \tilde{t}_L \cos \theta_t  - \tilde{t}_R \sin \theta_t \, , \qquad
\tilde{t}_2 = \tilde{t}_L \sin \theta_t  + \tilde{t}_R \cos \theta_t \, . \\[1ex]
\end{split}
\end{equation}
Here, the mixing angle $\theta_t$ is
\begin{equation}
\begin{split}
\\[-2.5ex]
\tan (2 \theta_t) &= - \frac{2 m_t \left| \mu \cot \beta \, e^{\iu \theta} - A_t^\ast - \frac{\epsilon_1}{\mu^\ast} v^2 c^2 _\beta \, e^{2\iu \theta} \right|}{m_{Q_3}^2 + \Delta_{\tilde{u}_L} - m_{u_3}^2 - \Delta_{\tilde{u}_R}} \label{mixingangle} \, . \\[1ex]
\end{split}
\end{equation}
Since the right-handed sbottom does not exist in the low-energy effective theory of natural SUSY,
the lighter eigenstate $\tilde{b}_1$ of the sbottom mass-squared matrix is simply given by the left-handed sbottom $\tilde{b}_L$.

\subsection{EDMs in the BMSSM} \label{sec:BMSSMEDMs}

\begin{figure}[!h]
\hspace{0cm}
 \begin{minipage}{0.33\hsize}
  \begin{center}
   \includegraphics[clip, width=7cm]{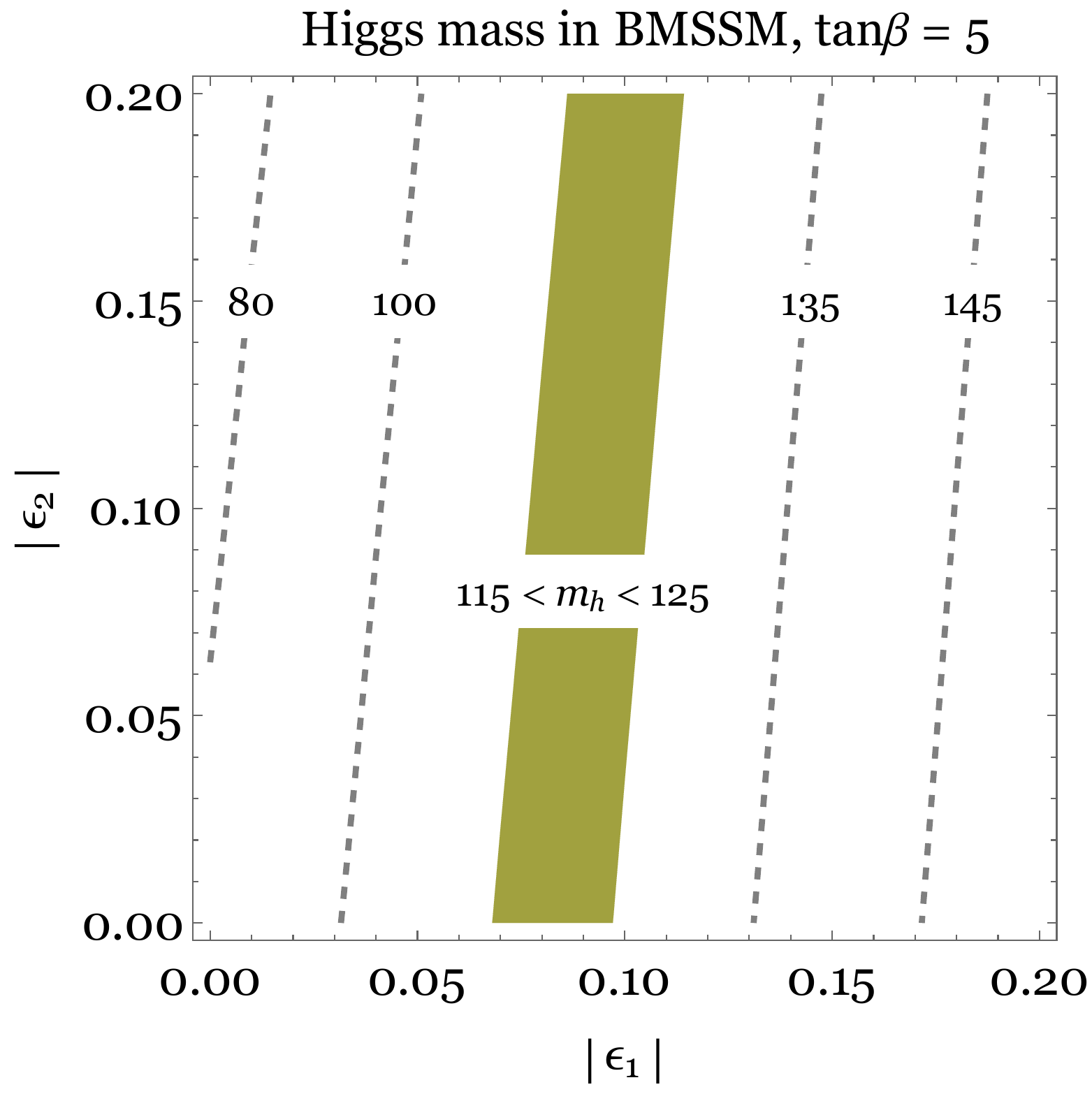}
  \end{center}
 \end{minipage}
\hspace{2.5cm}
\begin{minipage}{0.33\hsize}
  \begin{center}
   \includegraphics[clip, width=7cm]{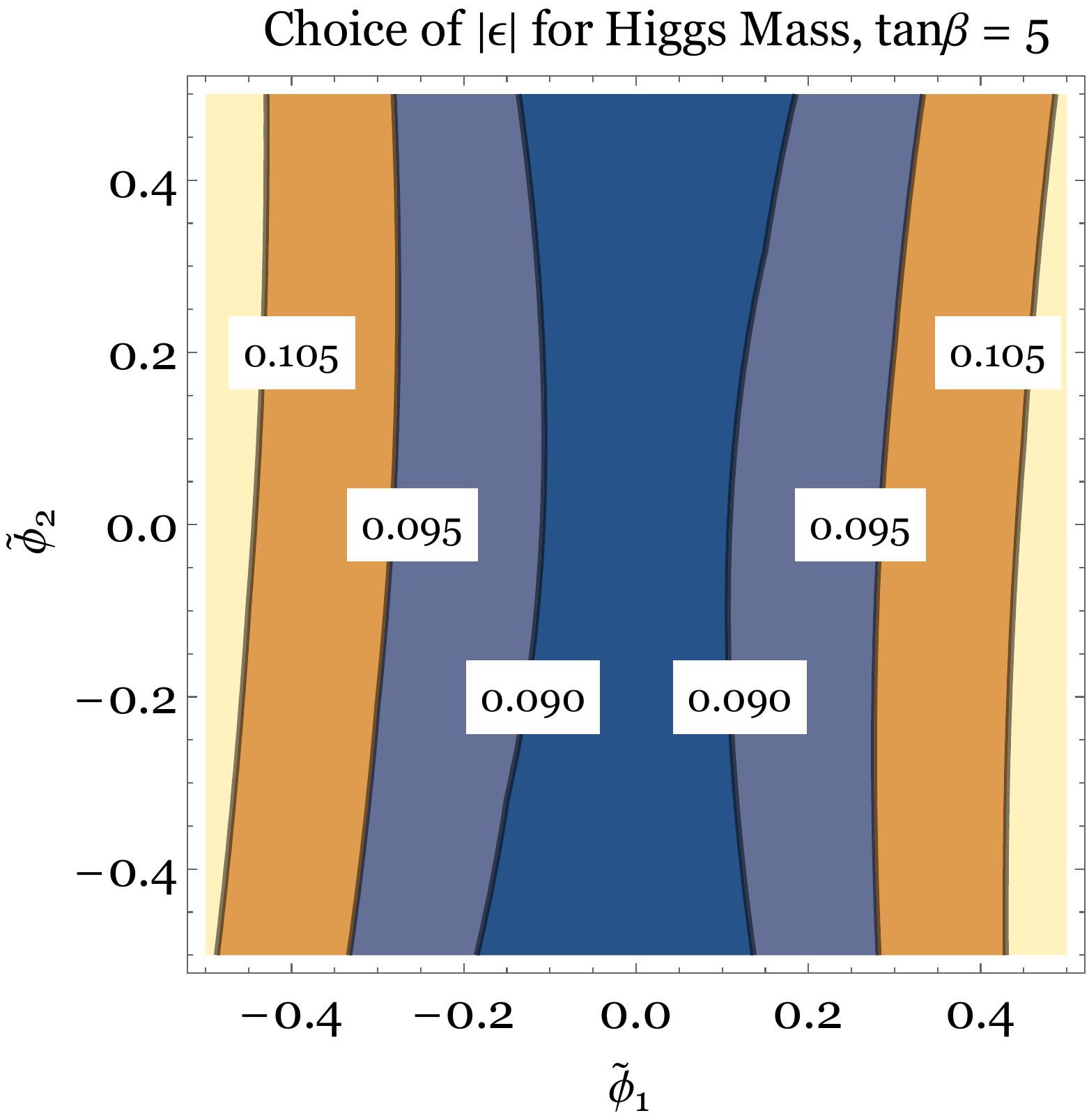}
  \end{center}
 \end{minipage}
\caption{The lightest neutral Higgs boson mass (in unit of GeV) as a function of $|\epsilon_{1,2}|$ (left) and choice of $|\epsilon|$
to realize the correct Higgs mass (right) in the BMSSM scenario. We take $\tan \beta =5$ and $m_A = 400 \, \rm GeV$.
For the left panel, we assume $\tilde{\phi}_{1,2} \equiv \pi  - \phi_{1,2} = 0.1$. The dashed lines denote the tree-level Higgs mass.
The green shaded region may give the correct Higgs mass, taking account of radiative corrections from top/stop loops.
For the right panel, we assume $|\epsilon_1| = |\epsilon_2|$ and the tree-level Higgs mass at $120 \, \rm GeV$,
relying on the radiative corrections to explain the rest.}
\label{fig: BMSSMHiggsmass}
\end{figure}%

\begin{figure}[!h]
\hspace{0cm}
 \begin{minipage}{0.33\hsize}
  \begin{center}
   \includegraphics[clip, width=7.5cm]{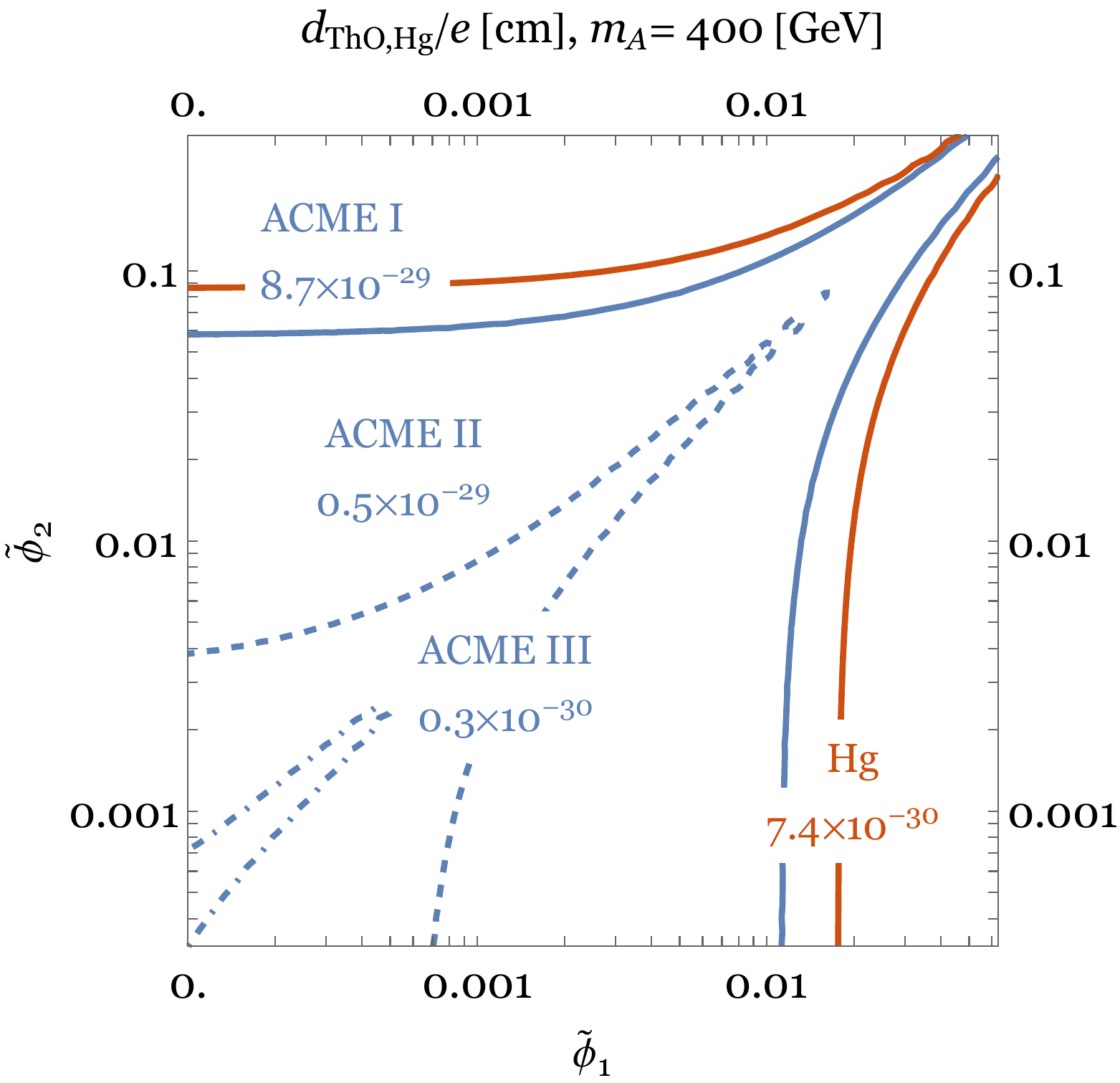}
  \end{center}
 \end{minipage}
\hspace{2.5cm}
\begin{minipage}{0.33\hsize}
  \begin{center}
   \includegraphics[clip, width=7.5cm]{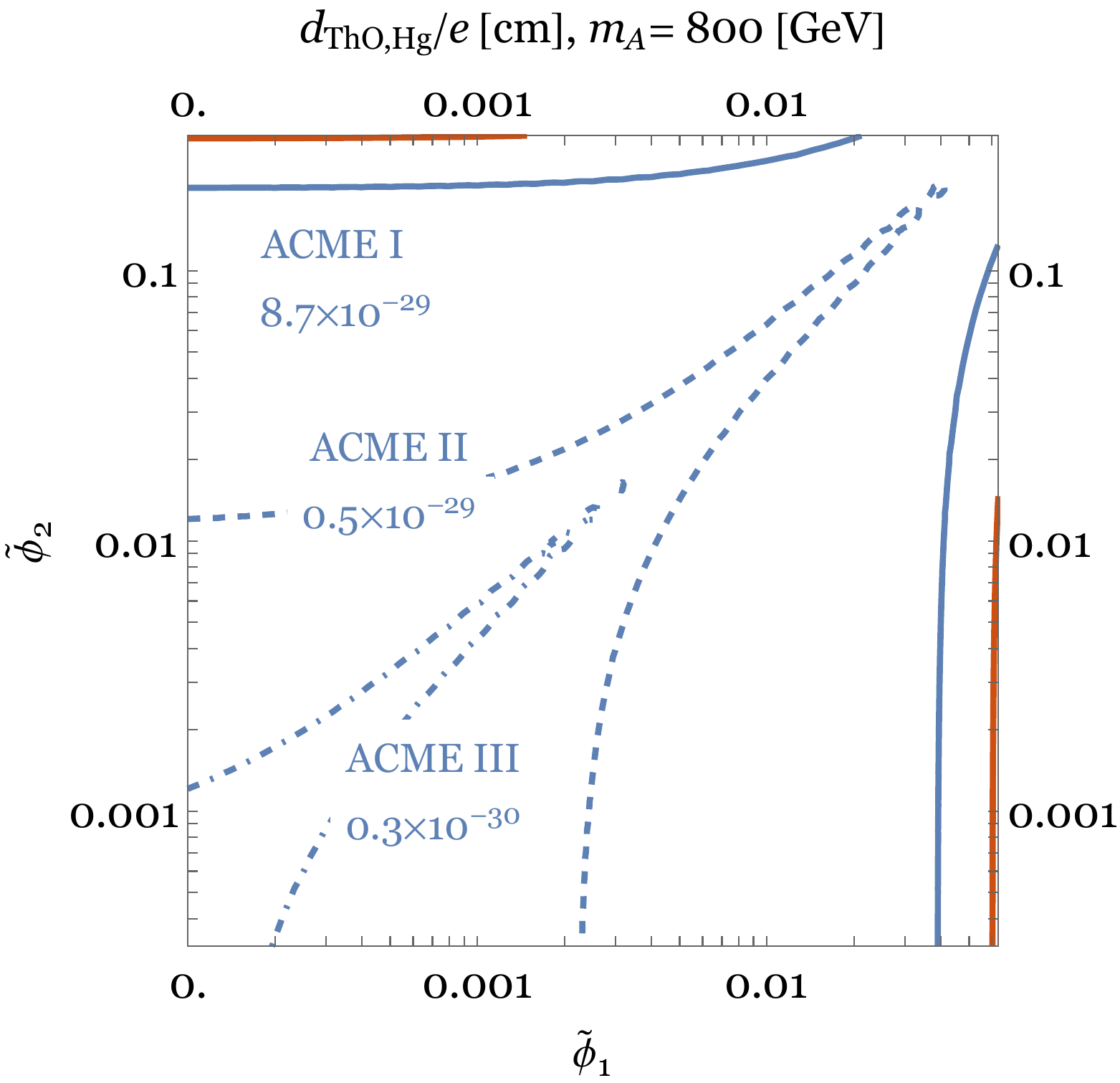}
  \end{center}
 \end{minipage}
\caption{EDM constraints on the CP-violating phases $\tilde{\phi}_{1,2}$ in the BMSSM scenario.
We take $m_A = 400 , 800$ GeV for the left and right panels respectively.
Regions of parameter space to outside the solid blue and red contours are excluded by measurements of ThO and Hg (the case (i)), respectively. The mercury EDM constraint for the case (ii) is weak and not shown in the figures.
The blue dashed and dot-dashed contours (``ACME II'' and ``ACME III'') are future projections.
We have fixed $|\epsilon_{1,2}| = 0.1$, $\tan \beta = 5$ and $\mu = 350$ GeV in these figures.}
\label{fig:BMSSMEDMfigure}
\end{figure}%

We first investigate the parameter space of the BMSSM to realize the correct Higgs boson mass.
Figure~\ref{fig: BMSSMHiggsmass} shows
the lightest neutral Higgs boson mass (in unit of GeV) as a function of $|\epsilon_{1,2}|$ (left) and choice of $|\epsilon|$
to realize the correct Higgs mass (right) in the BMSSM scenario. We take $\tan \beta =5$ and $m_A = 400 \, \rm GeV$.
For the left panel, we assume $\tilde{\phi}_{1,2} \equiv \pi  - \phi_{1,2} = 0.1$. The dashed lines denote the tree-level Higgs mass.
The green shaded region may give the correct Higgs mass, taking account of radiative corrections from top/stop loops.
We can see that the Higgs boson mass is easily lifted up with mild values of $|\epsilon|$.
For the right panel, we assume $|\epsilon_1| = |\epsilon_2|$ and the tree-level Higgs mass at $120 \, \rm GeV$,
relying on the radiative corrections to explain the rest.
The size of contributions from the new Higgs interactions also depends on their CP phases.

Let us now present numerical analyses of the EDMs in the BMSSM scenario.
We assume only nonzero CP phases in the new Higgs interactions and
set all the CP phases of the MSSM to zero to simplify our analyses.
Due to smallness of the masses of the Standard Model particles,
the dominant contributions to the EDMs come from the Barr-Zee diagrams with loops of $W$ bosons and top quarks. We find that the $W$ boson is the dominant contribution to the electron EDM, while top quark loops dominate the quark CEDMs and so are crucial for the mercury EDM. Such top quark loops have been computed in the MSSM in \cite{Pilaftsis:2002fe,Ellis:2008zy} and studied in the BMSSM context in \cite{Blum:2010by,Altmannshofer:2011rm}; they may be easily computed from the general formulas of \S\ref{twoloopfermions}. The $W$ loop contributions are computed with the general formula (\ref{eq:Wtwoloopresult}) following from the results in \cite{Abe:2013qla}, and to the best of our knowledge their importance for the BMSSM has not been previously pointed out in the literature. We have also checked the chargino contributions, using various expressions collected in Appendix \ref{sec:coupling} which are valid in the BMSSM; we have found that they are subdominant to the top and $W$ loops.

Figure~\ref{fig:BMSSMEDMfigure} shows EDM constraints on the CP-violating phases $\tilde{\phi}_{1,2}$ in the BMSSM scenario.
We take $m_A = 400 , 800$ GeV for the left and right panels respectively.
Regions of parameter space to the right of and above the solid blue and red contours are excluded
by measurements of ThO and Hg (the case (i)), respectively.
The mercury EDM constraint for the case (ii) is weak and not shown in the figures.
The blue dashed and dot-dashed contours (``ACME II'' and ``ACME III'') are future projections.
We have fixed $|\epsilon_{1,2}| = 0.1$, $\tan \beta = 5$ and $\mu = 350$ GeV in these figures.
We can see that the current constraints lead to about 10 percent tuning in the CP phases of the new Higgs interactions
and the situation is improved to less than one percent in future projections.

%
%

\section{Conclusions}\label{conclusion}

In this paper we have studied implications of EDM measurements on the parameter space of CP-violating natural SUSY. We have found that significant constraints on order-one CP-violating phases for superpartners near the TeV scale already exist from the measurements of ThO and Hg. Currently the mercury constraints, being sensitive to chromoelectric dipole moments of quarks induced by new physics with QCD charge (like stops), are somewhat stronger or comparable depending on which choice of nuclear physics calculation we follow, but the ThO constraints are crucial for new physics with only electroweak interactions. Moreover, if the ACME collaboration delivers results at the level of their estimated future reach, these will overtake mercury EDMs as stronger constraints even for the case of stops.

We have studied a few different scenarios and sources of CP-violating phases. Bounds on stops are strongest in the MSSM, where large $A$-terms are required to explain the heavy mass of the Higgs boson. However, this scenario is already rather fine-tuned, and we found that if some other new physics lifts the Higgs mass, there are still important constraints even on the smaller $A$-terms induced under RG running by the gluino mass. We have also found that the relative phase of gaugino masses and $\mu$ is strongly constrained by EDMs induced by charginos running in loops. We expect these charginos to be at or below the TeV scale for several reasons, including naturalness, gauge coupling unification, and the possibility that they constitute a fraction of the dark matter in our universe. As a result, these EDM constraints, which come entirely from the electron EDM, are important not only in natural SUSY but also in split SUSY, and they have cousins in any new physics scenario with purely electroweak new matter. Finally, we have found that if we add new superpotential Higgs interactions (and associated soft SUSY-breaking terms) to explain why the Higgs is heavy, the phases of these new interactions can directly induce important EDMs. A relative phase can appear in the VEVs of the up- and down-type Higgs bosons, inducing scalar-pseudoscalar mixing. The associated EDM effects have been considered previously in \cite{Blum:2010by,Altmannshofer:2011rm}. We have found that the most important contributions come from loops of Standard Model particles interacting with the full Higgs sector of the 2HDM. This includes the top quark, which plays the dominant role in CEDMs of quarks, and the $W$ boson, which plays the dominant role in the electron EDM. The latter effect appears to have been omitted in previous studies. As in the other cases, we find that there are already strong constraints on the phases of the new physics in the natural parameter space. 

We have also made some comparisons of EDMs to other effects of new physics. For instance, stop EDMs depend on the same combination of parameters appearing in $b \to s\gamma$, and are a stronger constraint unless the CP-violating phase is quite small. Most of the two-loop EDMs we have studied are also associated with modifications to the rate of $h \to \gamma \gamma$ or $h \to gg$. Dark matter direct detection can depend on some of the same parameters appearing in chargino-induced EDMs, if neutralinos are the dark matter. An interesting consequence of all of this is that if a nonzero EDM is measured in the future, for each hypothesis about its underlying origin, there is usually a clear prediction of a {\em lower bound} on the rate of some other new physics effect. Quantifying such lower bounds would be an important step in clarifying and understanding the physics giving rise to any observed EDM.

It would be timely to revisit which favored models of supersymmetry breaking have a SUSY CP problem, and how completely the problem is solved in the viable models. Although conventional wisdom holds that the CP problem is more readily solved in gauge mediation than in gravity mediation, it has been claimed that certain models of modulus mediation have enough structure to evade the CP problem \cite{Conlon:2007dw}. A recent analysis of one incarnation of modulus mediation argued that the electron EDM should not be larger than about $5 \times 10^{-30}\, e\, {\rm cm}$ \cite{Ellis:2014tea}---precisely the level that ACME II will probe. Meanwhile, gauge mediation, which has traditionally dodged flavor and CP problems with little effort, faces a renewed challenge in that the new physics that either lifts the Higgs mass directly or generates a large $A$-term (e.g.~\cite{Basirnia:2015vga}) must not reintroduce the problem. 

We should not view the SUSY CP problem as simply another engineering hurdle to solve, or module to tack on to a theory. We believe that it is worth asking for any given model: how large are the expected phases, if we don't add extra layers of ingenuity to squash them? Are there subleading effects that might generate phases at the $10^{-2}$ or $10^{-3}$ level---the size of loop factors, say---even in models that solve the CP problem to leading order? For example, in the chargino sector, a null result at ACME II could already imply that $\arg(M_2 \mu) \lesssim 10^{-2}$ over the entire region of interest for {\em either} naturalness or dark matter. This is a constraint that is of great interest for split supersymmetry as well as for natural SUSY. The coming years will bring important new experimental progress, and we should be prepared for a positive answer as well as a negative one.

\section*{Acknowledgments}\label{sec:ackno}

We would especially like to thank John Doyle for asking an appealingly simple question to which we are giving a less than simple answer. We thank John Doyle, Jerry Gabrielse, and Cris Panda for discussions about their exciting progress on ACME. We would also like to thank Teppei Kitahara, David Pinner, and Kohsaku Tobioka for useful discussions. We thank Martin Jung and Maxim Pospelov for correspondence related to uncertainties in the mercury EDM after the first version of our preprint appeared.
YN is supported by a JSPS Fellowship for Research Abroad.
MR is supported in part by the NSF Grant PHY-1415548. MR thanks the Institute for Advanced Study for funding and hospitality while this work was completed.

\footnotesize

\appendix

%
%

\section{Relevant couplings\label{sec:coupling}}

\subsection{Gauge boson and higgs couplings to Standard Model fermions}
\label{gVASPclarification}

To clarify the conventions used in \S\ref{twoloopfermions}, we tabulate the couplings $g^{V,A,S,P}$ for gauge and Higgs bosons to Standard Model fermions. First, we have the case where $V_\mu$ is the photon of QED. In that case we have
\be
A_\mu: \quad -g^V_f \mapsto Q_f e, \quad g^A_f \mapsto 0,
\ee
where $Q_f$ is the charge of the particle and $e > 0$ is the QED coupling constant. Next consider the $Z$ boson:
\be
Z_\mu: \quad g^V_f \mapsto -\frac{g}{2\cos \theta_W} (T_3 - 2 Q_f \sin^2 \theta_W), \quad g^A_f \mapsto \frac{g}{2 \cos \theta_W} T_3.
\ee
Finally we have the $W$ boson, linking any two fermions $f$ and $f'$ that lie in the same doublet:
\be
W_\mu: \quad g^V_{ff'} \mapsto -\frac{g}{2\sqrt{2}}, \quad g^A_{ff'} \mapsto \frac{g}{2\sqrt{2}}.
\ee
Note that because these couplings are real, $g^V_{f'f} = (g^V_{ff'})^* = g^V_{ff'}$.

Next we consider the Higgs bosons of the MSSM. These couple to Standard Model fermions through the Yukawa couplings
\be
{\cal L} \supset -y_e e^c e_L^- H_d^0 + y_e e^c \nu_L H_d^- - y_d d^c d_L H^0_d + y_d d^c u_L H_d^- - y_u u^c u_L H^0_u + y_u u^c d_L H_u^+ + {\rm h.c.}
\ee
in notation where the left-handed Weyl fermion fields are $Q = (u_L \, d_L)^T, u^c, d^c, L = (\nu_L \, e^-_L )^T, e^c$ and the Higgs doublets are $H_u = (H^+_u \, H^0_u)^T, H_d = (H_d^0 \, H_d^-)^T$. In the BMSSM, where the Higgs bosons mix with each other, the physical Higgs boson couplings with a Standard Model fermion $f$ are given by
\begin{equation}
\begin{split}
{\cal L}_{H \bar{f} f} = 
\sum_{i=1}^3 \left( g_{H_i \bar{f}f}^S {\bar f} f + \iu g_{H_i \bar{f}f}^P {\bar f} \gamma_5 f \right) H_i  \, , \label{fermionint} \\[1ex]
\end{split}
\end{equation}
where
\begin{equation}
\begin{split}
\\[-2.5ex]
&g_{H_i \bar{f}f}^S = -\frac{m_f}{v} \frac{1}{\sin \beta} \left( c_\alpha O_{1i} + s_\alpha O_{2i} \right) \, , \quad
g_{H_i \bar{f}f}^P = \frac{m_f}{v} \frac{\cos \beta_0}{\sin \beta} \, O_{3i} \, , \quad \text{for $T^f_3 = +1/2$} \, , \\[2ex]
&g_{H_i \bar{f}f}^S = -\frac{m_f}{v} \frac{1}{\cos \beta} \left( -s_\alpha O_{1i} + c_\alpha O_{2i} \right) \, , \quad
g_{H_i \bar{f}f}^P = \frac{m_f}{v} \frac{\sin \beta_0}{\cos \beta} \, O_{3i} \, , \quad \text{for $T^f_3 = -1/2$} \, . \\[1ex]
\end{split}
\end{equation}
In the MSSM and BMSSM at tree level we have $\beta_\pm = \beta_0 = \beta$, but in general this relationship is corrected.

The charged Higgs boson couplings with Standard Model fermions are given by
\begin{equation}
\begin{split}
\\[-2.5ex]
{\cal L}_{H^\pm f_\uparrow f_\downarrow} = \frac{g}{\sqrt{2} m_W} \sum_{(f_\uparrow , f_\downarrow) = (u,d), \, (\nu, \ell)}
H^+ \overline{f_\uparrow}
\left( m_{f_\uparrow} \, g_{H^+ \overline{f_\uparrow}f_\downarrow}^L \frac{1 - \gamma_5}{2}
+ m_{f_\downarrow} \, g_{H^+ \overline{f_\uparrow}f_\downarrow}^R \frac{1 + \gamma_5}{2} \right) f_\downarrow + {\rm h.c.} \, , \label{chargedfermionint} \\[1ex]
\end{split}
\end{equation}
where
\begin{equation}
\begin{split}
\\[-2.5ex]
&g_{H^+ \overline{f_\uparrow}f_\downarrow}^L = \cot\beta + i \eta \, , \qquad
g_{H^+ \overline{f_\uparrow}f_\downarrow}^R = \tan \beta - i \eta \, . \\[1ex]
\end{split}
\end{equation}


\subsection{Loop-induced wrong Higgs Yukawa coupling} \label{app:wronghiggsformula}

The explicit calculation of a stop/higgsino loop gives
\cite{Pilaftsis:2002fe}
\begin{equation}
\begin{split}
J_b \equiv \frac{y'_b}{y_b} \approx \frac{|y_t|^2}{16 \pi^2} 
\, A_t^\ast \mu^\ast  I (m_{\tilde{t}_1}^2, m_{\tilde{t}_2}^2, |\mu|^2) = \frac{|y_t|^2}{16 \pi^2} 
\, |A_t \mu| \, I (m_{\tilde{t}_1}^2, m_{\tilde{t}_2}^2, |\mu|^2) e^{- \iu \phi_t}\, , \\[1ex]
\end{split}
\end{equation}
where the loop function $I$ is
\begin{equation}
\begin{split}
I (p,q,r) = \frac{pq \log (p/q) + qr \log (q/r) + pr \log (r/p) }{(p-q) (q-r) (p-r)} \, . \\[1ex]
\end{split}
\end{equation}
Then, the bottom Yukawa coupling $y_b$ is related to the bottom quark mass as
\begin{equation}
\begin{split}
\\[-2.5ex]
y_b = \frac{g m_b}{\sqrt{2} m_W \cos \beta \left( 1 + J_b \tan \beta \right)} \, . \\[1ex]
\end{split}
\end{equation}
The physical Higgs boson couplings with a bottom quark are given by
\begin{equation}
\begin{split}
g_{H_i \bar{b}b}^S = -\frac{m_b}{v} \frac{1}{y_b \left( 1 + J_b \tan \beta \right)}
\biggl\{ &{\rm Re} (y_b) \, \frac{- s_\alpha \delta_{1i} + c_\alpha \delta_{2i}}{\cos \beta} 
+ {\rm Re} (y_b J_b) \, \frac{c_\alpha \delta_{1i} + s_\alpha \delta_{2i}}{\cos \beta} \\[1ex]
&- {\rm Im} (y_b) \, \delta_{3i} \tan \beta
+ {\rm Im} (y_b J_b) \, \delta_{3i} \biggl\} \, , \\[2ex]
g_{H_i \bar{b}b}^P = -\frac{m_b}{v} \frac{1}{y_b \left( 1 + J_b \tan \beta \right)}
\biggl\{ &-{\rm Im} (y_b) \, \frac{- s_\alpha \delta_{1i} + c_\alpha \delta_{2i}}{\cos \beta} 
- {\rm Im} (y_b J_b) \, \frac{c_\alpha \delta_{1i} + s_\alpha \delta_{2i}}{\cos \beta} \\[1ex]
&- {\rm Re} (y_b) \, \delta_{3i} \tan \beta
+ {\rm Re} (y_b J_b) \, \delta_{3i} \biggl\} \, . \label{MSSMfourfermi} \\[1ex]
\end{split}
\end{equation}
These results enter into the calculation of four-fermion operators that contribute to the measured EDMs.

\subsection{Stop couplings with Higgs bosons}\label{StopHiggs}

We here summarize the relevant stop interactions with the neutral Higgs bosons.
The trilinear couplings are given by
\begin{equation}
\begin{split}
\\[-2.5ex]
- \mathcal{L}_{H \tilde{t}^\ast \tilde{t}} \, = &\,\, \left( \kappa_u^L h_u + \kappa_d^L h_d \right) \tilde{t}^\ast_L \tilde{t}_L
+ \left( \kappa_u^R h_u + \kappa_d^R h_d \right) \tilde{t}^\ast_R \tilde{t}_R  \\[1.5ex]
&+ \left\{ \left( \tilde{\kappa}_u h_u + \tilde{\kappa}_d h_d + \iu \tilde{\kappa}_A A \right) \tilde{t}^\ast_R \tilde{t}_L + \rm h.c. \right\} \, ,
\end{split}
\end{equation}
where
\begin{equation}
\begin{split}
\\[-2.5ex]
&\kappa_u^L = \frac{2m_t^2}{s_\beta v} - \frac{2s_\beta m_Z^2}{v} \left( \frac{1}{2} - Q_t s_W^2 \right) \, , \qquad
\kappa_d^L = \frac{2c_\beta m_Z^2}{v} \left( \frac{1}{2} - Q_t s_W^2 \right) \, , \\[1.5ex]
&\kappa_u^R = \frac{2m_t^2}{s_\beta v} - \frac{2s_\beta m_Z^2}{v}  Q_t s_W^2 \, , \qquad
\kappa_d^R = \frac{2c_\beta m_Z^2}{v} Q_t s_W^2 \, , \\[1.5ex]
&\tilde{\kappa}_u = \left(  \frac{m_t}{s_\beta v} A_t
+ \frac{\epsilon_1^\ast}{\mu} m_t v \frac{c_\beta^2}{s_\beta} e^{-2\iu \theta} \right) e^{-\iu \delta_t} \, , \quad \tilde{\kappa}_d = \left( - \frac{m_t}{s_\beta v} \mu^\ast e^{-\iu \theta}
+ 2 \frac{\epsilon_1^\ast}{\mu} m_t v c_\beta e^{-2\iu \theta} \right) e^{-\iu \delta_t}
\, , \\[1.5ex]
&\tilde{\kappa}_A = \left\{ \left(  \frac{m_t}{s_\beta v} A_t
- \frac{\epsilon_1^\ast}{\mu} m_t v \frac{c_\beta^2}{s_\beta} e^{-2\iu \theta} \right) c_\beta
+ \left( \frac{m_t}{s_\beta v} \mu^\ast e^{-\iu \theta} - 2 \frac{\epsilon_1^\ast}{\mu} m_t v c_\beta e^{-2\iu \theta}  \right) s_\beta
\right\} e^{-\iu \delta_t} \, .  \\[1ex]
\end{split}
\end{equation}

In terms of mass eigenstates, the stop interactions with the Higgs bosons $H_i \, (i = 1,2,3)$
are written as a $2 \times 2$ matrix for the two stop eigenstates $\tilde{t}_a \, (a =1,2)$.
The diagonal part which are relevant for our calculations is given by
\begin{equation}
\begin{split}
\mathcal{L}_{H \tilde{t}^\ast \tilde{t}} &= 
\, \sum_{a=1}^2 \sum_{i=1}^3 \Gamma _{H_i {\tilde{t}_a}^\ast \tilde{t}_a} H_i {\tilde{t}_a}^\ast \tilde{t}_a +\cdots \\[1ex]
&\equiv \, \sum_{a=1}^2 \sum_{i=1}^3 \, \biggl[
\Gamma_{h_u {\tilde{t}_a}^\ast \tilde{t}_a} \left( c_\alpha O_{1i} + s_\alpha O_{2i} \right)  +\Gamma_{h_d {\tilde{t}_a}^\ast \tilde{t}_a} 
\left( -s_\alpha O_{1i} + c_\alpha O_{2i} \right) 
+ \Gamma_{A {\tilde{t}_a}^\ast \tilde{t}_a}  O_{3i} \biggr] H_i {\tilde{t}_a}^\ast \tilde{t}_a +\cdots 
\label{Higgsstopint} \, .
\end{split}
\end{equation}
The matrix $O$ diagonalizes the Higgs boson mass-squared matrix as presented in section \S\ref{naturalSUSY}. The coefficients are given by:
\begin{equation}
\begin{split}
\\[-2.5ex]
&\Gamma_{h_u \tilde{t}^\ast \tilde{t}} =
\begin{pmatrix}
s_{2t} {\rm Re} (\tilde{\kappa}_u) -  c_t^2 \kappa_u^L - s_t^2 \kappa_u^R
&\!\!\!\!\! - c_{2t} {\rm Re} (\tilde{\kappa}_u) -  s_t c_t \left( \kappa_u^L - \kappa_u^R \right) + \iu \, {\rm Im} (\tilde{\kappa}_u) \\[1ex]
- c_{2t} {\rm Re} (\tilde{\kappa}_u) -  s_t c_t \left( \kappa_u^L - \kappa_u^R \right) - \iu \, {\rm Im} (\tilde{\kappa}_u) &
- s_{2t} {\rm Re} (\tilde{\kappa}_u) -  s_t^2 \kappa_u^L - c_t^2 \kappa_u^R
\end{pmatrix} \, ,  \\[2ex]
&\Gamma_{h_d \tilde{t}^\ast \tilde{t}} =
\begin{pmatrix}
s_{2t} {\rm Re} (\tilde{\kappa}_d) -  c_t^2 \kappa_d^L - s_t^2 \kappa_d^R
&\!\!\!\!\! - c_{2t} {\rm Re} (\tilde{\kappa}_d) -  s_t c_t \left( \kappa_d^L - \kappa_d^R \right) + \iu \, {\rm Im} (\tilde{\kappa}_d) \\[1ex]
- c_{2t} {\rm Re} (\tilde{\kappa}_d) -  s_t c_t \left( \kappa_d^L - \kappa_d^R \right) - \iu \, {\rm Im} (\tilde{\kappa}_d) &
- s_{2t} {\rm Re} (\tilde{\kappa}_d) -  s_t^2 \kappa_d^L - c_t^2 \kappa_d^R
\end{pmatrix} \, ,  \\[2ex]
&\Gamma_{A \tilde{t}^\ast \tilde{t}} =
\begin{pmatrix}
- s_{2t} {\rm Im} (\tilde{\kappa}_A) & c_{2t} {\rm Im} (\tilde{\kappa}_A) + \iu \, {\rm Re} (\tilde{\kappa}_A) \\[1ex]
c_{2t} {\rm Im} (\tilde{\kappa}_A) - \iu \, {\rm Re} (\tilde{\kappa}_A)  & s_{2t} {\rm Im} (\tilde{\kappa}_A)
\end{pmatrix} \, . \\[1ex]
\end{split}
\end{equation}

\subsection{Chargino couplings with Higgs bosons}\label{CharginoHiggs}

In terms of the four-component spinor notation, $(\tilde{\chi}_j^-)^T = \left(
(\tilde{\chi}^-_j)_\alpha \,\, (\tilde{\chi}^+_j)^{\dagger \dot{\alpha}} \right)$ $(j = 1,2)$,
the chargino couplings with the Higgs bosons are in total presented as
\begin{equation}
\begin{split}
\\[-2.5ex]
\mathcal{L}_{H \tilde{\chi}^+\tilde{\chi}^-} = - \frac{g}{2 \sqrt{2}} \sum_{i=1}^3 H_i \sum_{j,k = 1}^2
 \left( a_{H_i \tilde{\chi}_j^+ \tilde{\chi}^-_k} \bar{\tilde{\chi}}_j^- \tilde{\chi}_k^-
+ \iu \, b_{H_i \tilde{\chi}_j^+ \tilde{\chi}^-_k} \bar{\tilde{\chi}}_j^- \gamma_5 \tilde{\chi}_k^- \right)  \, , \label{charginoHiggs} \\[1ex]
\end{split}
\end{equation}
where the coefficients are given by
\begin{align}
a_{H_i \tilde{\chi}_j^+ \tilde{\chi}^-_k} = &\,\, \left( - s_\alpha O_{1i} + c_\alpha O_{2i} \right) \left( e^{- \iu \theta}  {C^{R}_{2k}}^\ast C_{1j}^L
+ e^{\iu \theta}  {C^{R}_{2j}} {C_{1k}^L}^\ast \right)  
+\left(  c_\alpha O_{1i} + s_\alpha O_{2i} \right) \left( {C^{R}_{1k}}^\ast C_{2j}^L +  {C^{R}_{1j}} {C_{2k}^L}^\ast \right) \nonumber \\
&- \iu \, O_{3i}  \Bigl[ \sin \beta \left( e^{-\iu\theta} {C^{R}_{2k}}^\ast C_{1j}^L - e^{\iu\theta} {C^{R}_{2j}} {C_{1k}^L}^\ast \right)
+ \cos \beta \left( {C^{R}_{1k}}^\ast C_{2j}^L -  {C^{R}_{1j}} {C_{2k}^L}^\ast \right) \Bigr] \nonumber \\
&- \frac{\sqrt{2} v \sin \beta}{g} \left( - s_\alpha O_{1i} + c_\alpha O_{2i} \right) \left( \frac{\epsilon_1}{\mu^\ast} e^{\iu \theta} {C^{R}_{2k}}^\ast C_{2j}^L + \frac{\epsilon_1^\ast}{\mu} e^{-\iu \theta} {C^{R}_{2j}} {C_{2k}^L}^\ast \right)\nonumber \\
&- \frac{\sqrt{2} v \cos \beta}{g} \left( c_\alpha O_{1i} + s_\alpha O_{2i} \right) \left( \frac{\epsilon_1}{\mu^\ast} e^{\iu \theta}
{C^{R}_{2k}}^\ast C_{2j}^L + \frac{\epsilon_1^\ast}{\mu}e^{-\iu\theta} {C^{R}_{2j}} {C_{2k}^L}^\ast \right) \nonumber \\
&- \iu \frac{\sqrt{2} v\sin^2 \beta}{g} O_{3i} \left( \frac{\epsilon_1}{\mu^\ast} e^{\iu \theta} {C^{R}_{2k}}^\ast C_{2j}^L - \frac{\epsilon_1^\ast}{\mu} e^{-\iu \theta} {C^{R}_{2j}} {C_{2k}^L}^\ast \right)\nonumber \\
&- \iu \frac{\sqrt{2} v \cos^2 \beta}{g} O_{3i} \left( \frac{\epsilon_1}{\mu^\ast} e^{\iu \theta}
{C^{R}_{2k}}^\ast C_{2j}^L - \frac{\epsilon_1^\ast}{\mu}e^{-\iu\theta} {C^{R}_{2j}} {C_{2k}^L}^\ast \right) \, ,
\end{align}
\begin{align}
b_{H_i \tilde{\chi}_j^+ \tilde{\chi}^-_k} = &\,\, \iu \left( - s_\alpha O_{1i} + c_\alpha O_{2i} \right) \left( e^{- \iu \theta}  {C^{R}_{2k}}^\ast C_{1j}^L
- e^{\iu \theta}  {C^{R}_{2j}} {C_{1k}^L}^\ast \right)+\iu \left(  c_\alpha O_{1i} + s_\alpha O_{2i} \right) \left( {C^{R}_{1k}}^\ast C_{2j}^L -  {C^{R}_{1j}} {C_{2k}^L}^\ast \right) \nonumber \\
&+ \, O_{3i}  \Bigl[ \sin \beta \left( e^{-\iu\theta} {C^{R}_{2k}}^\ast C_{1j}^L + e^{\iu\theta} {C^{R}_{2j}} {C_{1k}^L}^\ast \right)
+ \cos \beta \left( {C^{R}_{1k}}^\ast C_{2j}^L +  {C^{R}_{1j}} {C_{2k}^L}^\ast \right) \Bigr] \nonumber \\
&- \iu \frac{\sqrt{2} v \sin \beta}{g} \left( - s_\alpha O_{1i} + c_\alpha O_{2i} \right) \left( \frac{\epsilon_1}{\mu^\ast} e^{\iu \theta} {C^{R}_{2k}}^\ast C_{2j}^L - \frac{\epsilon_1^\ast}{\mu} e^{-\iu \theta} {C^{R}_{2j}} {C_{2k}^L}^\ast \right) \nonumber \\
&- \iu \frac{\sqrt{2} v \cos \beta}{g} \left( c_\alpha O_{1i} + s_\alpha O_{2i} \right) \left( \frac{\epsilon_1}{\mu^\ast} e^{\iu \theta}
{C^{R}_{2k}}^\ast C_{2j}^L - \frac{\epsilon_1^\ast}{\mu}e^{-\iu\theta} {C^{R}_{2j}} {C_{2k}^L}^\ast \right) \nonumber \\
&+ \frac{\sqrt{2} v \sin^2 \beta}{g} O_{3i} \left( \frac{\epsilon_1}{\mu^\ast} e^{\iu \theta} {C^{R}_{2k}}^\ast C_{2j}^L + \frac{\epsilon_1^\ast}{\mu} e^{-\iu \theta} {C^{R}_{2j}}  {C_{2k}^L}^\ast \right) \nonumber \\
&+ \frac{\sqrt{2} v \cos^2 \beta}{g} O_{3i} \left( \frac{\epsilon_1}{\mu^\ast} e^{\iu \theta}
{C^{R}_{2k}}^\ast C_{2j}^L + \frac{\epsilon_1^\ast}{\mu}e^{-\iu\theta} {C^{R}_{2j}} {C_{2k}^L}^\ast \right)\, .
\end{align}

\subsection{$W$ boson couplings with charginos and neutralinos}\label{Wcharginoneutralino}

The interaction of a chargino, neutralino and $W$ boson is expressed as
\begin{equation}
\begin{split}
\\[-2.5ex]
\mathcal{L}_{\tilde{\chi}^0 \tilde{\chi}^\pm W^\mp} = - g \, \overline{\tilde{\chi}^+_i} \gamma^\mu
\left( G^L_{ij} \frac{1-\gamma_5}{2} + G^R_{ij} \frac{1+\gamma_5}{2}  \right) \tilde{\chi}_j^0 \, W^+_\mu + { \rm h.c.} \, , \\[1ex]
\end{split}
\end{equation}
where
\begin{equation}
\begin{split}
\\[-2.5ex]
G^L_{ij} = -  {C^L_{1i}}^\ast N_{2j} + \frac{1}{\sqrt{2}}  {C^L_{2i}}^\ast N_{4j} \, , \qquad
G^R_{ij} = -  {C^R_{1i}}^\ast N_{2j}^\ast - \frac{1}{\sqrt{2}}  {C^R_{2i}}^\ast N_{3j}^\ast \, . \\[1ex]
\end{split}
\end{equation}

\subsection{Charged Higgs couplings with charginos and neutralinos}\label{ChargedHiggs}

The charged Higgs couplings with charginos and neutralinos are 
\begin{equation}
\begin{split}
\\[-2.5ex]
\mathcal{L}_{H^\pm \tilde{\chi}^0\tilde{\chi}^\mp} = \frac{g}{\sqrt{2}} \sum_{i=1}^2 \sum_{j=1}^4 H^+
\overline{\tilde{\chi}_j^0}  \left( g^L_{H^+ \tilde{\chi}_j^0 \tilde{\chi}^-_i} \frac{1 - \gamma_5}{2}
+ g^R_{H^+ \tilde{\chi}_j^0 \tilde{\chi}^-_i} \frac{1 + \gamma_5}{2}  \right) \tilde{\chi}_i^- + {\rm h.c.} \, , \label{chargedHiggs} \\[1ex]
\end{split}
\end{equation}
where the coefficients are
\begin{equation}
\begin{split}
\\[-2.5ex]
g^L_{H^+ \tilde{\chi}_j^0 \tilde{\chi}^-_i} &= (s_\beta - \iu c_\beta \eta) \left( {C^R_{2i}}^\ast 
\left(  N_{2j} + N_{1j} t_W \right) - \sqrt{2} \, {C^R_{1i}}^\ast N_{3j}  \right)  
+ \frac{2\epsilon_1 v}{g \mu^\ast}  (c_\beta + \iu s_\beta \eta) \, {C^R_{2i}}^\ast
\left( s_\beta N_{3j} + e^{\iu \theta} c_\beta N_{4j} \right) \, , \\[1.5ex]
g^R_{H^+ \tilde{\chi}_j^0 \tilde{\chi}^-_i} &= - (c_\beta + \iu s_\beta \eta) \left( {C^L_{2i}}^\ast 
\left(  N_{2j}^\ast + N_{1j}^\ast t_W \right) + \sqrt{2} \, {C^L_{1i}}^\ast N_{4j}^\ast \right)+ \frac{2\epsilon_1^\ast v}{g \mu}  (s_\beta - \iu c_\beta \eta) \, {C^L_{2i}}^\ast
\left( s_\beta N_{3j}^\ast + e^{-\iu \theta} c_\beta N_{4j}^\ast \right) \, . \\[1ex]
\end{split}
\end{equation}

\section{Selected formulas for EDM contributions\label{sec:edmformulas}}

\subsection{The $W$ EDM contribution}

The chargino/neutralino loops coupled to the $W$ boson through the couplings in \S\ref{Wcharginoneutralino} contribute to the $W$ EDM which in turn induces fermion EDMs at two loops \cite{Kadoyoshi:1996bc,Giudice:2005rz} as shown in the bottom left diagram of Figure~\ref{fig:BarrZee}. The $W$ EDM contribution with chargino/neutralino loops is then given by
\begin{equation}
\begin{split}
\\[-2.5ex]
\frac{d_f}{e} \biggr|^{W \rm EDM}_{\tilde{\chi}^0 - \tilde{\chi}^\pm} = \frac{T^f_3 \alpha^2}{8 \pi^2 s_W^4}
\sum_{i =1}^2 \sum_{j = 1}^4 {\rm Im} \left( {G^L_{ij}} G^{R \ast}_{ij} \right) \frac{m_f m_{\tilde{\chi}_i} m_{\tilde{\chi}^0_j}}{m_W^4}
\int^1_0 \frac{dx}{1-x} J \left( 0, \frac{x r_{W \tilde{\chi}_i} + (1-x) r_{W \tilde{\chi}_j^0} }{x(1-x)} \right) \, . \\[1ex]
\end{split}
\end{equation}
Here, $r_{W \tilde{\chi}_i}  = (m_{\tilde{\chi}_i}/m_W)^2$, $r_{W \tilde{\chi}_j^0} = ( m_{\tilde{\chi_j}^0}/m_W )^2$, and the loop function $J(z, z')$ is
\begin{equation}
\begin{split}
\\[-2.5ex]
J(z, z') = \frac{1}{z - z'} \left( \frac{z \log z}{z-1} -  \frac{z' \log z'}{z'-1} \right) \, . \\[1ex]
\end{split}
\end{equation}

\subsection{The charged Higgs and top/bottom loop contributions}

The charged Higgs contribution to the EDM for a fermion $f$ with top/bottom quark loops is given by
\begin{equation}
\begin{split}
\\[-2.5ex]
\frac{d_f}{e} \biggr|_{H^\pm , \, tb} = \left( \frac{3 g^2}{32 \pi^2}  \right) \left( \frac{g^2}{32\pi^2 m_W} \right) 
\left( \frac{m_e}{m_W} \right) |V_{tb}|^2 \, {\rm Im}
\left( g_{H^+ \overline{f_\uparrow}f_\downarrow}^{L \, \ast} g_{H^+ \overline{f_\uparrow}f_\downarrow}^R \right)
\left( Q_t F_t + Q_b F_b \right) \, , \\[1ex]
\end{split}
\end{equation}
where $V_{tb}$ is a component of the Cabibbo-Kobayashi-Maskawa (CKM) matrix and the loop functions are
\begin{equation}
\begin{split}
\\[-2.5ex]
&F_t = \int^\infty_0 d Q^2 \int^1_0 dx \,\,  \frac{m_t^2 (1-x) (2-x) Q^2}{(m_{H^+}^2 + Q^2) (m_t^2 + x Q^2) (m_W^2 + Q^2)} \, , \\[2ex]
&F_b = \int^\infty_0 d Q^2 \int^1_0 dx \,\,  \frac{m_t^2 x (2-x) Q^2}{(m_{H^+}^2 + Q^2) (m_t^2 + x Q^2) (m_W^2 + Q^2)} \, . \\[1ex]
\end{split}
\end{equation}

\subsection{CEDM of the bottom quark} \label{app:CEDMbottomquark}

Here we present the calculation of the $b$-quark CEDM as shown in Figure~\ref{fig:OneLoopCEDM}. The generic interactions of a chargino or neutralino (denoted as $\chi$ collectively) with
a fermion $f$ and a sfermion $\tilde{f}'$ are given by
\begin{equation}
\begin{split}
\mathcal{L}_{\chi f \tilde{f}'} = g^L_{\chi f \tilde{f}'} \left( \bar{\chi} P_L f \right) \tilde{f}^{'\ast} 
+ g^R_{\chi f \tilde{f}'} \left( \bar{\chi} P_R f \right) \tilde{f}^{'\ast}  + {\rm h.c.}  \, ,  \\[1ex] 
\end{split}
\end{equation}
where $P_{L,R} = \frac{1\mp \gamma_5}{2}$ and the relevant couplings are 
\begin{equation}
\begin{split}
\\[-2.5ex]
&g^L_{\tilde{\chi}^\pm_i b \tilde{t}_1} = - g \, C^L_{1i} \cos \theta_t
- \frac{\sqrt{2} m_t}{v s_\beta} e^{-\iu \delta_t} C^L_{2i} \sin \theta_t \, , \\[2ex]
&g^L_{\tilde{\chi}^\pm_i b \tilde{t}_2} = - g \, C^L_{1i} \sin \theta_t
+ \frac{\sqrt{2} m_t}{v s_\beta} e^{-\iu \delta_t} C^L_{2i} \cos \theta_t \, ,  \\[2ex] 
&g^R_{\tilde{\chi}^\pm_i b \tilde{t}_1} = \frac{\sqrt{2} m_b}{v c_\beta} \, C^R_{2i} \cos \theta_t \, , \qquad
g^R_{\tilde{\chi}^\pm_i b \tilde{t}_2} = \frac{\sqrt{2} m_b}{v c_\beta} \, C^R_{2i} \sin \theta_t \, ,  \\[2ex] 
&g^L_{\tilde{\chi}^0_i b \tilde{b}_1} =  \frac{g}{\sqrt{2}}  \, N_{2i} - \sqrt{2} \left( Q_b + \frac{1}{2} \right)  N_{1i} \, , \qquad
g^R_{\tilde{\chi}^0_i b \tilde{b}_1} = - \frac{\sqrt{2} m_b}{v c_\beta} N^\ast_{3i} \, .  \\[2ex] 
\end{split}
\end{equation}
Then, the contributions to the CEDM of the bottom quark from chargino and neutralino loops are given by
\begin{equation}
\begin{split}
\\[-2.5ex]
&\tilde{d}_b \bigr|_{\tilde{\chi}^\pm} = \frac{1}{16 \pi^2} \sum_{i=1}^2 \sum_{a=1}^2 \frac{m_{\tilde{\chi}_i}}{m_{\tilde{t}_a}^2} 
{\rm Im} \left[ \left( g^R_{\tilde{\chi}^\pm_i b \tilde{t}_a} \right)^\ast g^L_{\tilde{\chi}^\pm_i b \tilde{t}_a} \right] 
B (m_{\tilde{\chi}_i}^2 / m_{\tilde{t}_a}^2 ) \, , \\[3ex]
&\tilde{d}_b \bigr|_{\tilde{\chi}^0} = \frac{1}{16 \pi^2} \sum_{i=1}^4 \frac{m_{\tilde{\chi}^0_i}}{m_{\tilde{b}_1}^2} 
{\rm Im} \left[ \left( g^R_{\tilde{\chi}^0_i b \tilde{b}_1} \right)^\ast g^L_{\tilde{\chi}^0_i b \tilde{b}_1} \right] 
B (m_{\tilde{\chi}^0_i}^2 / m_{\tilde{b}_1}^2 )  \, , \\[1ex] 
\end{split}
\end{equation}
where
\begin{equation}
\begin{split}
\\[-2.5ex]
&B(z) = \frac{1}{2(1-z)^2} \left( 1 + z + \frac{2z\log z}{1-z} \right) \, . \\[1ex] 
\end{split}
\end{equation}

\section{Likelihood analysis of the mercury EDM constraint \label{sec:mercuryuncertainty}}

\begin{figure}[!t]
\hspace{0cm}
 \begin{minipage}{0.33\hsize}
  \begin{center}
   \includegraphics[clip, width=7cm]{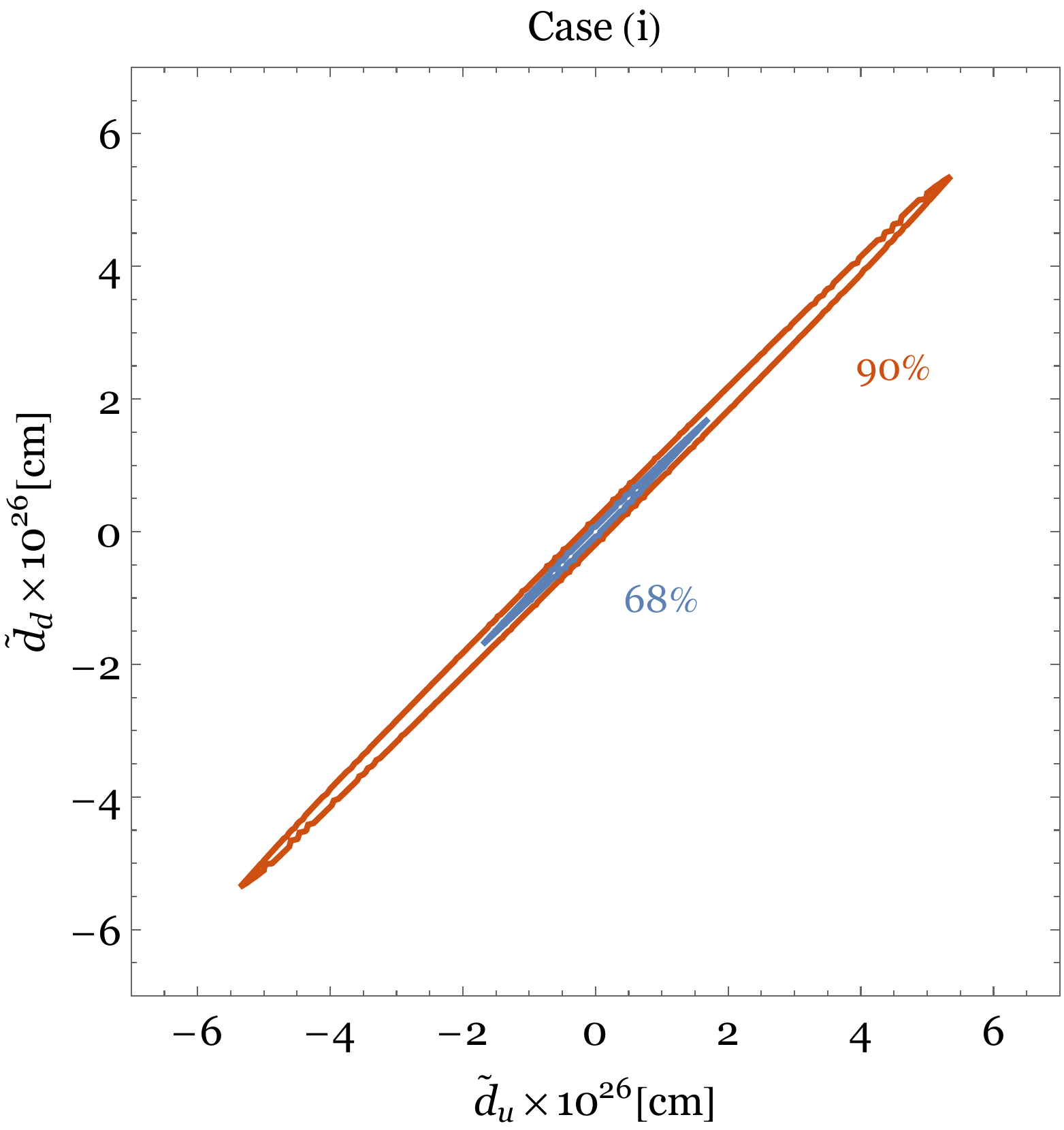}
  \end{center}
 \end{minipage}
\hspace{2.5cm}
\begin{minipage}{0.33\hsize}
  \begin{center}
   \includegraphics[clip, width=7cm]{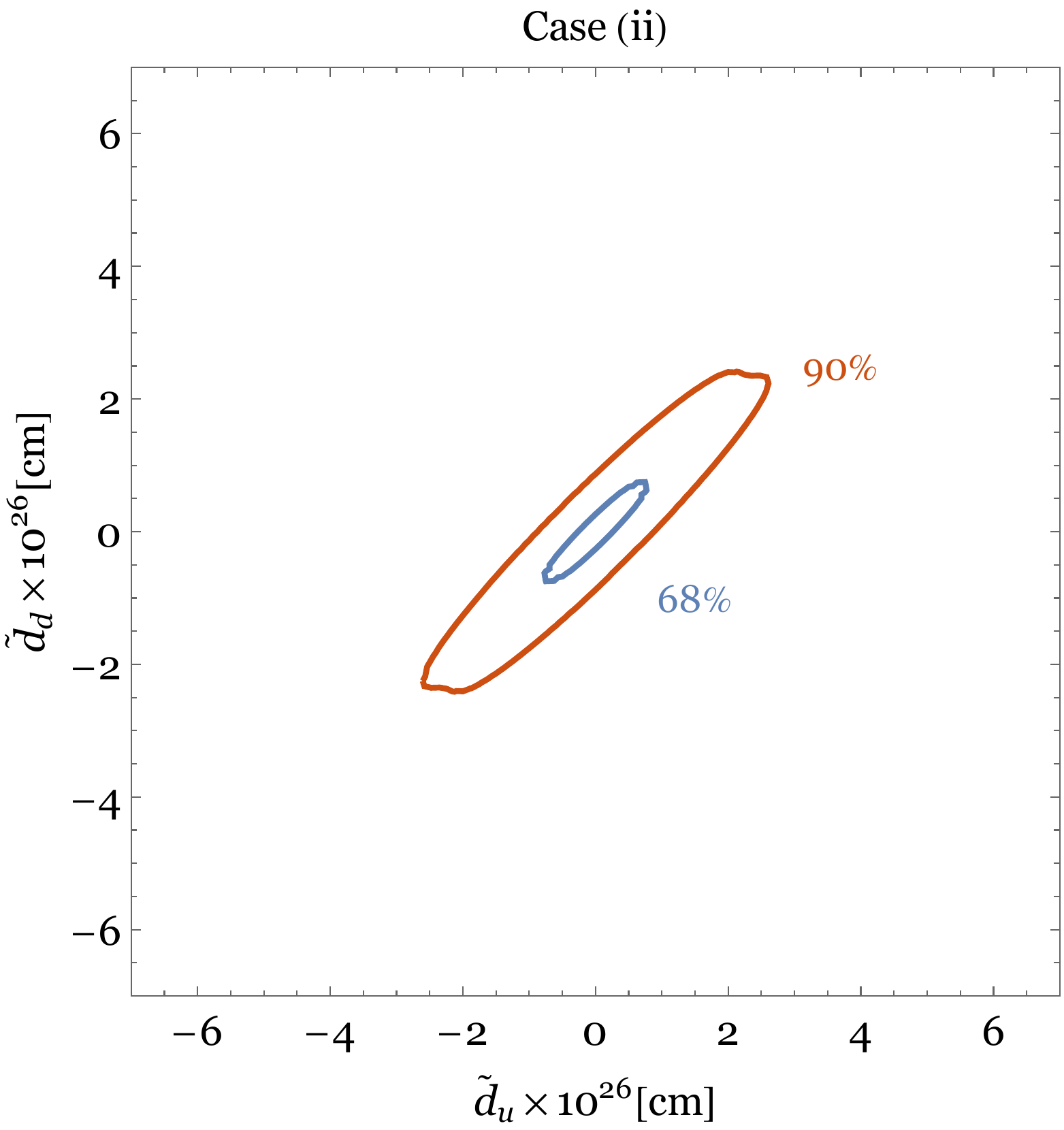}
  \end{center}
 \end{minipage}
\vspace{0.5cm}
\caption{The mercury EDM constraint on the quark CEDMs.
The left and right panels correspond to the case (i) and (ii).
The outer regions of the contours are excluded.
The blue and orange contours denote
the constraints at 68\% and 90\% C.L. respectively.}
\label{fig:mercuryconst}
\end{figure}%

The mercury EDM suffers from large theoretical uncertainty and it may not be appropriate to consider the mercury EDM constraint
in terms of the central value for each contribution quoted from the literature in \eqref{mercuryEDM}.
In this appendix, we present a likelihood analysis of the constraint as the best effort
to extract implications on physics beyond the standard model from the mercury EDM measurement.
Here, we only consider the most important contributions from the quark CEDMs
and derive a constraint on the CEDMs at some probability.
For the theoretical calculation of the mercury EDM, we follow the discussion of ref.\cite{Jung:2013hka}.

The mercury EDM is related to the Schiff moment $S$ as $d_{\rm Hg} = -2.46 \times 10^{-17} e \, {\rm cm} \times
\frac{S}{e \, {\rm fm^3}}$ and 
the Schiff moment can be parametrized in terms of the CP-odd pion-nucleon interactions,
$\mathcal{L}_{\pi NN} \supset \bar{g}^{(0)} \bar{N} \tau^a N \pi^a + \bar{g}^{(1)} \bar{N} N \pi^0$,
as
\begin{equation}
\begin{split}
S \approx 13.17 \times
\left[ (a_0 + b) \, \bar{g}^{(0)}  + a_1 \, \bar{g}^{(1)}   \right] , 
\end{split}
\end{equation}
where the uncertainty of the overall coefficient is negligible.
The coefficients $a_0, a_1, b$ are determined by nuclear calculations.
So far, several groups have presented results which span a wide range of values.
Here we use the results of two recent calculations,
\begin{equation}
\begin{split}
{\rm (i)} \qquad &a_0 = ( 0.002 - 0.010 ) \, {e \, {\rm fm^3}} \, , \qquad a_1 = (0.057 - 0.090) \, {e \, {\rm fm^3}}
\quad \quad (\!\text{\cite{deJesus:2005nb}}), \\[1ex]
{\rm (ii)} \qquad &a_0 = (0.009 - 0.041) \, {e \, {\rm fm^3}} \, , \qquad a_1 = (-0.027 - 0.005) \, {e \, {\rm fm^3}} 
\quad \,\, (\!\text{\cite{Ban:2010ea}}),
\label{a0a1b}
\end{split}
\end{equation}
and $b = (0.002 - 0.013) \, {e \, {\rm fm^3}}$ for both cases. Notice that the two results are not very compatible with each other, so it seems inappropriate to simply average them, as is sometimes done in the literature. Rather, the discrepancy suggests that at least one of the approximation schemes used is getting the physics wrong in a systematic way. Further theoretical work on the structure of the mercury nucleus will be needed to clarify the situation.

We next relate the CP-odd pion-nucleon interactions to the quark CEDMs by using the QCD sum rule technique.
We follow the calculation of ref.~\cite{Pospelov:2001ys} where $\bar{g}^{(0)}, \bar{g}^{(1)}$ are given in terms of the quark CEDMs
$\tilde{d}_{u,d}$ and some undetermined parameters such as a the choice of generalized nucleon interpolating current
and the infrared cutoff.
We parametrize the CP-odd pion-nucleon interactions in terms of the quark CEDMs as
\begin{equation}
\begin{split}
\\[-2.5ex]
&\bar{g}^{(0)} = {c}^{(0)} \times 10^{-12} \, \frac{\tilde{d}_u + \tilde{d}_d}{10^{-26} \, \rm cm}
\frac{|\langle \bar{q} q \rangle |}{(225 \, \rm MeV)^3}, \\[1ex] 
&\bar{g}^{(1)} = {c}^{(1)} \times 10^{-12} \, \frac{\tilde{d}_u - \tilde{d}_d}{10^{-26} \, \rm cm}
\frac{|\langle \bar{q} q \rangle |}{(225 \, \rm MeV)^3}. \\[1ex] 
\end{split}
\end{equation}
To estimate the numerical coefficients $c^{(0)}, c^{(1)}$, we assume flat probability profiles for the undetermined parameters
with a certain range presented in ref.~\cite{Pospelov:2001ys}
(with the Borel parameter fixed at $M^2 = 0.8 \, \rm GeV^2$) and investigate the distributions of the coefficients $c^{(0)}, c^{(1)}$.
The means of $c^{(0)}, c^{(1)}$ are $(0.47, 1.48)$ and the standard deviations are given by $(0.37, 8.21)$.

Let us now derive a constraint on the quark CEDMs by a likelihood analysis.
We use the above distributions of the coefficients $c^{(0)}, c^{(1)}$
and further assume flat probability profiles for the coefficients $a_0, a_1, b$ within the range denoted in \eqref{a0a1b}
to obtain the distribution of the coefficients of $\tilde{d}_u$, $\tilde{d}_d$ in the expression of the mercury EDM
$d_{\rm H_g} = c_u \tilde{d}_u + c_d \tilde{d}_d$.
Then, we fit the distribution by a two-dimensional normal distribution.
With appropriate normalization, we find the probability distribution $P_{\rm theory} (c_u, c_d)$.
The case (i) in \eqref{a0a1b} shows a stronger correlation between $c_u$ and $c_d$ than the case (ii) because the range of values in $a_0$ is smaller than that in $a_1$ for the case (i) while $a_0$ and $a_1$ are comparable
for the case (ii). This fact is reflected to the final constraint.
We define the likelihood function of $\tilde{d}_u$, $\tilde{d}_d$ as 
\begin{equation}
\begin{split}
\\[-2.5ex]
L (\tilde{d}_u , \tilde{d}_d) = \int_{-\infty}^{\infty} d c_u \int_{-\infty}^{\infty} d c_d \,
P_{\rm exp} ({\rm data} | d_{\rm H_g} = c_u \tilde{d}_u + c_d \tilde{d}_d)
\times P_{\rm theory} (c_u, c_d), \\[1ex] 
\end{split}
\end{equation}
where $P_{\rm exp} ({\rm data} | d_{\rm H_g})$ comes from the present constraint of the mercury EDM
\eqref{mercuryEDMupperbound} at $2 \sigma$.
Then, we evaluate the delta log-likelihood $\Delta \ln L \equiv \ln L (0 , 0) - \ln L (\tilde{d}_u , \tilde{d}_d)$.

Figure~\ref{fig:mercuryconst} shows the mercury EDM constraint on the quark CEDMs.
The left and right panels correspond to the case (i) and (ii) respectively.
The outer regions of the contours are excluded.
The blue and orange contours denote
the constraints at 68\% and 90\% C.L. respectively.
Since $a_0$ and $a_1$ are comparable for the case (ii) and
the down quark CEDM contribution to the mercury EDM is canceled between the terms of $\bar{g}^{(0)}$ and $\bar{g}^{(1)}$,
the constraint on $\tilde{d}_d$ for the case (ii) is milder than that for the case (i).

\small

\bibliography{ref}
\bibliographystyle{utphys}
\end{document}